\documentclass[useAMS,usenatbib,usegraphicx]{mn2e}

\usepackage{aas_macros}

\usepackage{amsmath}
\usepackage{amssymb}

\usepackage{mathbbol}

\usepackage{dsfont}

\title[Dwarf Galaxy Planes in the Local Group]{Dwarf Galaxy Planes: the discovery of symmetric structures in the Local Group}
\author[Pawlowski, Kroupa \& Jerjen]{Marcel S. Pawlowski$^{1}$\thanks{E-mail:
mpawlow@astro.uni-bonn.de}, Pavel Kroupa$^{1}$, Helmut Jerjen$^{2}$ \\
$^{1}$Argelander Institute for Astronomy, University of Bonn, Auf dem H\"{u}gel 71, D-53121 Bonn, Germany\\
$^{2}$Research School of Astronomy and Astrophysics, Australian National University, Mt Stromlo Observatory,\\ Cotter Rd., Weston ACT 2611, Australia
}
\begin{document}

\date{Accepted 2013 July 23.  Received 2013 July 23; in original form 2013 May 14}

\pagerange{\pageref{firstpage}--\pageref{lastpage}} \pubyear{2012}

\maketitle

\label{firstpage}

\begin{abstract}
Both major galaxies in the Local Group (LG) are surrounded by thin planes of mostly co-orbiting satellite galaxies, the vast polar structure (VPOS) around the Milky Way (MW) and the Great Plane of Andromeda (GPoA) around M31. 
We summarize the current knowledge concerning these structures and compare their relative orientations by re-determining their properties in a common coordinate system. 
The existence of similar, coherent structures around both major LG galaxies motivates an investigation of the distribution of the more distant non-satellite galaxies in the LG. This results in the discovery of two planes (diameters of 1--2\,Mpc) which contain almost all nearby non-satellite galaxies. The two LG planes are surprisingly symmetric. They are inclined by only 20 degrees relative to the galactic disc of M31, are similarly thin (heights of $\approx 60$\,kpc) and have near-to-identical offsets from the MW and from M31. They are inclined relative to each other by 35 degrees.
Comparing the plane orientations with each other and with additional features reveals indications for an intimate connection between the VPOS and the GPoA. They are both polar with respect to the MW, have similar orbital directions and are inclined by about $45 \pm 7$\ degrees relative to each other. The Magellanic Stream approximately aligns with the VPOS and the GPoA, but also shares its projected position and line-of-sight velocity trend with a part of the dominating structure of non-satellite dwarf galaxies. In addition, the recent proper motion measurement of M31 indicates a prograde orbit of the MW-M31 system, the VPOS and the GPoA. The alignment with other features such as the Supergalactic Plane and the over-density in hypervelocity stars are discussed as well. We end with a short summary of the currently proposed scenarios trying to explain the LG galaxy structures as either originating from cosmological structures or from tidal debris of a past galaxy encounter. We emphasise that there currently exists no full detailed model which satisfactorily explains the existence of the thin symmetric LG planes.
\end{abstract}


\section{Introduction}
\label{sect:intro}

\subsection{The Milky Way satellite system}

Signs for the existence of what now is called the vast polar structure (VPOS) of satellite object around the MW were first reported by \citet{LyndenBell1976} and \citet{Kunkel1976}. \citet{LyndenBell1976} discovered that a number of MW satellite galaxies (Draco, LMC, SMC, Ursa Minor), globular clusters (Palomar 1 and 14) and streams align with the orbital plane of the Magellanic Clouds.
\citet{LyndenBell1982} identified a second possible group consisting of the dwarf spheroidal (dSph) galaxies Fornax, Leo I, Leo II and Sculptor, which all lie along a common great-circle as seen from the MW centre, and termed this second association the Fornax-Leo-Sculptor (FLS) stream. 

These early discoveries of coherent structures, made before the advent of the currently prevailing cosmological model based on dark energy and cold dark matter ($\Lambda$CDM), were thought to be related to planes of tidal debris from which new stellar systems formed. Thus, initially the planar structures were seen as an indication that many of the MW satellites are what today are termed tidal dwarf galaxies (TDGs, e.g. \citealt{Barnes1992}, \citealt*{Elmegreen1993}, \citealt{Duc1998}, \citealt*{Wetzstein2007}, \citealt{Bournaud2010}, \citealt{Duc2011}). Under this assumption, \citet{Kroupa1997} modelled dark matter free MW satellite galaxies. One of these models, when compared to the Hercules satellite galaxy discovered later, turned out to be one of the few successful predictions concerning satellite galaxies \citep{Kroupa2010}.

However, later studies have focussed on comparing the MW satellite galaxies to the expectations for primordial dwarf galaxies residing in dark matter sub-haloes, revealing a number of unsuccessful predictions of the $\Lambda$CDM model.These include the predicted central dark matter peak \citep[core/cusp problem][]{Dubinski1991}, the large predicted total number of satellites \citep[missing satellite problem, ][]{Klypin1999,Moore1999}, the predicted existence of very concentrated, massive satellites \citep[missing bright satellites or too big to fail problem, ][]{Bovill2011,BoylanKolchin2011}, and the predicted internal dark matter distribution \citep{Strigari2008}.

\citet*{Kroupa2005} analysed the spatial distribution of the 11 brightest, 'classical' MW satellites and compared it with the expected distribution derived from $\Lambda$CDM models. They found that all MW satellites reside within a single thin plane, and that this distribution is inconsistent with a near-isotropic one expected for cosmological dark matter sub-haloes. From this they concluded that the MW satellite galaxies could be better understood as tidal dwarf galaxies with a common origin in a galaxy interaction instead of dark matter dominated primordial dwarf galaxies of cosmological origin, therefore returning the discussion back to the TDG scenario. More sophisticated studies of the spatial distribution of MW satellite galaxies, including the fainter MW satellites discovered in the meantime, confirmed the existence of the planar distribution (\citealt*{Metz2007}, \citealt*{Metz2009}, \citealt{Kroupa2010}, \citealt*{Pawlowski2012a}).

With increasing evidence that the satellite planarity is indeed significant \citep{Metz2007} and that the satellites preferentially orbit within the plane (\citealt*{Metz2008}, \citealt{Pawlowski2013}), the attention has shifted towards identifying possible mechanisms which could give rise to flattened and even coherently orbiting sub-halo populations within a $\Lambda$CDM cosmology. One suggested mechanism is the accretion of primordial dwarf galaxies in groups \citep{LiHelmi2008,DOnghiaLake2008,Deason2011}, such that the galaxies would then orbit their host in a common direction. \citet{Metz2009b} have refuted these claims by demonstrating that observed dwarf galaxy associations are much too extended to be able to form thin VPOS-like planes. In addition, as almost all MW satellites lie close to the same plane they would have had to be accreted as a single group. However, \citet*{Wang2013} demonstrate that the majority of the 11 most-massive (in stellar mass) satellites in their high-resolution $\Lambda$CDM simulations must have been accreted individually. Based on the dynamical friction time scale of the MW satellite galaxies, \citet*{Angus2011} also argue that a recent accretion of the satellites does not work out.

Another attempt to reconcile the planar MW satellite distribution with expectations from cosmological models was the suggestion that luminous sub-haloes are accreted along dark matter filaments \citep{Libeskind2005,Libeskind2010,Lovell2011}. This claim is disputed by \citet{Pawlowski2012b}, who demonstrate that the distribution of orbital poles of sub-haloes that have been accreted onto a host via filaments does not reproduce quasi-planar distributions with a coherent rotation comparable to that of the VPOS. One of the underlying inconsistencies of the filamentary accretion scenario is that cosmological dark matter filaments are more extended than the virial radius of the host galaxy dark matter halo \citep{Vera-Ciro2011}, such that they are orders of magnitude too wide to be responsible for the formation of a structure only a few 10 kpc thin, a fact that has already been pointed out by \citet{Kroupa2005}. Nevertheless, dark matter filaments have prevailed as a frequently mentioned mechanism related to the formation of thin planes of satellite galaxies (\citealt*{Keller2012}, \citealt{Tully2013}, \citealt{Wang2013}).

Recent studies populating simulated dark matter sub-haloes with luminous satellites via semi-analytic galaxy formation models demonstrate that the positional flattening of the MW satellite system is unlikely in a $\Lambda$CDM context \citep{Starkenburg2013,Wang2013}. These studies over-estimate the agreement of the simulated results with the observed situation. One reason is that they only consider the flattening in the satellite positions, but not the comparable flattening in velocity space \citep{Metz2008,Fouquet2012,Pawlowski2013}, even though the sub-halo velocities can be extracted from the simulations. Furthermore, these studies only consider the distribution of the brightest satellites, but the less luminous ones follow the same polar structure \citep{Metz2009,Kroupa2010}.

In addition to the MW satellite galaxies, those globular clusters classified as young halo objects \citep{Mackey05} follow a planar distribution which aligns with that of the satellite galaxies \citep{Keller2012,Pawlowski2012a}. The preferred alignment of stellar and gaseous streams within the MW halo with the VPOS provides additional evidence that many of the MW satellite objects orbit within the VPOS \citep{Pawlowski2012a}.

\subsection{The M31 satellite system}

The importance of VPOS-like satellite galaxy planes has been stressed by \citet{Wang2013}, who write: '[a] larger sample of satellites around other galaxies will test the tidal formation hypothesis of Pawlowski et al. (2012) in which highly flattened configurations are easily achieved and should therefore be the norm. If, on the other hand, the CDM model is a realistic description of nature, then the average satellite configurations should be only moderately flattened [...]'.
Such an additional sample of satellites around another galaxy is the nearby M31 system. Several early searches for a preferred planar distribution of M31 satellites have been carried out \citep{McConnachie2006,Koch2006,Metz2007}. These initial studies were hampered by the small number of known satellite galaxies but have already identified possible planar structures. 

An analysis of the spatial distribution of satellite galaxies should be easier for our neighbouring galaxy than for the MW. In contrast to the searches for satellites around the MW not the entire sky must be surveyed for M31 satellite galaxies but only the region towards M31. The Pan-Andromeda Archaeological Survey \citep[PAndAS, ][]{McConnachie2009} is such a survey covering an area of 150 kpc radius around M31 in projection. It has resulted in the discovery of numerous M31 satellite galaxies. In addition to the positions of galaxies on the sky, their distances have to be known in order to make the discovery of structures in their full spatial distribution possible. One of the first large catalogues of distances to Local Group dwarf galaxies, measured with the tip of the red giant branch method, was provided by \citet{McConnachie2005}. Recently, accurate distances to the M31 satellites were determined in a homogeneous way by \citet{Conn2011,Conn2012}. The resulting dataset of M31 satellite galaxies with coordinates and distances allowed a detailed analysis of their spatial distribution \citep{Ibata2013,Conn2013}. This resulted in the discovery of a vast thin plane of satellites around M31 at very high statistical significance \citep{Conn2013}. About half of M31's satellite galaxies can be associated with this structure, which is seen edge-on from the MW. The line-of-sight velocities of the satellites in the structure indicate that most of them follow a common orbital sense \citep{Ibata2013}. Like the VPOS, the satellite plane around M31 is rotating around its host galaxy.

\subsection{The satellite planes and the Local Group}

In any scenario addressing the formation of thin planes of satellite galaxies, a causal connection of the planes to a larger-scale structure requires that structure to have a similarly narrow spatial extend as the planes. The distribution of the galaxies within the LG has been investigated in search for a preferred plane in several studies  \citep{Hartwick2000,Sawa2005,Pasetto2007}. They have in common that a preferred planar direction is generally found. However, they either focus on the overall distribution of all (isolated) LG galaxies, resulting in structures with a thickness of several 100 kpc \citep{Hartwick2000,Pasetto2007}, or they define a thin plane by visual inspection of the galaxy distribution only, resulting in many distant LG galaxies being outliers from the reported plane \citep{Sawa2005}.

The aim of the current study is to compile our present knowledge of the satellite galaxy planes in the LG. We do this by re-determining the plane parameters in a common coordinate system in order to facilitate and encourage the comparison with other structures. We furthermore discuss the distribution of non-satellite galaxies in the LG in the context of the VPOS and GPoA and suggest the existence of two very symmetric planes of dwarf galaxies in the LG. At least one of these is essentially connecting the MW and M31 in position and velocity space. Almost all presently known nearby galaxies in the LG can be associated to one of several planes with heights of only a few 10 kpc each.
We discuss the relative orientations of all planes and compare them with a number of prominent features, including the Magellanic Stream, the orbital pole of the MW-M31 system deduced from the recent proper motion measurement for M31 and the hypervelocity star over-density in the MW halo. These comparisons indicate that many of these features might be intimately related. 

Sect. \ref{sect:methods} presents the dataset used in the following analysis and describes the employed methods. In Sects. \ref{sect:MW} and \ref{sect:M31} we re-analyse the planes found in the satellite galaxy distributions around the MW and M31, respectively. In Sect. \ref{sect:LG} we expand upon the previous works and investigate the distribution of known non-satellite dwarf galaxies in the LG, which reveals two symmetric planes of galaxies. Sect. \ref{sect:other} discusses the remaining dwarf galaxies not associated with any of the planes and briefly mentions a possible second preferred plane of M31 satellites. The results are discussed in Sect. \ref{sect:discussion}, in particular analysing the (mutual) orientations of the found planar galaxy structures and their relation to other features. Possible avenues to be explored in order to find an explanation for the structured LG dwarf galaxy populations are then discussed in Sect. \ref{sect:origins}, and finally the conclusions are given in Sect. \ref{sect:conclusions}.

\section{The dataset and methods}
\label{sect:methods}

Table \ref{tab:symbols} compiles descriptions of frequently used symbols which will be introduces in the following section.

\begin{table}
 \caption{Symbol definitions}
 \label{tab:symbols}
 \begin{center}
 \begin{tabular}{@{}ll}
  \hline
  Symbol & Description \\
  \hline
$d_{\mathrm{MW}}$ & Distance of a galaxy from the centre of the MW \\
$d_{\mathrm{M31}}$ & Distance of a galaxy from the centre of M31 \\
$d_{\mathrm{LG}}$ & Distance from midpoint between MW and M31 \\
$\mathbf{r}_{0}$ & Centroid of a galaxy sample \\
$\mathbf{n}$ & Normal direction to a plane-fit, expressed in Galactic\\
             & Longitude $l$\ and Latitude $b$ \\
$D_{\mathrm{MW}}$ & Offset (minimum distance) of a plane from the MW \\
$D_{\mathrm{M31}}$ & Offset (minimum distance) of a plane from M31 \\
$\Delta$ & RMS height of galaxy sample from its best-fitting plane \\
$c/a$ & Ratio of short to long axis of a galaxy sample \\
$b/a$ & Ratio of intermediate to long axis of a galaxy sample \\
\hline
 \end{tabular}
 \end{center}
 \small \medskip
Symbols (first column) frequently used in this paper, their description (second column).
\end{table}

\subsection{The Local Group Galaxy Dataset}
\label{subsect:dataset}

\begin{table}
 \tiny
 \caption{Positions of galaxies in the Local Group}
 \label{tab:galpos}
 \begin{center}
 \begin{tabular}{@{}lccccc}
 \hline 
Name &  $x$ &  $y$ &  $z$ & $- \Delta r_{\mathrm{Sun}}$ & $+ \Delta r_{\mathrm{Sun}}$ \\
 & [kpc] & [kpc] & [kpc] & [kpc] & [kpc] \\
 \hline

The Galaxy &  193 & -312 &  144 &    0 &    0\\ 
Canis Major &  181 & -318 &  143 &    1 &    1\\ 
Sagittarius dSph &  210 & -309 &  138 &    2 &    2\\ 
Segue (I) &  173 & -321 &  162 &    2 &    2\\ 
Ursa Major II &  162 & -300 &  163 &    4 &    5\\ 
Bootes II &  199 & -313 &  183 &    1 &    1\\ 
Segue II &  161 & -298 &  123 &    2 &    2\\ 
Willman 1 &  165 & -304 &  176 &    6 &    8\\ 
Coma Berenices &  182 & -316 &  187 &    4 &    4\\ 
Bootes III &  194 & -305 &  189 &    2 &    2\\ 
LMC &  192 & -353 &  117 &    2 &    2\\ 
SMC &  209 & -350 &   99 &    3 &    4\\ 
Bootes (I) &  207 & -312 &  206 &    2 &    2\\ 
Draco &  188 & -249 &  187 &    6 &    6\\ 
Ursa Minor &  170 & -260 &  197 &    3 &    4\\ 
Sculptor &  187 & -321 &   59 &    5 &    6\\ 
Sextans (I) &  156 & -369 &  202 &    4 &    4\\ 
Ursa Major (I) &  132 & -292 &  223 &    4 &    5\\ 
Carina &  168 & -408 &  104 &    6 &    6\\ 
Hercules &  277 & -261 &  223 &   12 &   13\\ 
Fornax &  151 & -363 &   10 &   12 &   13\\ 
Leo IV &  178 & -396 &  273 &    6 &    7\\ 
Canes Venatici II &  176 & -293 &  303 &    4 &    4\\ 
Leo V &  171 & -404 &  296 &   10 &   10\\ 
Pisces II &  208 & -190 &   11 &    0 &    0\\ 
Canes Venatici (I) &  195 & -275 &  358 &   10 &   10\\ 
Leo II &  115 & -370 &  359 &   14 &   14\\ 
Leo I &   69 & -431 &  336 &   15 &   16\\ 
Andromeda & -193 &  312 & -144 &   25 &   26\\ 
M32 & -202 &  328 & -157 &   74 &   82\\ 
Andromeda IX & -211 &  291 & -114 &   24 &   25\\ 
NGC 205 & -208 &  349 & -153 &   26 &   27\\ 
Andromeda I & -171 &  264 & -169 &   24 &   24\\ 
Andromeda XVII & -163 &  285 &  -87 &   26 &   38\\ 
Andromeda XXVII & -215 &  370 & -104 &   45 &   47\\ 
Andromeda III & -145 &  273 & -187 &   24 &   25\\ 
Andromeda XXV & -197 &  371 &  -78 &   44 &   46\\ 
Andromeda XXVI & -163 &  338 &  -49 &   41 &   43\\ 
Andromeda V & -256 &  290 &  -58 &   28 &   29\\ 
Andromeda XI & -153 &  234 & -213 &   17 &   17\\ 
Andromeda XIX & -130 &  345 & -233 &  147 &   31\\ 
Andromeda XXIII & -278 &  220 & -163 &   45 &   47\\ 
Andromeda XX &  -73 &  298 & -191 &   53 &   42\\ 
Andromeda XIII & -212 &  299 & -274 &   19 &   20\\ 
Andromeda X & -188 &  205 &  -63 &   39 &   25\\ 
Andromeda XXI & -108 &  414 & -128 &   26 &   23\\ 
Andromeda XXXII & -202 &  344 &   -7 &   48 &   52\\ 
NGC 147 & -142 &  257 &  -22 &   27 &   29\\ 
Andromeda XXX & -153 &  261 &  -11 &   77 &   32\\ 
Andromeda XIV & -177 &  246 & -291 &  181 &   22\\ 
Andromeda XII & -248 &  381 & -299 &  135 &   39\\ 
Andromeda XV & -165 &  137 & -116 &   34 &   80\\ 
Andromeda II & -173 &  131 & -173 &   18 &   18\\ 
NGC 185 & -121 &  201 &  -10 &   25 &   26\\ 
Andromeda XXIX &  -28 &  279 & -230 &   70 &   78\\ 
Triangulum & -292 &  189 & -277 &   22 &   23\\ 
Andromeda XXIV & -169 &  143 &  -24 &   32 &   34\\ 
Andromeda VII &  -66 &  396 &   12 &   34 &   36\\ 
IC 10 & -200 &  382 &   98 &   43 &   45\\ 
Andromeda XXXI &   44 &  401 &  -74 &   41 &   43\\ 
LGS 3 & -164 &  154 & -359 &   24 &   25\\ 
Andromeda VI &   10 &  295 & -320 &   25 &   26\\ 
Andromeda XXII & -331 &  249 & -372 &  141 &   30\\ 
Andromeda XVI &  -51 &   25 &  -98 &   30 &   44\\ 
Andromeda XXVIII &  173 &  297 & -113 &   58 &  171\\ 
IC 1613 &  -53 &  -26 & -514 &   41 &   43\\ 
Phoenix &  190 & -461 & -243 &   19 &   20\\ 
NGC 6822 &  578 & -125 &   -1 &   17 &   17\\ 
Cetus &  140 &  -94 & -578 &   24 &   25\\ 
Pegasus dIrr &  129 &  353 & -490 &   29 &   30\\ 
Leo T &  -63 & -484 &  432 &   19 &   20\\ 
WLM &  249 &  -57 & -751 &   34 &   35\\ 
Andromeda XVIII & -286 &  750 & -209 &   44 &   40\\ 
Leo A & -281 & -453 &  776 &   43 &   45\\ 
Aquarius &  943 &  201 & -413 &   39 &   40\\ 
Tucana &  664 & -674 & -509 &   48 &   50\\ 
Sagittarius dIrr & 1140 &   56 & -155 &   85 &   92\\ 
UGC 4879 & -778 &  -48 & 1071 &   25 &   25\\ 

 \hline
 \end{tabular}
 \end{center}
 \small \medskip
Positions of the LG galaxies in $x$-, $y$- and $z$-coordinates of the Cartesian coordinate system defined in Sect. \ref{subsect:dataset}. The position uncertainties are along the line connecting the Sun (situated at $\mathbf{r}_{\mathrm{Sun}} = [184,-312, 144]$~kpc) and the respective galaxy. They are given as $- \Delta r_{\mathrm{Sun}}$, denoting the 1$\sigma$\ radial distance uncertainty towards the Sun, and $+ \Delta r_{\mathrm{Sun}}$, denoting the 1$\sigma$\ radial distance uncertainty away from the Sun.
\end{table}

The analysis presented in the following is based on the catalogue of nearby galaxies as compiled by \citet{McConnachie2012} (see also \citealt{Mateo1998}). It includes information on all known galaxies within 3 Mpc from the Sun, which have distance estimates based on resolved stellar populations. We use the galaxy positions, radial distances and line-of-sight velocities of the LG galaxies as provided by the most-recent online version of the tables by \citet{McConnachie2012}\footnote{We make use of the data tables updated on June 17, 2013, as provided online at https://www.astrosci.ca/users/alan/Nearby\_Dwarfs\_Database.html.}. To this we add the recently-published line-of-sight velocity for Andromeda XXIX \citep{Tollerud2013} for which no velocities are provided in the catalogue yet. For consistency with the catalogue provided by \citet{McConnachie2012} we treat the Canis Major over-density as a MW satellite \citep[e.g.][]{Martin2004}, but note that it might be a substructure of the MW caused by the warp of the Galactic disc \citep[e.g.][]{Momany2006}. Due to its close proximity to the centre of the MW, excluding this object from the analysis of the VPOS would not significantly affect our results.

The catalogue is most likely still incomplete because it was not constructed based on a deep, homogeneous all-sky survey (which does not exist yet). In the following analysis, this prevents us from determining the significances of the known and suggested galaxy planes in a meaningful way. We therefore will not attempt a detailed statistical analysis like that carried out for the M31 satellites by \citet{Conn2013}. However, as the catalogue compiles all known, nearby dwarf galaxies, it provides us with the most-complete picture of our neighbourhood to date and is therefore the best available dataset to search for possible associations of dwarf galaxies in planar structures. Future discoveries of additional galaxies will then provide observational tests of these suggested structures.

Throughout this paper, we adopt a Cartesian coordinate system ($x$, $y$, $z$), with the $z$-axis pointing toward the Galactic north pole, the $x$-axis pointing in the direction from the Sun to the Galactic centre, and the $y$-axis pointing in the direction of the Galactic rotation. We chose the origin of the coordinate system to be the midpoint between the MW and M31, which we denote the centre of the LG. We assume a distance of 8.5 kpc between the Sun and the centre of the MW, such that M31 is at a distance of 788 kpc from the MW centre. We decided to chose the midpoint and not the LG barycentre for simplicity. The exact mass distribution in the LG is still uncertain, in particular the reported mass-ratio $M_\mathrm{M31}/M_\mathrm{MW}$\ between M31 and the MW varies between about 0.8 and 2.0 \citep{vdMarel2008}. As the two galaxies have approximately similar total (halo) masses in any case, the midpoint between the two galaxies can be assumed to approximate the barycentre. All positions given relative to this origin can be converted to a MW- or M31-origin by subtracting the respective galaxy's position in $[x, y, z]$\ coordinates, i.e. $\mathbf{r}_{\mathrm{MW}} = [ 193, -312, 144]$~kpc or $\mathbf{r}_{\mathrm{M31}} = [-193, 312, -144]$~kpc. The positions of the LG galaxies in this Cartesian coordinate system are compiled in Tab. \ref{tab:galpos}.

We chose this MW-based coordinate system to ease the comparison of the different galaxy structures. In particular, directions (for example of normal vectors) can then also be expressed in Galactic longitude and latitude. This allows the comparison of different datasets and shall encourage the reader to compare the found dwarf galaxy planes with additional data not included in this study. 
As an example, we express the orientations of the galactic discs of the MW and M31 via their spin directions. The MW spin points into the direction of the negative $z$-axis ($[x,y,z] = [0,0,-1]$), which in Galactic coordinates corresponds to $b = -90^{\circ}$. We determine M31's spin vector direction by adopting the same parameters as \citet{Conn2013}: a position angle $\theta = 39.^{\circ}8$\ and an inclination of $i = 77.^{\circ}5$\ \citep{deVaucouleurs1958}. This results in a M31 spin direction pointing to $(l,b) = (241^{\circ}, -30^{\circ})$\ \citep[e.g.][]{Gott1978,Raychaudhury1989}, i.e. $(x,y,z) = (-0.420, -0.757, -0.500)$.

In the following sections, all galaxies of the \citet{McConnachie2012} catalogue which are within 1.5 Mpc of the origin will be considered LG members. This radius was chosen because it is about half way to the next nearby galaxy groups such as Sculptor \citep[fig. 9 of ][]{Jerjen1998}. Thus, beyond $\approx 1.5$~Mpc the LG's gravitational influence can be expected to become less important (assuming a similar mass for the LG and these galaxy groups). In addition, this distance corresponds to approximately two times the current distance between the MW and M31, the most prominent distance scale in the LG. It is also similar to the radius of the zero-velocity surface around the LG. The observationally inferred radius is $\approx 1$~Mpc \citep{Karachentsev2009}, while the turnaround radius of $1.56^{+ 0.08}_{-0.07}$~Mpc predicted from the LG mass estimated via the timing argument \citep{vdMarel2008} coincides well with our adopted radial cut. Finally, there is a gap in the distribution of galaxies in our sample at about this radius. The galaxy furthest from the origin but within our adopted radial cut is UGC 4879, which lies at a distance of 1.3 Mpc from our adopted origin, while NGC 3109, the next galaxy further away, is already at a distance of 1.6 Mpc.

\begin{figure}
 \centering
 \includegraphics[width=88mm]{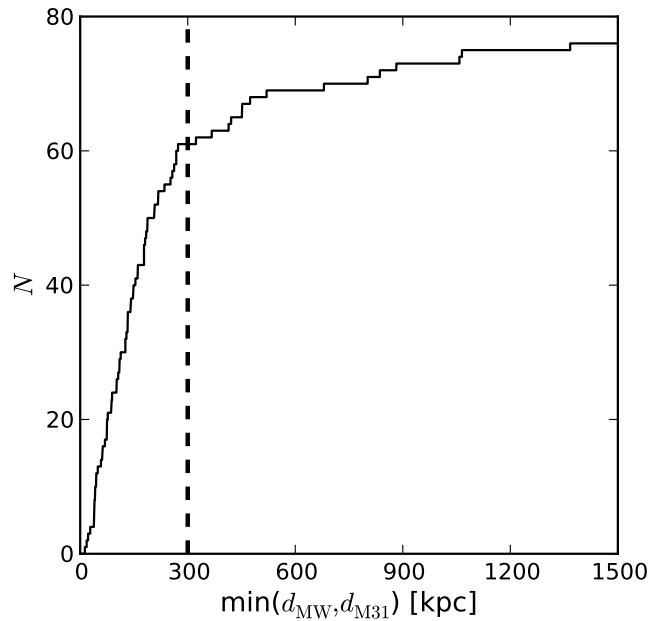}
 \caption{Cumulative distribution of dwarf galaxies within a certain distance from their host, which is the closest major LG galaxy (MW or M31). Galaxies left of the dashed vertical line (within 300 kpc of either the MW or M31) are categorized as satellite galaxies, galaxies at larger distances are categorized as non-satellite LG galaxies.}
 \label{fig:cumulativedwarfdist}
\end{figure}

With this distance cut, our sample consists of 78 galaxies. We split these up into three categories: \textit{hosts}, \textit{satellites} and \textit{non-satellites}. The two most-massive galaxies in the LG, the MW and M31, are considered to be hosts. They each harbour a large number of satellite galaxies. To determine which galaxies we consider to be satellites, we introduce another distance criterion. For each galaxy (except the MW and M31), the distance to the MW ($d_{\mathrm{MW}}$) and to M31 ($d_{\mathrm{M31}}$) is determined. The minimum of these two values is the distance to the nearest host ($d_{\mathrm{host}} = \rm{min}(d_{\mathrm{MW}}, d_{\mathrm{M31}})$). In Fig. \ref{fig:cumulativedwarfdist} we plot the cumulative distribution of galaxies against $d_{\mathrm{host}}$. The vertical dashed line indicates our distance criterion for satellite galaxies: all galaxies closer than 300 kpc to either host galaxy are considered to be satellites, all other galaxies are considered to be relatively isolated non-satellite galaxies. This distance of 300 kpc was chosen for three reasons: (1) The cumulative distance distribution shows that there is a small gap in distances close to this value (no known galaxy lies at a distance between 270 and 320 kpc). (2) At this distance the slope of the cumulative distribution changes, becoming shallower for larger radii\footnote{This change might, in part, be due to the fact that the surveys searching for dwarf galaxies in the LG focus on the volumes around the two host galaxies}. (3) The radius corresponds to the viral radii of the dark matter haloes assumed to surround the MW (308 kpc) and M31 (300 kpc) \citep{vdMarel2012a}.

To conclude, our sample consists of two host galaxies, 15 non-satellite galaxies, 27 MW satellites and 34 M31 satellites.

\subsection{Plane-fitting technique}
\label{subsect:planefitting}

The best-fitting plane for a set of $N$\ galaxy positions $\mathbf{r}_i$\ is determined following the method described in \citet{Metz2007}. The positions of the galaxies are not weighted, i.e. each galaxy has the same weight in the plane fit. 

At first, the centroid $\mathbf{r}_0$\ of all positions is calculated as
\begin{displaymath}
\mathbf{r}_0 = \frac{1}{N} \sum^{N}_{i = 1} \mathbf{r}_i.
\end{displaymath}
Then the moments of inertia tensor $\mathbf{T}_0$ around this centroid is constructed as
\begin{displaymath}
\mathbf{T}_0 = \sum^{N}_{i = 1} \left[ \left(\mathbf{r}_i - \mathbf{r}_0\right)^2 \cdot \mathds{1} - \left(\mathbf{r}_i - \mathbf{r}_0\right) \cdot \left(\mathbf{r}_i - \mathbf{r}_0\right)^{\mathrm{T}} \right],
\end{displaymath}
where $\mathds{1}$\ is the unit matrix and $\mathbf{r}^{\mathrm{T}}$\ is the transposed version of the vector $\mathbf{r}$.
The eigenvalues and eigenvectors of $\mathbf{T}_0$\ are determined. The eigenvector corresponding to the largest eigenvalue is the normal to the plane containing the centroid\footnote{\citet{Metz2007} wrongly state that the eigenvector corresponding to the smallest eigenvalue gives the normal of the best-fitting plane.}, while the eigenvectors corresponding to the intermediate and smallest eigenvalue correspond to the intermediate and longest axis of the distribution.
This method has been tested against the one used in \citet{Kroupa2010} and \citet{Pawlowski2012a} and both were found to give the same results, with the method used in this analysis being much more efficient.

In addition to the centroid $\mathbf{r}_0$\ and the normal vector to the best-fitting plane $\mathbf{n}$, we determine $\Delta$, the root-mean-square (RMS) thickness perpendicular to the best-fitting plane. We furthermore measure the RMS extend of the distribution along the intermediate and the long axis. With this, we determine the RMS axis ratios between shortest and longest ($c/a$) and intermediate and longest axis ($b/a$) of the distribution. A small value of $c/a$\ indicates either an oblate distribution (a thin plane) if $b/a$\ is large, or a narrow prolate distribution (filament-like) if $b/a$ is similarly small ($c/a \approx b/a$).
Finally, we also determine the offset  (distance perpendicular to the plane) of each galaxy in our LG sample from the plane.

We want to caution the reader that the assumption of a perfect planar alignment of the satellites might be too simplistic. A coherent distribution of satellite galaxies might be affected by precession, which is more significant for the satellites close to a host galaxy than for those far away, or a satellite plane might be 'bend' if it is initially offset from the centre of mass, to name only two possibilities.

\subsection{Effects of galaxy distance uncertainties}
\label{subsect:posuncert}

All dwarf galaxy positions are determined from their heliocentric distance modulus, as reported by \citet{McConnachie2012}, and the distance uncertainties are determined from the reported distance modulus uncertainties. We assume that the MW centre is at a distance of 8.5 kpc from the Sun, but the exact distance is not important for our analysis because a change in this value simply translates all galaxy positions by the same distance, preserving their mutual orientations. It only changes the position of the MW and results in a minor change in the position of the origin of our coordinate system, i.e. the reported distances between the planes and the MW would change. However, these changes are on the order of less than 1 kpc, which is negligible for the typical distances of many 100 kpc in the LG.

The uncertainties in the position of the galaxies are dominated by their uncertain radial distance from the Sun. We assume that the angular position on the sky is accurate. To determine the effect of the distance uncertainties, in the following all parameters are determined by sampling the galaxy distances 1000 times. For each realisation, the galaxy distances are generated by starting with their most-likely distance as reported in the catalogue by \citet{McConnachie2012}. This distance is then increased (or decreased) by a value sampled from a Gaussian distribution with a width set to the positive (negative) $1\sigma$\ distance uncertainty. We therefore assume that the distance uncertainties follow a Gaussian distribution, which can be asymmetric around its peak to account for differing positive and negative uncertainties.

The parameters determined in the analysis are stored for each of the 1000 realisations. Unless mentioned otherwise, the parameters reported in the following (e.g. in Tables \ref{tab:allplanes}, \ref{tab:dwarfdist} and \ref{tab:angles}) are the mean values determined by averaging over the values of all realisations. We report the standard deviation of the parameters around this average as the uncertainties. 

For the normal vector directions, we determine the spherical standard distance $\Delta_{\mathrm{sph}}$\ \citep{Metz2007,Pawlowski2012b} of the normal directions for the $k = 1000$ realisations relative to the normal direction determined for the most-likely galaxy positions. The spherical standard distance is a measure for the clustering of vectors. It is defined as
$$
\Delta_{\mathrm{sph}} = \sqrt{ \frac{ \sum_{i=1}^{k} \left[ \arccos \left( \left| \langle \mathbf{n} \rangle \cdot \mathbf{n}_{i} \right| \right) \right] ^{2} }{ k } },
$$
where $\mathbf{n}_i$\ are the normal direction unit vectors, $\langle \mathbf{n} \rangle$\ is the normal vector determined by fitting a plane to the most-likely galaxy positions and '$\cdot$' denotes the scalar product of the vectors. Note that the formula deals with axial data and therefore includes the absolute value of the scalar product, in contrast to the case in which vectorial data such as an orbital pole (direction of angular momentum) is used.

\subsection{4-galaxy-normal density plots}
\label{subsect:4dwarfnormals}

The plane-fitting method described in Sect. \ref{subsect:planefitting} determines the parameters of a plane fitted to a pre-chosen group of galaxies. This method is not suitable to determine whether the chosen group defines the most-prominent planar arrangement in a sample of galaxies. We therefore also investigate whether, in a given galaxy sample, there are signs for a dominant plane defined by a sub-sample only. This allows a consistency check by comparing whether this sub-sample constituting the dominant plane is similar to the chosen group of galaxies. 

This task is approached with a method based on constructing planes for many small sub-samples of galaxies. We draw all possible combinations of four galaxies from a given galaxy sample. For each combination, we fit a plane as described in Sect. \ref{subsect:planefitting} and record the normal-axis direction $\mathbf{n}_{4}$\ to the plane (which we call the 4-galaxy-normal) and the axis ratios. For a total sample of $N$\ galaxies, there are $\frac{N!}{4! (N-4)!}$ possible combinations of 4 galaxies each. If several galaxies in a sample lie within a common plane, different combinations of four galaxies from this plane will result in very similar 4-galaxy-normal directions. Thus, when plotting the density of 4-galaxy-normal directions on a sphere we can identify the dominant plane orientation by looking for an over-density of normal directions. The density distribution for all possible normal directions $\mathbf{n}_4$\ is plotted in a Galactic coordinate system, weighting each 4-galaxy-normal with the logarithm of the ratio between the shortest and the sum of the intermediate and long axis
\begin{displaymath}
\mathrm{weight} = \log\left(\frac{a+b}{c}\right).
\end{displaymath}
This weighting emphasizes those normal directions which are associated to plane-like distributions, i.e. short dimensions along the short axes $c$\ and large dimensions along the two remaining axes $a$\ and $b$\ of the distribution. 

By determining which galaxies contribute to an over-density in the 4-galaxy-normal distribution we can identify galaxies as likely members of the dominant plane. Galaxies which do not contribute at all to an over-density can be excluded, because they can not lie within the respective plane: there is no combination with any other galaxies which would give a 4-galaxy-normal direction close to the plane normal. 

To determine which galaxies contribute to a 4-galaxy-normal density peak, we proceed as follows. 
For a 4-galaxy-normal which lies within a defined angle (typically $15^{\circ}$) of a density peak, we determine the four contributing galaxies. Each of these galaxies is counted as contributing the 4-galaxy-normal's weight to the peak. This is repeated for all 4-galaxy-normals which are close to the density peak. In the end, for each galaxy the sum of all its associated plane weights is determined. We then express this as a relative weight for each galaxy by normalising the weight contribution of the most-dominant galaxy to one. The resulting relative weights are plotted on a common axis for the satellite galaxies' contribution to different peaks, revealing which galaxies contribute most to which peak direction.

To account for the uncertainties in the galaxy positions, 100 realisations sampling from the galaxy distance uncertainties as described in Sect. \ref{subsect:posuncert} are generated for each galaxy sample. Each of these 100 realizations contributes equally to the 4-galaxy-normal density plots and the determination of the dwarf galaxy contributions to the peak directions.

We have also tested this method (without weighting) using all possible combinations of only three galaxies, which always define a perfect plane. The resulting normal distribution plots look similar and also reveal the same preferred normal directions (4-galaxy-normal density peaks). However, as three points always define a plane, this method can not weight the normal directions. We have therefore chosen to use combinations of four points. It is possible to extend the analysis to combinations of more points, but this then de-emphasizes planes consisting of only few satellites, in particular when we deal with small-number samples of dwarf galaxies like the only 15 non-satellite galaxies.

The method we use is similar to the one employed recently by \citet{Conn2013}, but there are important differences. As we investigate the distribution of dwarf galaxies in the whole LG, we do not assume a fixed point which the planes have to contain. In contrast to this, \cite{Conn2013}, interested only in the satellite galaxies of M31, forced all planes to run through the centre of M31. 
Another difference is that we only consider combinations of four galaxies, while \citet{Conn2013} have used combinations of 3 to 7 galaxies in the plane construction. They found that the smaller combination sizes (3-4 satellites per plane fit) are particularly useful for identifying the thinnest planes, which thus supports our approach.
Finally, \citet{Conn2013} have used a different weight, the inverse of the thickness of the fitted planes. We, however, investigate the distribution of dwarf galaxies not only around a host, but also within the much larger LG. We therefore decided to weight by the axis ratios of the fitted planes because this is a scale-free representation of the thinness of a planar distribution. As discussed in Sect. \ref{sect:M31}, our approach reproduces the findings of \citet{Conn2013}, giving further confidence in the agreement of both methods.

\section{The vast polar structure (VPOS) around the MW}
\label{sect:MW}

\begin{figure*}
 \centering
 \includegraphics[width=180mm]{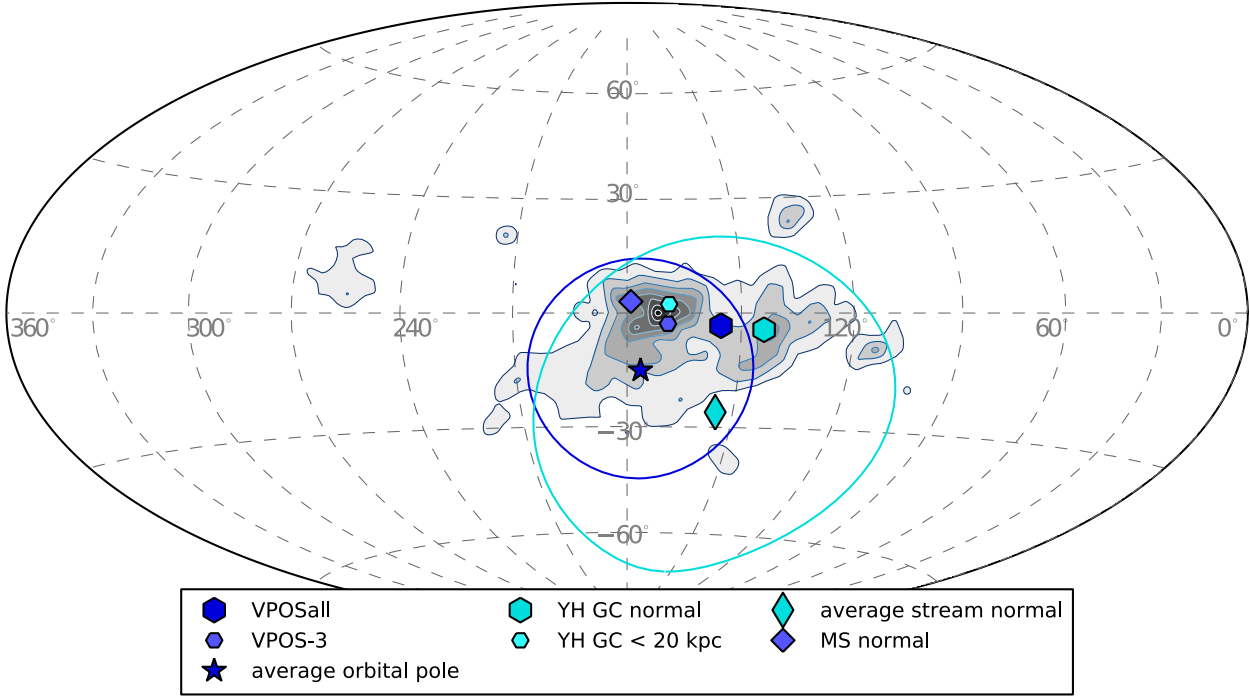}
 \caption{Density distribution of the 4-galaxy-normal directions for our sample of 27 MW satellite galaxies, determined and weighted as explained in Sect. \ref{subsect:4dwarfnormals}. They show one distinct peak (Peak 1) at $(l,b) \approx (170^{\circ}, 0^{\circ})$, close to the MW equator, indicating a preferred polar orientation of the planes. A second, smaller peak (Peak 2) is found at $(l,b) \approx (140^{\circ}, -5^{\circ})$, also close to the equator. 
Note that normal directions are axial, so each 4-galaxy-combination has two normals in opposite directions. For clarity and easier comparison, we only plot the 4-galaxy-normals in the centre of the plot between $l = 90^{\circ}$\ and $l = 270^{\circ}$. 
 Several other directions are also plotted: 
The direction of the normal to the plane fitted to all 27 MW satellites (VPOSall, dark-blue hexagon) lies in-between the two peaks, while the normal to the plane fitted to all satellites but Leo I, Hercules and Ursa Major (I)  (VPOS-3, smaller, light-blue hexagon) coincides with the more-pronounced peak. The Magellanic stream normal \citep[blue diamond, ][]{Pawlowski2012a} is close to this peak, too. This is also the case for the average direction of the orbital poles of the MW satellites for which proper motions have been measured \citep[dark-blue star, ][]{Pawlowski2013}. The blue circle around it indicates the concentration of the orbital poles around their average direction, as measured by the spherical standard distance (see Sect. \ref{subsect:posuncert}) of the orbital pole distribution, which is $\Delta_{\mathrm{sph}} = 29^{\circ}.3$. The normal to the plane fitted to all 30 young-halo globular clusters of the MW \citep[cyan hexagon, ][]{Pawlowski2012a} coincides well with the second peak, but restricting the sample to the 20 young-halo GCs within 20 kpc of the MW results in a plane normal which is close to the major peak \citep[small light cyan hexagon, ][]{Pawlowski2012a}. Finally, the average of the normal directions fitted to streams in the MW halo \citep[cyan diamond, ][]{Pawlowski2012a} also points into the general direction of the two peaks, but the spherical standard distance of these stream normals, illustrated by the cyan circle, is large ($\Delta_{\mathrm{sph}} = 46^{\circ}$).}
 \label{fig:MW4dwarfnormal}
\end{figure*}

\begin{figure}
 \centering
 \includegraphics[width=88mm]{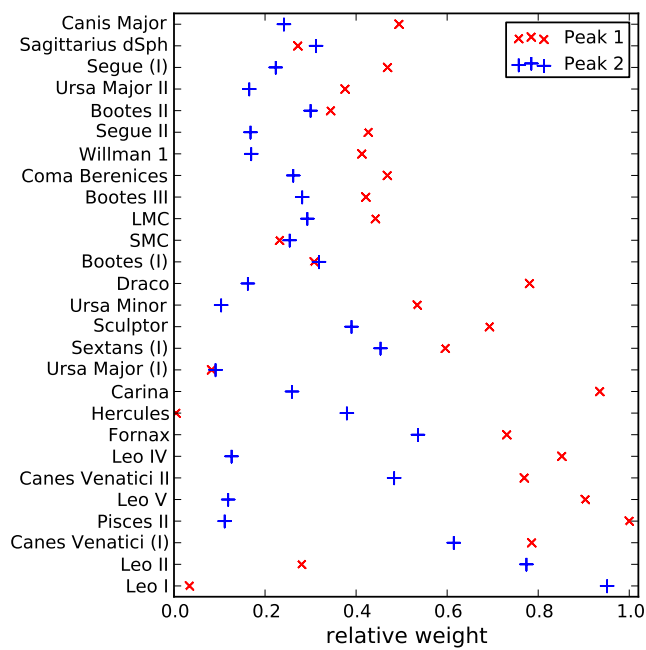}
 \caption{Relative contributions of the different MW satellites to the regions within $15^\circ$\ of the 4-galaxy-normal peaks in Fig. \ref{fig:MWpeakcontribution}.
 }
 \label{fig:MWpeakcontribution}
\end{figure}

Within our sample of galaxies, 27 objects are satellites of the MW with a maximum radius from the MW of $d_{\mathrm{MW}} \approx 260$~kpc. A fit to all of them results in a best-fitting plane of RMS height $\Delta = 29.3 \pm 0.4$~kpc, which is offset from the MW centre by $D_{\mathrm{MW}} = 7.9 \pm 0.3$~kpc. The distribution has axis ratios of $c/a = 0.301 \pm 0.004$\ and $b/a = 0.576 \pm 0.007$. The normal $\mathbf{n}$\ to the best-fitting plane points to $(l, b) = (155^{\circ}.6, -3^{\circ}.3)$, the uncertainty in this direction is given by the spherical standard distance of the normal directions for the different galaxy distance samples. It is $\Delta_{\mathrm{n}} = 1^{\circ}.1$. We adopt this plane fit as the 'VPOSall'.
This plane fit reproduces the earlier results of \citet{Kroupa2010} ($\Delta = 28.9$~kpc, $D_{\mathrm{MW}} = 8.2$~kpc, $\mathbf{n}$\ pointing to $(l, b) = (156^{\circ}.4, -2^{\circ}.2)$), even though the current sample includes updated radial distances for some of the satellites, four additional objects (Canis Major, Segue I and II and Bootes III) and lacks Pisces I.

Does the 4-galaxy-normal technique confirm this plane orientation as the preferred one? In Fig. \ref{fig:MW4dwarfnormal} we plot the direction of the normal vector of the plane fitted to all 27 satellites. In addition, we include the normal vector to the plane of all 30 young halo globular clusters (YH GC), that of only those YH GCs closer than 20 kpc from the MW centre, and the average stream normal direction, as reported in \citet{Pawlowski2012a}. We also plot the average orbital pole of the eight classical MW satellite galaxies which co-orbit in the VPOS \citep{Pawlowski2013}, which represents the normal direction to the average orbital plane of the MW satellites in the VPOS. The contours indicate the density distribution of the 17550 different possible 4-galaxy-normal directions. There is a pronounced density peak (Peak 1) at $(l,b) \approx (175^\circ, 0^\circ)$. This is close, but inclined by about $20^{\circ}$, to the normal of the plane fitted to all satellite positions. As second, much shallower peak (Peak 2) at $(l,b) \approx (145^\circ, -5^\circ)$\ coincides with the position of the YH GC normal. 

Interestingly, these two peaks approximately agree with the two 'stream' axis directions already discussed by \citet{LyndenBell1982}: his Magellanic stream axis, pointing to $(l,b) = (185^\circ, 3^\circ)$, and his FLS stream axis, pointing to $(l,b) = (135^\circ, -3^\circ)$. That the outer YH GCs coincide with the second peak / FLS stream axis direction has already been noticed by \citet{Majewski1994}. However, the first peak in Fig. \ref{fig:MW4dwarfnormal} is much more pronounced than the second.

As discussed in Sect. \ref{subsect:4dwarfnormals}, we can check which satellite galaxies contribute to the peaks. Fig. \ref{fig:MWpeakcontribution} plots the contributions of the different satellite galaxies to the 4-galaxy-normals in the regions $15^\circ$\ around the two peaks. Overall, the weight contributions to Peak 1 are larger than those to Peak 2, because the second peak is less pronounced than the first.

Almost all satellite galaxies contribute to Peak 1, with three marked exceptions: Leo I, Hercules and Ursa Major (I). For these three satellites there is almost no combination with any three of the other satellites which describes a plane with a normal pointing close to Peak 1. They are also the three satellite galaxies which have the largest vertical distance ($44.4 \pm 1.4$, $71.5 \pm 2.2$\ and $53.3 \pm 1.4$~kpc, respectively) to the best-fitting plane to all 27 MW satellites. All other satellites have distances of less than 40 kpc from the best-fitting plane. We have checked that the 4-galaxy-normal distribution for the 24 MW satellites without the three outlying ones does not show Peak 2 any more, while Peak 1 is still present.

It might therefore be worthwhile to exclude Leo I, Hercules and Ursa Major (I) from the plane fit. We have done so and fitted a plane to the remaining 24 satellites only, referring to this as the 'VPOS-3'. This plane is much thinner than the VPOSall, $\Delta = 19.9 \pm 0.3$~kpc, and slightly more offset from the MW centre, $D_{\mathrm{MW}} = 10.4 \pm 0.2$~kpc. The distribution has axis ratios of $c/a = 0.209 \pm 0.002$\ and $b/a = 0.536 \pm 0.006$. The normal $\mathbf{n}$\ to this plane now points close to the 4-galaxy-normal density peak in Fig. \ref{fig:MW4dwarfnormal}, into the direction of $(l, b) = (169^{\circ}.5, -2^{\circ}.8)$\ and the standard deviation of this direction for the different galaxy distance samples is $0^{\circ}.4$. Thus, by excluding the three outlying galaxies, the orientation of the satellite galaxy plane fit changes by $14^\circ$, but the best-fitting plane is polar in both cases. The VPOS-3 normal direction is much closer to the direction of the average orbital pole of the MW satellites \citep{Metz2009,Pawlowski2013}, to the normal of the best-fitting plane to the inner YH GCs and to the Magellanic Stream normal \citep{Pawlowski2012a}. These alignments might be seen as indications that the VPOS-3 is a better representation of the satellite structure surrounding the MW.

A similar analysis of the contributions to Peak 2 remains inconclusive. All MW satellite galaxies contribute to this peak to some degree. Excluding the five satellites contributing the least (Ursa Major (I), Ursa Minor, Pisces II, Leo V and Leo IV) results in a plane fit which has a normal pointing to $(l, b) = (141^{\circ}.6, -6^{\circ}.4)$\ with an uncertainty of $0^{\circ}.7$. It is offset from the MW centre by $D_{\mathrm{MW}} = 1.0 \pm 0.3$~kpc, has a RMS height of $\Delta = 22.3 \pm 0.4$~kpc and axis ratios of $c/a = 0.239 \pm 0.005$, $b/a = 0.590 \pm 0.008$.

Among the seven satellites contributing the most to Peak 2 (Leo I, Leo II, Canes Venatici, Fornax, Canes Venatici II, Sextans and Sculptor) are all galaxies which \citet{LyndenBell1982} identified to be in the FLS stream (Fornax, Leo I, Leo II and Sculptor), and also Sextans, which was reported to be in the FLS stream by \citet{Majewski1994}. Fitting a plane to these gives a normal pointing to $(l, b) = (141^{\circ}.8, -4^{\circ}.6)$\ with an uncertainty of $0^{\circ}.1$. It is offset from the MW centre by $D_{\mathrm{MW}} = 5.3 \pm 0.1$~kpc, has a RMS height of $\Delta = 7.4 \pm 0.2$~kpc and axis ratios of $c/a = 0.052 \pm 0.001$, $b/a = 0.458 \pm 0.008$. Most of these seven satellites, however, substantially contribute to Peak 1, too.

In a similar manner we can select the seven satellites contributing most to Peak 1 (Pisces II, Carina, Leo V, Leo IV, Canes Venatici, Draco and Canes Venatici II). Fitting a plane to these gives a normal pointing to $(l, b) = (171^{\circ}.5, -0^{\circ}.4)$\ with an uncertainty of $0^{\circ}.1$. It is offset from the MW centre by $D_{\mathrm{MW}} = 8.6 \pm 0.1$~kpc, has a RMS height of $\Delta = 5.6 \pm 0.1$~kpc and axis ratios of $c/a = 0.046 \pm 0.001$, $b/a = 0.619 \pm 0.009$.

It will require a more-complete census of the satellite galaxy population in the southern hemisphere, such as the Stromlo Missing Satellites Survey \citep{Jerjen2010,Jerjen2012}, to reveal whether the two-peak structure in the 4-galaxy-normal distribution becomes more pronounced. More tightly constrained proper motions for the dSphs will then allow to test whether the VPOS consists of two separate polar streams (the satellite orbital poles would cluster around the two peaks), whether the VPOS is better interpreted as one structure with a few unrelated objects (most orbital poles would point into one preferred direction), or whether the VPOS is one dynamical structure with an opening angle defined by the two peaks (the orbital poles would be distributed in between the two peaks). For many satellites, the current uncertainties in proper motion determinations result in orbital pole directions which are uncertain to $\approx 15^{\circ}$\ or more \citep{Pawlowski2013}, and are therefore still inconclusive.

For the following discussion, we adopt the parameters for the VPOSall (fitted to all 27 MW satellites) and the VPOS-3 (fitted to all MW satellites except Leo I, Hercules and Ursa Major). As the normal to the VPOSall lies in-between the two peaks in Fig. \ref{fig:MW4dwarfnormal}, we focus on this fit. If the two peaks indeed suggest the existence of two separate planar distributions around the MW, these planes would be inclined by $\approx 30^\circ$\ with respect to each other, and by $\approx 15^\circ$\ with respect to the VPOSall. Therefore, the error in the plane orientation we make by adopting the VPOSall is only $\approx 15^\circ$\ if there are indeed two planes.
We also consider the VPOS-3, as its normal direction coincides with the dominant peak of the 4-galaxy-normal distribution, and it also agrees better with a number of additional features. In particular, it is aligned with the Magellanic Stream and the average orbital pole of the MW satellites, indicating that at least a number of satellites orbit preferentially within this plane. 
The resulting parameters of the VPOSall and VPOS-3 plane fits are compiled in Table \ref{tab:allplanes}, the distances of individual galaxies to the best-fitting planes are given in Table \ref{tab:dwarfdist}.

Do any of the non-satellite galaxies in the LG lie close to the satellite galaxy plane around the MW? The galaxy closest to the VPOSall is Phoenix, which has a distance of only $48 \pm 4$~kpc. WLM, the next-nearest galaxy, already has a distance of more then twice this value ($104 \pm 5$~kpc). Interestingly, both Phoenix and WLM are closer to the VPOS-3. Phoenix then has a distance of only $16 \pm 2$\ and WLM of $27 \pm 4$~kpc, which is quite remarkable given the VPOS-3's RMS height of only 20 kpc. Thus, Phoenix and WLM, which have distances from the MW of 415 and 930 kpc, respectively, are within $3^{\circ}$\ of the VPOS-3. All remaining dwarf galaxies are offset by more than 100 kpc from the VPOS-3, but due to their large distances from the MW some have relatively small angular distances from the VPOS-3 ($9^{\circ}$\ for Cetus, $13^{\circ}$\ for Andromeda XXVIII and Pegasus dIrr).

\section{The Great Plane of Andromeda (GPoA)}
\label{sect:M31}

\begin{figure*}
 \centering
 \includegraphics[width=180mm]{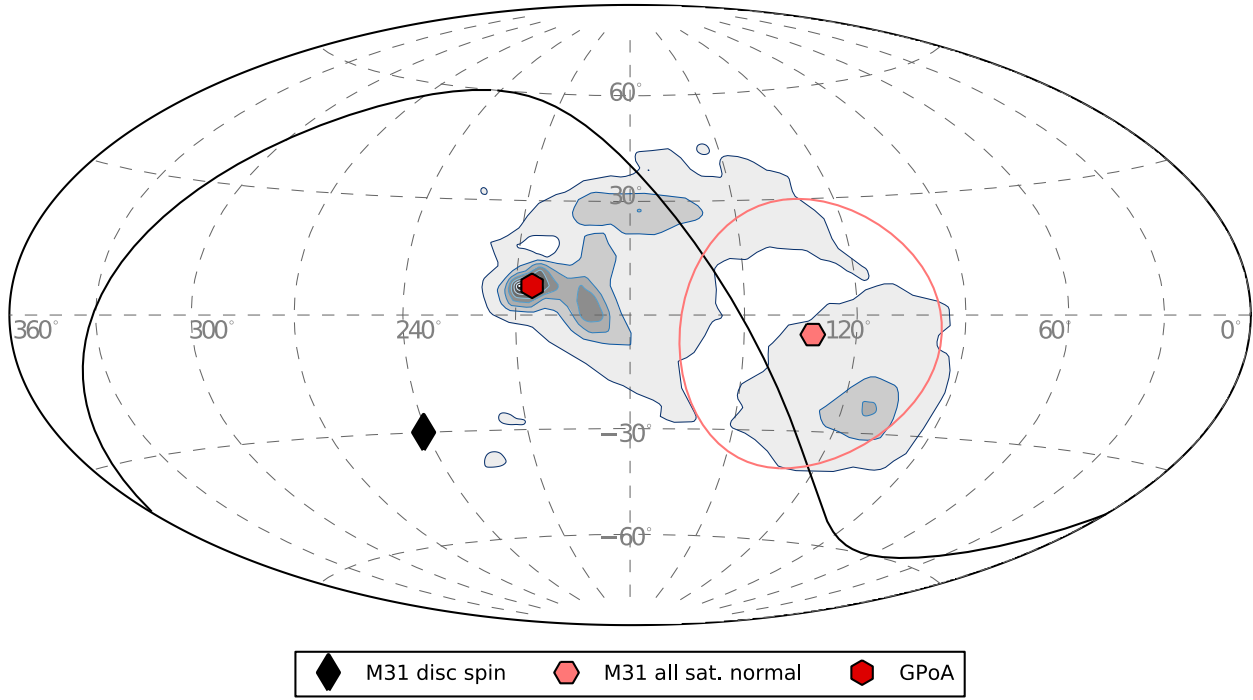}
 \caption{Density distribution of the 4-galaxy-normal directions for our sample of 34 M31 satellite galaxies, determined and weighted as explained in Sect. \ref{subsect:4dwarfnormals}. They show one very pronounced peak at $(l,b) \approx (205^{\circ}, 10^{\circ})$. 
As in Fig. \ref{fig:MW4dwarfnormal} only the 4-galaxy-normals in the centre of the plot between $l = 90^{\circ}$\ and $l = 270^{\circ}$\ are shown for clarity, not their mirrored counterparts.
 The spin direction of the galactic disc of M31 is indicated by the black diamond, and the M31 equator lies along the great circle $90^\circ$\ offset from this direction, plotted as a black line. Also shown are the directions of the normals for planes fitted to the whole sample of M31 satellites (light-red hexagon) and to the M31 satellite subsample defining the GPoA (dark-red hexagon). The light-red circle indicates the spherical standard distance of the normal direction distribution for all M31 satellites, resulting from varying the satellite distances. Its large extend ($\Delta_{\mathrm{sph}} = 35^{\circ}$) indicates that the best-fit plane for the full sample is only poorly defined, the full M31 satellite distribution is only mildly anisotropic.}
 \label{fig:M314dwarfnormal}
\end{figure*}

\begin{figure}
 \centering
 \includegraphics[width=88mm]{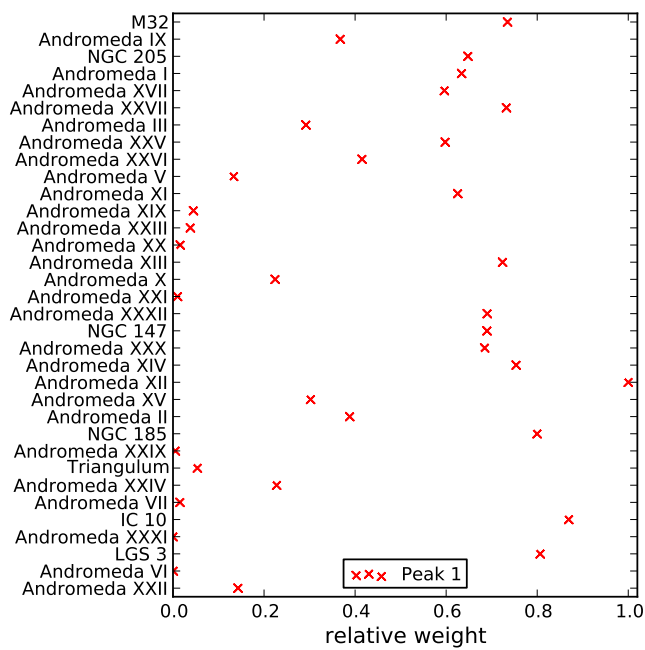}
 \caption{Contributions of M31 satellites to the region within $15^{\circ}$\ of the dominant 4-galaxy-normal peak at $(l,b) \approx (205^{\circ}, 10^{\circ})$\ in Fig. \ref{fig:M314dwarfnormal}. Compare with the similar plot shown in figure 13 of \citet{Conn2013}. Our and their dwarf contributions are comparable, despite the slightly different method and our more numerous but less homogeneous M31 satellite sample.
 }
 \label{fig:M31peakcontribution}
\end{figure}

In analogy to the MW satellite galaxies, we start by fitting a plane to all 34 dwarf galaxies that are considered to be M31 satellites. The parameters of the fit reveal that this distribution is only mildly anisotropic. The axis ratios are very similar to each other ($c/a = 0.6$\ and $b/a = 0.7$) and the RMS height of $\Delta = 77.5 \pm 4.3$~kpc is comparable to the RMS radius of the M31 satellite distribution $d_{\mathrm{M31}}^{\mathrm{RMS}} / \sqrt{3} = 93$~kpc. The direction of the normal vector to the plane fit is very uncertain, the best-fitting direction for the different galaxy distance realizations varies by $\Delta_{\mathrm{n}} = 35^\circ$. Interestingly, the average plane normal $\mathbf{n}$\ of the fits to the M31 satellites points to $(l, b) = (132^\circ, -5^\circ)$, roughly in the direction of the VPOSall normal and the normal of the plane defined by the YH GCs around the MW.

Following a detailed analysis of the M31 satellite galaxy positions, \citet{Ibata2013} and \citet{Conn2013} have identified a sub-sample of 15 out of their 27 M31 satellites which lie within a thin plane. Their analysis gives a very high significance for this discovery. The probability that a similar alignment occurs at random is only 0.13 per cent \citep{Ibata2013}. The structure's significance rises to 99.998 per cent when also taking into account the line-of-sight velocities which reveal that 13 of the 15 plane members co-orbit. Most M31 satellites in the northern part of the plane recede from the MW while most in the southern part approach the MW relative to M31. 
We therefore tentatively adopt their sample of galaxies: Andromeda I, Andromeda III, Andromeda IX, Andromeda XI, Andromeda XII, Andromeda XIV, Andromeda XVII, Andromeda XXV, Andromeda XXVI, NGC 147, NGC 185, Andromeda XIII, Andromeda XXVII and Andromeda XXX.

Andromeda XVI, which is in the plane sample by \citet{Ibata2013} and \citet{Conn2013}, has a distance of 323 kpc from M31 and is therefore considered a non-satellite according to our criteria (Sect.\ref{subsect:dataset}). For reasons of consistency of our distance criterion we exclude Andromeda XVI from the galaxy sample, but note that the plane fitting results do not change significantly if Andromeda XVI is included.

The plane fitted to the resulting 14 dwarf satellite galaxies has a normal vector $\mathbf{n}$\ pointing to $(l, b) = (206^{\circ}.2, 7^{\circ}.8)$, with a standard deviation of this direction of only $1^{\circ}.0$\ (see Fig. \ref{fig:M314dwarfnormal}). The RMS height of the plane members around the best-fitting plane is $\Delta = 14.2 \pm 0.2$~kpc and the plane is offset by $D_{\mathrm{M31}} = 4.1 \pm 0.7$~kpc from the centre of M31. The axis ratios of the dwarf galaxy distribution are $c/a = 0.125 \pm 0.014$\ and $b/a = 0.578 \pm 0.084$. This plane is inclined by $50^\circ.5$\ from the galactic disc of M31. Despite the differences in our dataset and disc fitting analysis, our plane fit is very similar to that of \citet{Conn2013}, who report a slightly smaller RMS plane height of $12.34^{+0.75}_{-0.43}$\ kpc and an inclination from M31's galactic disc of $51^\circ.7$.

The analysis of \citet{Ibata2013} and \citet{Conn2013} concentrates on those M31 satellites which are found within the PAndAS \citep{McConnachie2009} survey region. Which additional M31 satellites (i.e. dwarf galaxies within 300 kpc from M31) are close to this plane? The five closest which are not in the sample of \citet{Ibata2013} are NGC 205 ($0.9 \pm 0.7$~kpc from the best-fit plane), M32 ($4.5 \pm 1.9$~kpc), IC 10 ($12.7 \pm 2.4$~kpc) and LGS 3 ($18.7 \pm 2.7$~kpc), whose likely alignment is also mentioned by \citet{Conn2013}, and in addition the recently discovered satellite galaxy Andromeda XXXII ($17.5 \pm 2.0$~kpc) \citep{Martin2013}. All remaining satellites have a distance of more than $\approx 60$~kpc from this plane, about four times the plane's RMS height $\Delta$. 
We therefore add these four objects to the sample. The parameters of a plane fitted to this extended sample of 19 galaxies\footnote{Andromeda I, Andromeda III, Andromeda IX, Andromeda XI, Andromeda XII, Andromeda XIII, Andromeda XIV, Andromeda XVII, Andromeda XXV, Andromeda XXVI, Andromeda XXVII, Andromeda XXX, Andromeda XXXII, IC 10, LGS 3, M32, NGC 147, NGC 185, NGC 205. We keep Andromeda XVI excluded as we consider it a non-satellite, but including it does not change the results significantly.} are only minimally different from the fit to the 14 objects. The orientation of the best-fitting plane to the larger sample differs by only $0^{\circ}.4$, its normal $\mathbf{n}$\ points to $(l, b) = (205^{\circ}.8, 7^{\circ}.6)$, with a standard deviation of this direction of only $0^{\circ}.8$. The RMS height of the plane members around the best-fitting plane is slightly smaller for the larger sample ($\Delta = 13.6 \pm 0.2$~kpc) and the plane passes closer to the centre of M31, $D_{\mathrm{M31}} = 1.3 \pm 0.6$~kpc. The axis ratios are $c/a = 0.107 \pm 0.005$\ and $b/a = 0.615 \pm 0.058$. For the following discussion, we will adopt this sample and the resulting plane parameters as the GPoA.

In addition to Andromeda XVI, there are two other non-satellite dwarf galaxies which lie close to the GPoA: IC 1613 ($25 \pm 3$~kpc, $3^{\circ}$\ from the GPoA) and Phoenix ($14 \pm 9$~kpc, $1^{\circ}$). They have a distance of more than 500 and 800 kpc from M31, respectively. All remaining non-satellite dwarfs in our sample have offsets from the GPoA of more than 100 kpc, but Cetus and Andromeda XVIII are at angular distances of only $\approx 15^{\circ}$. Interestingly, the non-satellites Phoenix and Cetus are also close to the VPOS-3 plane.

The normal direction to the GPoA is also prominent as a strong peak in the density-contours of the 4-galaxy-normal distribution of the 34 M31 satellites (Fig. \ref{fig:M314dwarfnormal}). This is consistent with the similar analysis by \citet{Conn2013}, even though we consider a slightly different sample consisting of all currently known M31 satellite galaxies, but without objects at distances larger than 300 kpc from M31. The normal direction of the plane fitted to all M31 satellites does not coincide with a feature in the 4-galaxy-normal plot, which is another indication that the full M31 satellite population does not follow a single preferred plane.

Fig. \ref{fig:M31peakcontribution} shows how much the different satellites of M31 contribute to the 4-galaxy-normals within $15^{\circ}$ of the GPoA peak. Sorted according to their relative weighted contribution, the 21 satellites contributing the most are (those written in italics are in the GPoA satellite galaxy sample): \textit{Andromeda XII}, \textit{IC 10}, \textit{LGS 3}, \textit{NGC 185}, \textit{Andromeda XIV}, \textit{M32}, \textit{Andromeda XXVII}, \textit{Andromeda XIII}, \textit{Andromeda XXXII}, \textit{NGC 147}, \textit{Andromeda XXX}, \textit{NGC 205}, \textit{Andromeda I}, \textit{Andromeda XI}, \textit{Andromeda XXV}, \textit{Andromeda XVII},  \textit{Andromeda XXVI}, Andromeda II, \textit{Andromeda IX}, Andromeda XV, \textit{Andromeda III}. Thus, among the 21 galaxies contributing most to the peak are all 19 M31 satellites that make up our GPoA sample.

The plane fit parameters for the GPoA, which will be used for the later discussion, are compiled in Table \ref{tab:allplanes}. Before we investigate the possibility that some of the remaining M31 satellite galaxies constitute a second common plane we first turn our attention to the distribution of the non-satellite galaxies in the LG.

\section{Local Group Planes}
\label{sect:LG}

\begin{figure*}
 \centering
 \includegraphics[width=180mm]{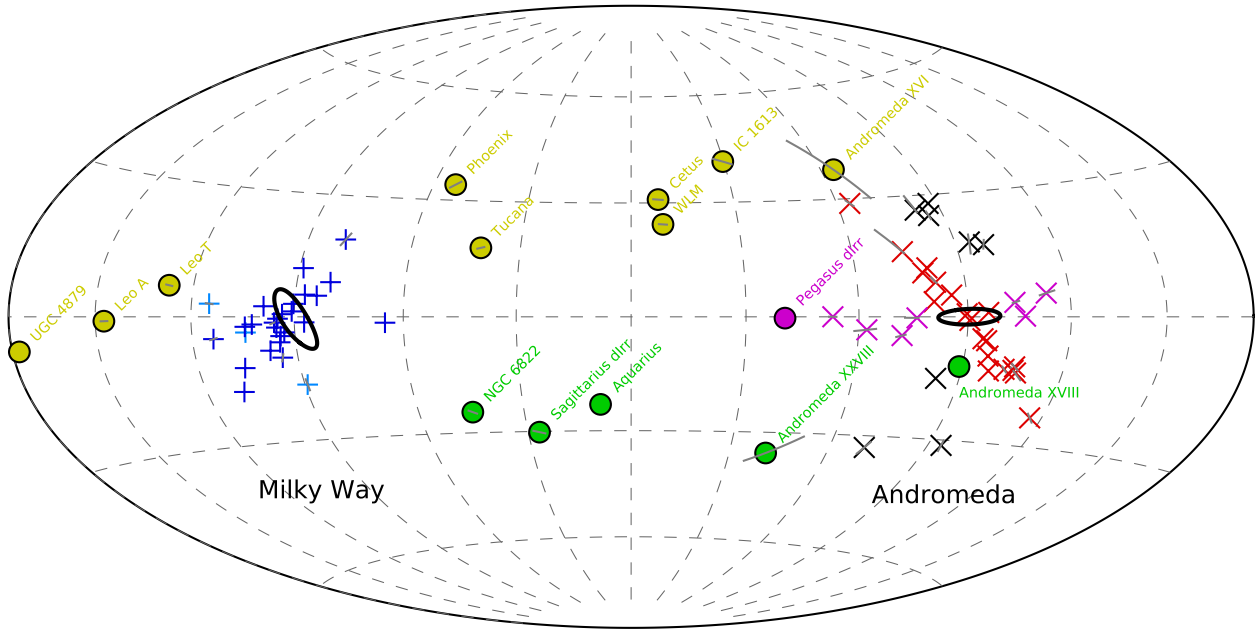}
 \caption{The distribution of LG galaxies as seen from the midpoint between the MW and M31. Note that in contrast to the previous plots, this is not plotted in Galactic coordinates $l$\ and $b$. Instead, the orientation of the coordinate system was chosen such that the MW and M31 lie on the equator and the normal to the plane fitted to all 15 non-satellite galaxies points to the north pole.
The positions and orientations of the MW and M31 discs are indicated by black ellipses. The Galactic disc of the MW is seen from the south, the Galactic south pole points to the upper right of the plot.
Satellite galaxies are plotted as crosses (+ for MW, $\times$\ for M31), non-satellites are plotted as filled circles. The one-sigma distance uncertainties for the galaxies result in position uncertainties in this projection, which are indicated by the grey lines. For most galaxies they are smaller than the symbols. 
Galaxies within a common plane are marked with the same color. All MW satellites are assumed to lie in the VPOSall are plotted in blue, while the M31 satellites assigned to the GPoA are plotted in red. 
Most of the non-satellite galaxies in the LG lie along one of two 'bands', one above and one below the plot's central axis.
The only LG galaxy not along one of the bands is the Pegasus dwarf irregular (dIrr). It is, however, very close to the plane of M31 itself. We have indicated this by marking the satellites close to the M31 disc plane, but not in the GPoA, in magenta.
 }
 \label{fig:LGASP}
\end{figure*}

Our galaxy sample contains 15 objects which we consider non-satellite galaxies as they are more distant than 300 kpc from both the MW and M31. Fitting a single plane to their distribution results in a best-fit normal vector pointing to $(l,b) = (227^{\circ}.2, -35^{\circ}.2)$, with an uncertainty of $1^{\circ}.98$. The best-fit plane runs through both the MW ($D_{\mathrm{MW}} = 7.2 \pm 6.5$\ kpc) and M31 ($D_{\mathrm{M31}} = 13.1 \pm 11.9$\ kpc). However, the fit results in an RMS height of $\Delta = 295.1 \pm 4.5$\ and axis ratios of $c/a = 0.469 \pm 0.005$\ and $b/a = 0.647 \pm 0.011$. Therefore, the distribution is not planar, but rather a triaxial ellipsoidal.

We can compare this 'plane' with the LG galaxy planes discussed in earlier works. \citet{Hartwick2000} determine the spatial distribution of 13 galaxies which they consider to be 'relatively isolated' LG galaxies. In contrast to our sample, their galaxies have LG distances of up of to 2.5 Mpc. They describe the galaxy distribution with a flat ellipsoid which has a short axis pointing to $(l,b) = (228^{\circ}.2^{+20^{\circ}.1}_{16^{\circ}.3}, -19^{\circ}.7^{+13^{\circ}.4}_{-7^{\circ}.4})$, which is relatively close to the normal vector we determined for our LG non-satellite galaxy sample.

\citet{Sawa2005} determine a planar distribution of LG galaxies by first investigating their positions on the sky, plotted in Galactic coordinates as seen from the Sun. They identify a ring-like distribution traced by most LG galaxies. To avoid parallax effects due to the projected view, they then look at the three-dimensional positions of the LG galaxies and identify a thin plane (they report a thickness of 50-100 kpc without stating how it was measured) of galaxies which they claim to be responsible for the ring-like distribution. This plane's normal points to $(l,b) = (206^{\circ}, -11^{\circ})$. A look at their figure 3 reveals that those galaxies agreeing best with their LG plane are mostly members of the GPoA, the non-satellites IC 1613 and Phoenix which lie very close to the GPoA, and the MW satellites, which also lie within the GPoA because it is seen edge-on from the MW (see Sect. \ref{sect:discussion}). Consequently, the normal direction of the \citet{Sawa2005} LG plane is close to the GPoA normal direction.

\citet{Pasetto2007} have also determined a best-fitting plane to the same sample of LG galaxies used by \citet{Sawa2005} by applying a principal component analysis technique. They report a plane normal direction of $(l,b) = (-136^{\circ}, -28^{\circ})$, corresponding to $(l,b) = (224^{\circ}, -28^{\circ})$\ in our notation of non-negative Galactic longitude, and a plane thickness estimate of 200 kpc without specifying how this thickness was measured. Using a second method which assumes that the line connecting the MW and M31 lies within the LG plane, they repeat their plane fit, resulting in a plane normal pointing to $(l,b) = (133^{\circ}, -27^{\circ})$. As this normal direction points close to the position of M31 ($[l,b]_{\mathrm{M31}} = [121^{\circ}, -22^{\circ}]$), it can not describe a plane including both the MW and M31. We therefore have to assume that the $l$-component of their second normal direction lacks a minus sign, which would agree with the statement by \citet{Pasetto2007} that the difference between their two planes is small. If this is the case, their second plane fit would have a normal pointing to $(l,b) = (227^{\circ}, -27^{\circ})$\ in our notation. Thus, their results agree well with our plane fitted to all non-satellite galaxies in the LG.

With a RMS height of almost 300 kpc, the single plane fitted to all non-satellite galaxies is much wider than the satellite galaxy planes around the MW and M31. Motivated by the GPoA, which consists of only a sub-sample of M31 satellites, we look for the possibility that there are sub-samples of non-satellite galaxies in the LG which lie in a thinner plane. Fig. \ref{fig:LGASP} shows an Aitoff projection of the distribution of all LG galaxies as seen from the midpoint between the MW and M31 (the origin of our Cartesian coordinate system). The angular coordinate system for this plot is chosen such that the normal-vector of the plane fitted to all 15 non-satellite galaxies defines the north pole, and such that the MW and M31 lie along the equator at longitudes of $L'=90^{\circ}$\ and $L'=270^{\circ}$, respectively. All non-satellite galaxies are plotted as filled points in Fig. \ref{fig:LGASP}, the MW satellite positions are indicated with plus signs and those of the M31 satellites with crosses.

Galaxies which lie within a common plane that contains or passes close to the midpoint between the MW and M31 will lie along a common great-circle in Fig. \ref{fig:LGASP}. This is, for example, the case for the M31 satellites in the GPoA (red symbols), because the GPoA is oriented such that it is seen edge-on from the MW and therefore also from the midpoint between the MW and M31. Two groupings are obvious for the non-satellites. Mostly contained in the upper half of the plot, the LG galaxies UGC 4879, Leo A, Leo T, Phoenix, Tucana, Cetus, WLM, IC 1613 and Andromeda XVI (plotted in yellow) lie along a common 'band' (below, this group will be referred to as LGP1). A second, smaller grouping can be identified in the lower half of the plot, consisting of NGC 6822, Sagittarius dIrr, Aquarius, Andromeda XXVIII and Andromeda XVIII (plotted in green, will be referred to as LGP2). Only the Pegasus dwarf irregular (dIrr) seems to be unrelated to these two bands, as it lies in-between them. It is, however, very close to a number of M31 galaxies (plotted in magenta, see Sect. \ref{sect:other}) which lie close to the plane defined by M31's galactic disc (black ellipse in the plot). 

Fitting planes to the two groups of non-satellite galaxies demonstrates that the galaxies indeed lie within two thin planes.

\subsection{Local Group Galaxy Plane 1 (LGP1)}
\label{subsect:LGP1}

\begin{figure}
 \centering
 \includegraphics[width=88mm]{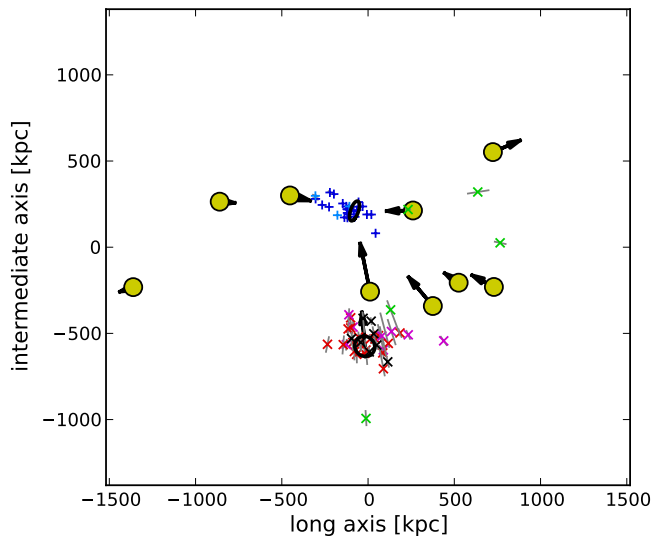}
 \caption{Face-on view of LG Plane 1 (LGP1). Galaxies assigned to this plane are plotted as yellow dots. The horizontal axis is parallel to the major axis of the distribution of galaxies in the plane, while the vertical axis is parallel to the intermediate axis. The black ellipses indicate the positions and orientations of the MW (upper) and M31 (lower). The black arrows show the direction and amount of the Galactocentric velocity $v_{\mathrm{GSR}}$\ of the plane members and M31, a length of 100 kpc represents a velocity of 100 $\mathrm{km~s}^{-1}$\ (note that the tangential velocities are unknown). All other LG galaxies are plotted as crosses, with their colour representing plane membership as in Fig. \ref{fig:LGASP}. The plot is centred on the centroid position $\mathbf{r}_0$\ of LGP1, which in this projection is close to the midpoint between the MW and M31.
 }
 \label{fig:LGP1faceon}
\end{figure}

For the first group, which we denote the Local Group Plane 1 (LGP1), we determine the following plane parameters. The normal vector $\mathbf{n}$\ points to $(l,b) = (220^{\circ}.4, -22^{\circ}.4)$, with an uncertainty of $0^{\circ}.4$. The plane is offset by $D_{\mathrm{MW}} = 177.4 \pm 2.1$\ kpc from the MW and by $D_{\mathrm{M31}} = 168.1 \pm 4.3$\ kpc from M31. It has a RMS height of $\Delta = 54.8 \pm 1.8$\ kpc, less than one fifth of the RMS height of the total non-satellite sample. The axis ratios of $c/a = 0.077 \pm 0.003$\ and $b/a = 0.445 \pm 0.005$ indicate a very thin, planar distribution. A face-on view of this plane is plotted in Fig. \ref{fig:LGP1faceon}.

The only galaxy associated with LGP1 which has a distance of more than two times the RMS plane height is Andromeda XVI, which is offset by $111 \pm 4$~kpc from the best-fit plane. Andromeda XVI lies within the GPoA around M31 (see Sect. \ref{sect:M31}), follows the GPoA velocity trend \citep[i.e. is co-orbiting with the majority of the GPoA satellites, see][]{Ibata2013}, and is only barely classified as a non-satellite by our radial cut at 300 kpc distance (it has a distance of about 320 kpc from M31). Therefore, it might indeed be unrelated to the LGP1, but rather belong to the GPoA. 

None of the MW satellite galaxies is closer than $\Delta$\ from LGP1, but the best-fit plane runs right through the M31 satellite galaxy LGS 3 ($9 \pm 5$~kpc) and passes close to Triangulum/M33 ($30 \pm 5$~kpc). Fitting a plane to the LGP1 sample without Andromeda XVI results in a better alignment of M33 ($11 \pm 5$~kpc) and adds the possible M33 satellite galaxy Andromeda XXII \citep{Chapman2013} to the well-aligning M31 satellites ($23 \pm 16$~kpc), but LGS 3 is then more offset ($47 \pm 5$~kpc).
Two of these galaxies are no GPoA members (Triangulum/M33 and Andromeda XXII), but LGS 3 is the southernmost known M31 satellite which lies within the GPoA. Its line-of-sight velocity, which is similar to the line-of-sight velocity of M31, does not follow the strong co-rotating trend of the majority of GPoA satellites. Therefore, we consider these three galaxies possible members of LGP1. A plane-fit to the larger sample, now consisting of 11 objects (Andromeda XVI excluded; Andromeda XXII, Triangulum/M33, and LGS 3 included) results in a normal vector $\mathbf{n}$\ pointing to $(l,b) = (222^{\circ}.5, -21^{\circ}.8)$, with an uncertainty of $0^{\circ}.4$. It is offset from the MW by $D_{\mathrm{MW}} = 182.2 \pm 2.4$\ and from M31 by $D_{\mathrm{M31}} = 204.2 \pm 4.5$\ kpc. The RMS height of the dwarf galaxies around the best-fit plane is remarkably small given that the plane diameter is 1--2 Mpc: $\Delta = 36.0 \pm 2.1$\ kpc and the axis ratios are $c/a = 0.055 \pm 0.003$\ and $b/a = 0.612 \pm 0.015$. 

For the following discussion, we will use the LGP1 satellite galaxy sample and plane fit parameters determined for the 9 non-satellite galaxies only, but keep the possibly associated M31 satellites in mind. The resulting fit parameters are compiled in Table \ref{tab:allplanes}.

\subsection{Local Group Galaxy Plane 2 (LGP2)}
\label{subsect:LGP2}

\begin{figure}
 \centering
 \includegraphics[width=88mm]{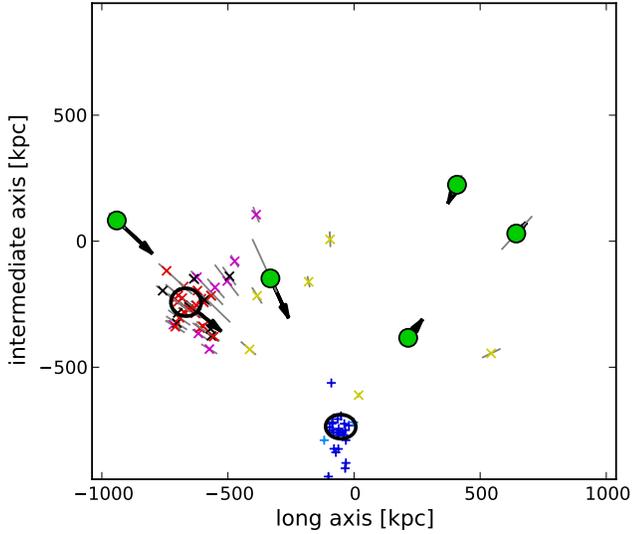}
 \caption{Same as Fig. \ref{fig:LGP1faceon}, but for LG plane 2. The centroid of this plane is offset considerably from the line connecting the MW (upper) and M31 (left).
 }
 \label{fig:LGP2faceon}
\end{figure}

Similarly, the second group of non-satellite galaxies seen in Fig. \ref{fig:LGASP}, which consists of NGC 6822, Sagittarius dIrr, Aquarius, Andromeda XXVIII and Andromeda XVIII, is denoted the Local Group Plane 2 (LGP2). The best-fit plane parameters are: normal vector $\mathbf{n}$\ pointing to $(l,b) = (242^{\circ}.3, -52^{\circ}.9)$, with an uncertainty of $1^{\circ}.7$, $D_{\mathrm{MW}} = 121.5 \pm 17.6$\ kpc, $D_{\mathrm{M31}} = 132.4 \pm 9.4$\ kpc, $\Delta = 65.5 \pm 3.1$\ kpc and axis ratios of $c/a = 0.110 \pm 0.004$\ and $b/a = 0.359 \pm 0.012$. A face-on view of this plane is plotted in Fig. \ref{fig:LGP2faceon}.

The LGP1 member Leo T lies close to LGP2 ($50 \pm 25$~kpc), indicating that the two LG planes intersect close to that galaxy. Nevertheless, we keep Leo T assigned to LGP1 because it is very close to the best-fit LGP1 ($4 \pm 3$~kpc) and together with the nearby LGP1 members Leo A and UGC 4879 traces a common trend in radial distance and line-of-sight velocity (Figs. \ref{fig:LGP1-LGP2} and \ref{fig:LGASP_longitude}).

All three most-recently discovered M31 satellites are close to LGP2: Andromeda XXX ($36 \pm 12$\,kpc), Andromeda XXXI ($43 \pm 8$\,kpc) and Andromeda XXXII ($9 \pm 7$\,kpc), but Andromeda XXX and XXXII are also members of the GPoA. Of the other outer M31 satellites, IC 10, Andromeda VII and Andromeda XXIV are within $2 \times \Delta$\ ($108 \pm 13$, $81 \pm 10$\ and $116 \pm 13$~kpc, respectively), but also several of the inner M31 satellites are close. Similarly, many MW satellites are close to the plane. Most satellite galaxies are, however, closer to one of the satellite galaxy planes than to LGP1 or LGP2, as can be studied in detail by comparing the dwarf galaxy distances from the plane fits compiled in Table \ref{tab:dwarfdist}.

The most-distant LGP2 member from the plane fit is Sagittarius dIrr, which is offset by $103 \pm 2$~kpc. Removing it results in the following plane parameters: normal vector $\mathbf{n}$\ pointing to $(l,b) = (235^{\circ}.9, -49^{\circ}.1)$, with an uncertainty of $0^{\circ}.6$, $D_{\mathrm{MW}} = 138.8 \pm 6.2$\ kpc, $D_{\mathrm{M31}} = 161.9 \pm 5.2$\ kpc, $\Delta = 5.5 \pm 6.2$\ kpc and axis ratios of $c/a = 0.010 \pm 0.011$\ and $b/a = 0.426 \pm 0.015$. Thus, the remaining four dwarf galaxies lie within a very thin plane, but given the small number of galaxies this might not be unexpected. The distant M31 satellite Andromeda XXXI is somewhat closer to this plane fit ($34.7 \pm 7.7$\,kpc).

\subsection{Comparing LGP1 and LGP2}
\label{subsect:LGPcompare}

\begin{figure}
 \centering
 \includegraphics[width=88mm]{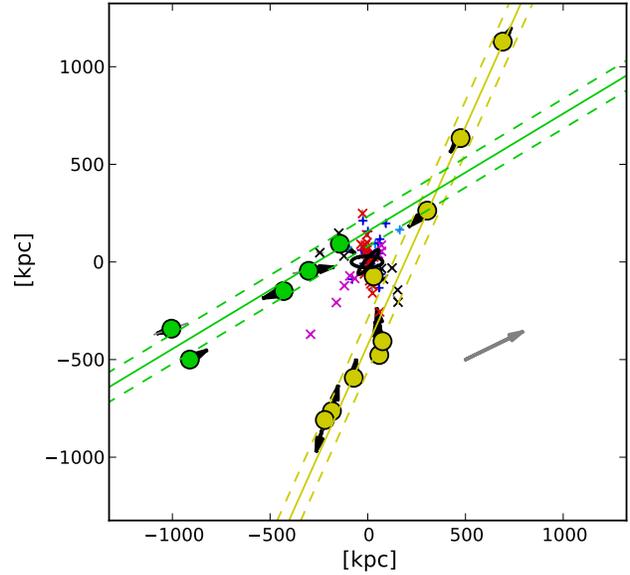}
 \caption{Edge-on view of both LG planes. The orientation of the MW and M31 are indicted as black ellipses in the centre. Members of the LGP1 are plotted as yellow points, those of LGP2 as green points. MW galaxies are plotted as plus signs (+), all other galaxies as crosses ($\times$), the colours code their plane membership as in Fig. \ref{fig:LGASP}.
The best-fit planes are plotted as solid yellow and green lines (for LGP1 and LGP2, respectively), and the dashed lines of the same colours indicate the planes' RMS heights $\Delta$. 
The view direction was determined from the cross-product of the two plane normal vectors, resulting in a projection in which both planes are seen edge-on. The view direction is along $(l,b) = (121.^{\circ}1, -21.^{\circ}4)$, and the Galactic north points up. The view, thus, is along the line connecting the MW and M31, such that the two major galaxies and their surrounding satellites overlap in the centre. Therefore both LG planes are parallel to the MW-M31 line. The two planes cross at about the position of Leo T (Leo A if removing Sagittarius dIrr from LGP2), which therefore might be a member of either plane. As Leo T (Leo A) falls onto the line extending from (connecting) the two nearby LGP1 members UGC 4879 and Leo A (Leo T), we nevertheless consider it to be a member of LGP1.
The black arrows indicate the line-of-sight velocities of the non-satellite galaxies in this projection as in Figs. \ref{fig:LGP1faceon} and \ref{fig:LGP2faceon}. The grey arrow in the lower right indicates the motion of the LG with respect to the CMB rest-frame (Sect. \ref{sect:discussion}). It points approximately into the direction where LGP1 and LGP2 intersect in this projection, but the major component of this velocity is directed along the MW-M31 line (perpendicular to the figure).
 }
 \label{fig:LGP1-LGP2}
\end{figure}

The two bands of galaxies seen in Fig. \ref{fig:LGASP} are indeed indicative of two thin, planar structures within which almost all known non-satellite galaxies in the LG are found (14 out of 15). As expected, the Pegasus dIrr lies in-between the two planes, having a distance of $291 \pm 5$~kpc from LGP1 and $306 \pm 6$~kpc from LGP2. 

The most striking property of these planes is their symmetry. Much of this symmetry is visible in Fig. \ref{fig:LGP1-LGP2}, which shows a view of the LG such that both LG planes are seen edge-on:

\begin{enumerate} 
\renewcommand{\theenumi}{(\arabic{enumi})}
\item Both planes have a similar RMS height (LGP1: $\Delta = 55$~kpc, LGP2: $\Delta = 66$~kpc), but are very wide (diameters of 1-2 Mpc). This is indicated by the dashed lines in Fig. \ref{fig:LGP1-LGP2}.
\item Both planes have a similar offset from the MW (LGP1: $D_{\mathrm{MW}} = 177$~kpc, LGP2: $D_{\mathrm{MW}} = 122$~kpc).
\item Both planes have similar offsets from M31 (LGP1: $D_{\mathrm{M31}} = 168$~kpc, LGP2: $D_{\mathrm{M31}} = 132$~kpc), which at the same time are similar to their offsets from the MW.
\item Thus, both planes are parallel to the line connecting the MW and M31. In Fig. \ref{fig:LGP1-LGP2}, the MW and M31 (whose positions and orientations are shown as black ellipses in the centre of the plot) as well as most of their satellite galaxies are within a 'wedge' formed by LGP1 and LGP2.
\item Both planes have a similar inclination from the galactic disc of M31 (LGP1: $20^{\circ}$, LGP2: $23^{\circ}$), but different inclinations from the Galactic disc of the MW (LGP1: $68^{\circ}$, LGP2: $37^{\circ}$). The black ellipse representing M31 in Fig. \ref{fig:LGP1-LGP2} is also seen almost edge-on and its orientation is similar to the two planes (major axis running from lower left to upper right).
\item Both planes cross the outer parts of the satellite galaxy distribution of the MW and M31. 
In particular the northernmost (IC 10) and southernmost (LGS 3) M31 satellites in the GPoA (respectively the uppermost and lowermost red cross in Fig. \ref{fig:LGP1-LGP2}) are each close to one of the planes (LGP2 and LGP1, respectively). LGS 3, the southernmost GPoA member lies close to LGP1 and does not follow the strong line-of-sight velocity trend of the GPoA. Similarly, IC 10 is the northernmost GPoA member, lies relatively close to the LGP2 and does not follow the strong line-of-sight velocity trend of the GPoA either.
\end{enumerate}

\begin{figure}
 \centering
 \includegraphics[width=88mm]{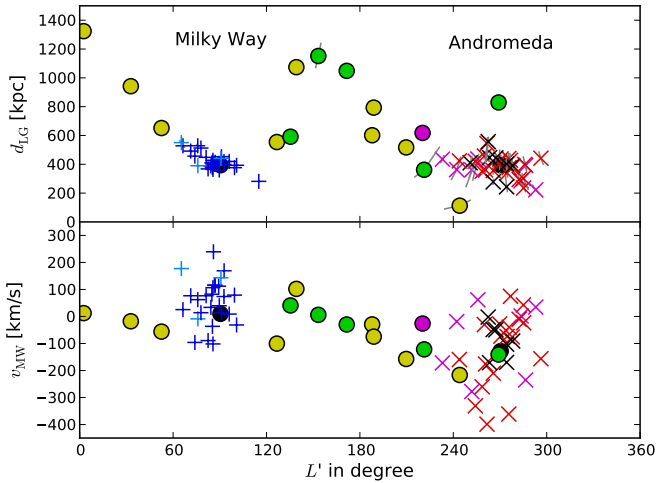}
 \caption{Other properties of the LG galaxies shown in Fig. \ref{fig:LGASP}. The symbols and colours are the same, and the positions on the horizontal axis are given by the longitude $L'$\ also used in Fig. \ref{fig:LGASP}. 
The \textit{upper panel} plots the radial distance $d_{\mathrm{LG}}$\ from the midpoint between the MW and M31 (i.e. the distance of each galaxy from the point of view adopted in Fig. \ref{fig:LGASP}). Uncertainties are again shown as grey lines. The LG galaxies follow a pronounced trend. From left to right, they are first found at large distances ($\approx 1$~Mpc), then approach the MW distance. To the right of the MW, the distances of non-satellites increase again to more than 1 Mpc, but then approach the M31 distance. Thus, the non-satellite galaxies are not only ordered in their projected position, but also follow a common radial distance behaviour. It is also interesting that no galaxies are known in the region spanning about $60^\circ$\ to the right of the M31 satellite system.
The \textit{lower panel} plots $v_{\mathrm{GSR}}$, the line-of-sight velocity with respect to the Galactic standard of rest. This is the only velocity component available for the non-satellite galaxies. It is measured from the position of the Sun, so the angle between this velocity direction and the line-of-sight adopted for Fig. \ref{fig:LGASP} varies for the different objects. There appears to be a trend of decreasing $v_{\mathrm{GSR}}$\ with increasing $L'$\ for the non-satellites, with a break at the MW position (see lower panel of Fig. \ref{fig:MagStream}, too).
 }
 \label{fig:LGASP_longitude}
\end{figure}

Looking for further indications of coherence in the structure of the LG planes, in the upper panel of Fig. \ref{fig:LGASP_longitude} we have plotted the radial distance from the midpoint between the MW and M31 against the longitude $L'$\ in the coordinate system of Fig. \ref{fig:LGASP}, approximately along the two bands. This reveals a seemingly ordered behaviour of the radial distances of the non-satellite galaxies: Starting with UGC 4879, the most-distant LG galaxy in our sample, on the left near $L' = 0^{\circ}$, the LG galaxy distance decreases systematically as as we move closer to the MW in $L'$. The MW satellites then follow the same trend of decreasing distance with increasing $L'$. To the right of the MW the LGP1 and LGP2 members seem to follow a similar radial behaviour. The galaxy distances now increase with increasing $L'$, towards a maximum of about 1.2 Mpc (Sagitarrius dIrr) between the MW and M31 position. At larger $L'$, the galaxy distances decrease almost monotonously towards M31 and its satellite galaxies. Only Andromeda XVI (rightmost yellow point) is much closer to the origin/point of view than M31, and Andromeda XVIII (rightmost green point) lies further away than M31. The face-on view of LGP1 in Fig. \ref{fig:LGP1faceon} reveals this arc-like distribution, too, which starts with UGC 4879 at the tip of the long axis, and then passes through the position of the MW from where it bends down to end at the position of M31.

The lower panel of Fig. \ref{fig:LGASP_longitude} plots the Galactocentric line-of-sight velocities of the LG galaxies against $L'$. Qualitatively, the trend of the line-of-sight velocities is similar to that of the distances in the upper panel: from left-to-right, the non-satellite galaxy velocities first become more negative (approaching the MW) with increasing $L'$, then rise and again drop almost monotonically between the MW and M31. Again, the LGP1 and LGP2 members follow a similar trend.
The lower panel of Fig. \ref{fig:LGASP_longitude} also shows the velocity trend of the GPoA satellites (red crosses). On the left of M31, most velocities are more-negative than those of M31, while the opposite is the case on the right side. The two outermost GPoA satellites, LGS 3 (leftmost red cross) and IC 10 (rightmost red cross) do not follow this trend, but they lie close to LGP1 and LGP2, respectively, and have velocities similar to the LGP1 and LGP2 members in the vicinity of M31. This might be another indication that they are better understood as LG plane members or as a connection between the LG planes and the GPoA.

\subsection{Consistency check}
\label{subsect:LGPconsistency}

\begin{figure*}
 \centering
 \includegraphics[width=180mm]{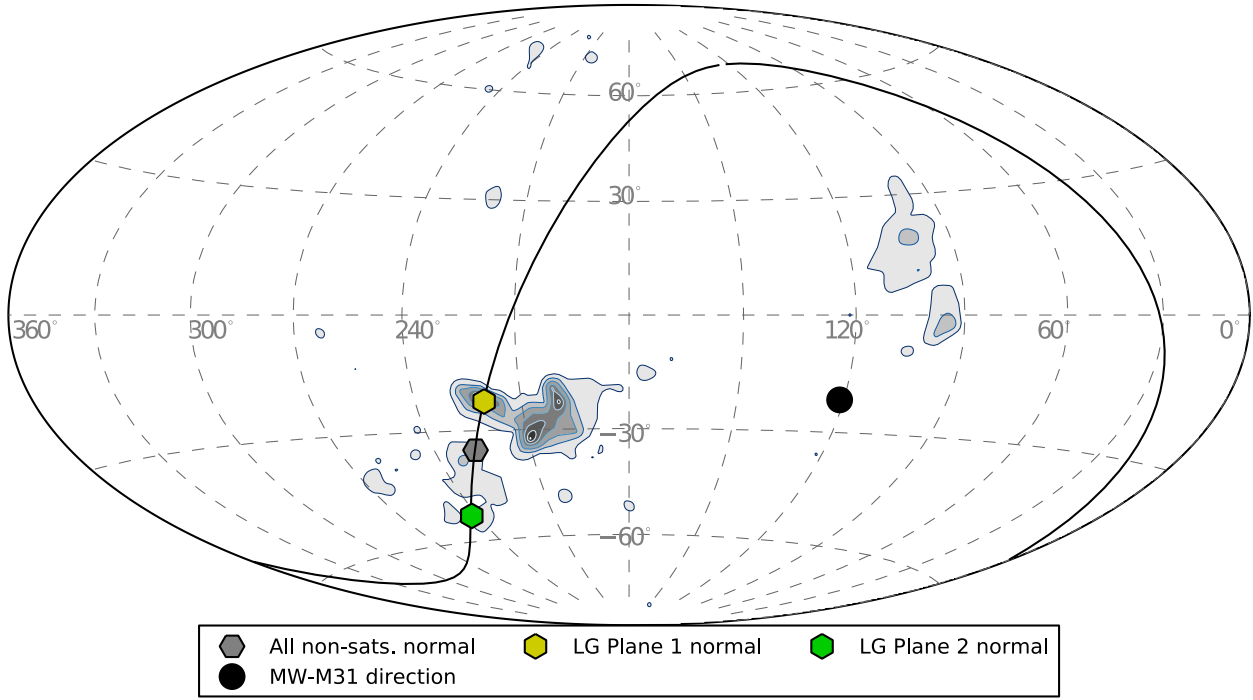}
 \caption{Density distribution of the 4-galaxy-normal directions for our sample of 15 non-satellite galaxies in the LG, determined and weighted as explained in Sect. \ref{subsect:4dwarfnormals}. They show two very pronounced peaks at $(l,b) \approx (220^{\circ}, -20^{\circ})$\ (Peak 1) and $(l,b) \approx (210^{\circ}, -30^{\circ})$\ (Peak 2), and a smaller one at $(l,b) \approx (200^{\circ}, -20^{\circ})$ (Peak 3), which are are close to each other. The contributions of the different non-satellite galaxies to these three peaks are shown in Fig. \ref{fig:LG4contribution}.
 As in Figs. \ref{fig:MW4dwarfnormal} and \ref{fig:M314dwarfnormal} only the 4-galaxy-normals in the centre of the plot between $l = 90^{\circ}$\ and $l = 270^{\circ}$\ are shown, not their mirrored counterparts.
The black point indicates this direction of the line connecting the MW and M31, with the black line being the great circle around this direction. It illustrates all possible directions of normal vectors for planes parallel to the line connecting the MW and M31. Also plotted are the normal to the plane fitted to all 15 non-satellite galaxies (grey hexagon) and the normals of the LG planes 1 and 2 (yellow and green hexagons, respectively, i.e. the same colours in which their members are marked in Figs. \ref{fig:LGASP} and \ref{fig:LGASP_longitude}). The normal to LG plane 1 coincides with Peak 1, and the 4-galaxy-normal contribution plot for this peak in Fig. \ref{fig:LG4contribution} demonstrates that only LG plane 1 members contribute to this peak.
 }
 \label{fig:LG4dwarfnormal}
\end{figure*}

\begin{figure}
 \centering
 \includegraphics[width=88mm]{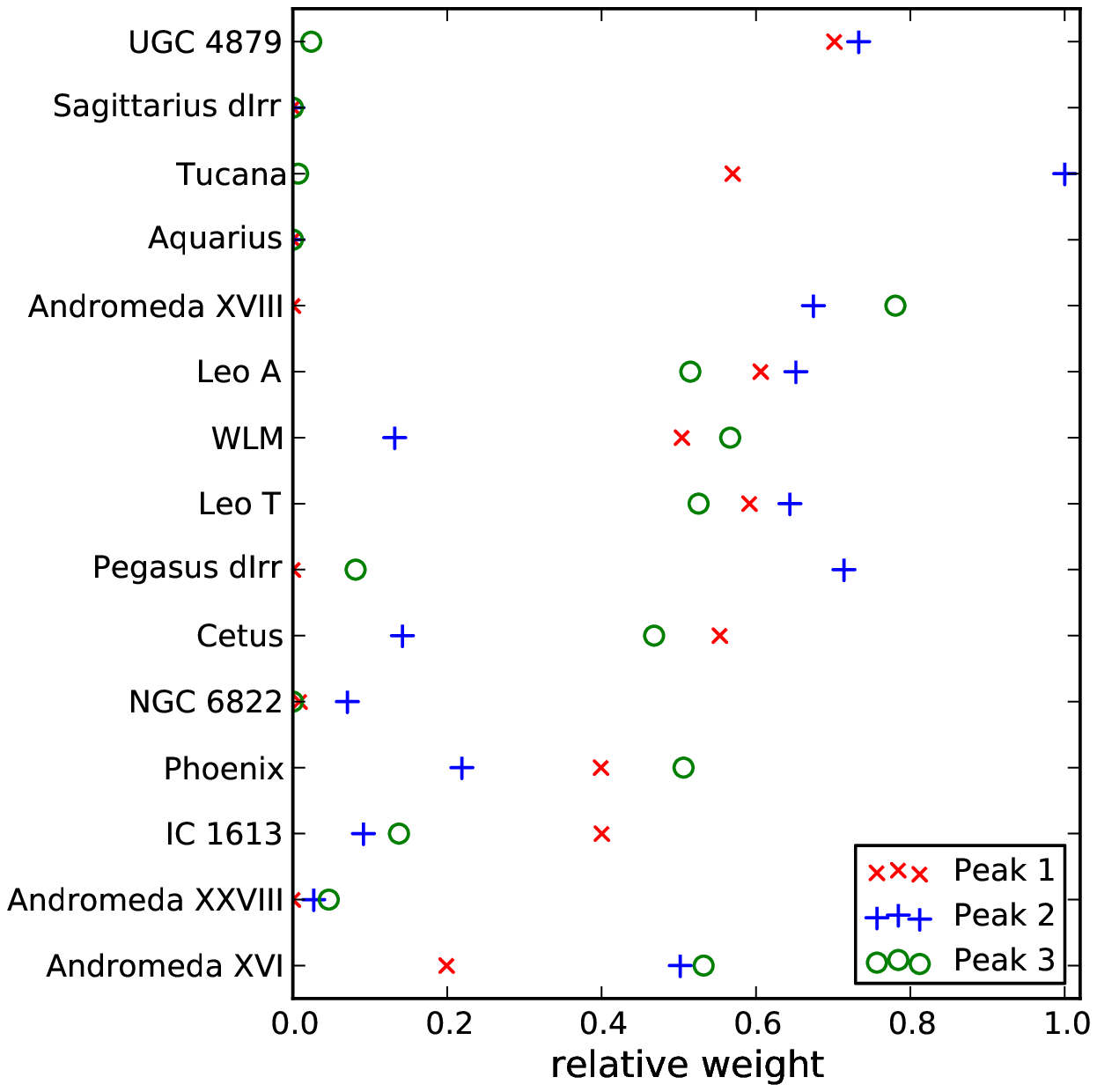}
 \caption{Relative weight contributions of the different non-satellite LG galaxies to the three peaks identified in Fig. \ref{fig:LG4dwarfnormal}. For each peak, the contributions of the 4-galaxy-normals pointing to within $5^{\circ}$\ (instead of $15^{\circ}$\ as used for Figs. \ref{fig:MWpeakcontribution} and \ref{fig:M31peakcontribution}) around the peak positions are shown. This smaller angle was chosen to avoid an overlap in the peak areas due to the closeness of the three peaks.
 }
 \label{fig:LG4contribution}
\end{figure}

The following discussion (Sect \ref{sect:discussion}) is restricted to the LGP1 and LGP2, but before proceeding we have to discuss whether there might be a different dominant plane in the LG non-satellite galaxy distribution. As a consistency test, Fig. \ref{fig:LG4dwarfnormal} shows the density distribution of all 1365 possible 4-galaxy-normal directions for the 15 non-satellite galaxies. As before, 1000 realisations of the galaxy positions varying their distances within the uncertainties have been combined. The plot reveals that the LGP1 normal direction coincides with a pronounced peak (Peak 1 at $[l,b] \approx [220^{\circ}, -20^{\circ}]$) in the density distribution, lending further support to our discovery. However, there is a second strong peak at $(l,b) \approx (210^{\circ}, -30^{\circ})$\ (Peak 2), and another nearby but smaller peak at $(l,b) \approx (200^{\circ}, -20^{\circ})$\ (Peak 3). 

We have determined which dwarf galaxies contribute to each of the three peaks (Fig. \ref{fig:LG4contribution}). As expected, the nine galaxies contributing significantly to Peak 1 are identical to those we assign to LGP1. The contributions to the other two peaks are dominated by only seven galaxies each. These are (sorted by their relative contribution): Tucana, UGC 4879, Pegasus dIrr, Andromeda XVIII, Leo A, Leo T and Andromeda XVI for Peak 2 and Andromeda XVIII, WLM, Andromeda XVI, Leo T, Leo A, Phoenix, and Cetus for Peak 3. Five out of the seven galaxies contributing to Peak 2 are members of LGP1, and even six of seven galaxies contributing to Peak 3 are LGP1 members. The galaxy samples contributing to Peaks 2 and 3 are therefore very similar to the LGP1 sample and trace a similar structure, as indicated by their proximity to the LGP1 peak. Fitting planes to the Peak 2 and Peak 3 galaxy samples confirms the similar orientation. For Peak 2, the fit results in a normal vector $\mathbf{n}$\ pointing to $(l,b) = (208^{\circ}.5, -31^{\circ}.0)$, with an uncertainty of $0^{\circ}.19$, $D_{\mathrm{MW}} =  128.1 \pm 1.8$\ kpc, $D_{\mathrm{M31}} = 61.7 \pm 4.1$\ kpc, $\Delta = 27.5 \pm 2.2$\ kpc and axis ratios of $c/a = 0.039 \pm 0.003$\ and $b/a = 0.754 \pm 0.010$. For Peak 3, the fit gives a normal vector $\mathbf{n}$\ pointing to $(l,b) = (200^{\circ}.6, -21^{\circ}.4)$, with an uncertainty of $0^{\circ}.35$, $D_{\mathrm{MW}} = 209.8 \pm 2.6$\ kpc, $D_{\mathrm{M31}} = 33.4 \pm 5.4$\ kpc, $\Delta = 21.5 \pm 2.1$\ kpc and axis ratios of $c/a = 0.037 \pm 0.003$\ and $b/a = 0.677 \pm 0.012$.

With the presently-known galaxies, the previously-determined LGP1 and LGP2 are therefore not the only possible planar structures in the LG. However, they contain all but one of the 15 non-satellite LG dwarf galaxies (which in turn might be related to the M31 disc plane, see the following Section) and at the same time exhibit a striking symmetry in their parameters. This is not the case for the two dwarf galaxy planes which give rise to Peak 2 and Peak 3, which each contain only 7 out of 15 galaxies.

In the future it will be important to study the statistical significance of the here discovered symmetries of the LG, but currently no conclusive meaningful test is available since a model is required as a null hypothesis. Furthermore, such a test has to take observational biases like the sky coverage of surveys searching for LG dwarf galaxies into account. Due to the very inhomogeneous nature of the galaxy data, this is currently not feasible. It would appear rather clear though that a distribution process which is inherently stochastic will not be able to deliver present-day positions of galaxies within the LG which end up being as symmetrically distributed as is observed.

\section{The remaining dwarf galaxies}
\label{sect:other}

Out of the 76 non-host galaxies in our LG sample, 16 galaxies are currently not associated with any plane. Except for Pegasus dIrr, all of them are M31 satellites. 
Some of these non-associated galaxies have already been discussed as possible LG plane members due to their closeness to the best-fit planes. For LGP1, the closest non-associated M31 satellites are Triangulum/M33 ($31 \pm 6$~kpc), Andromeda XXIII ($40 \pm 6$~kpc), Andromeda II ($60 \pm 4$~kpc) and Andromeda XXII ($70 \pm 18$~kpc). For LGP2, these are Andromeda XXI ($40 \pm 7$~kpc), Andromeda XXXI ($43 \pm 8$\,kpc) and Andromeda VII ($81 \pm 10$~kpc).

Is there any other planar structure to be found? Of the 16 non-associated galaxies, 5 are within less than $4^\circ$\ of the galactic disc plane of M31 (Andromeda V, Andromeda VI, Andromeda XIX, Andromeda XXIX and Pegasus dIrr). In particular the non-satellite Pegasus dIrr lies perfectly within the plane defined by M31's disc orientation. Another three (Andromeda X, Andromeda XX and Andromeda XXIV) are within seven to 16 degree of M31's disc plane. These eight objects are marked with magenta symbols in the Figures. Except for Andromeda V, which has a distance of $83 \pm 2$~kpc from LGP2, none of these galaxies are closer than 140 kpc from the two LG planes\footnote{This is expected because the two LG planes are inclined by only $20^{\circ}$\ to M31's galactic disc plane and offset from M31's centre by more than 100 kpc. Therefore, the planes do not come close to M31's galactic disc plane.}. All non-GPoA members are further than $\approx 60$~kpc ($\approx 4 \Delta$) from the GPoA.

For an object with a randomly chosen position, the chance to be within 4 degree of an independently given plane is 7 per cent. The probability to find at least five out of 16 objects within 4 degree of such a plane is about 0.4 per cent, assuming that the objects are randomly distributed. This is not the case here, because all galaxies within the GPoA have already been excluded from the sample, such that the probability will be considerably higher. More importantly, the existence of the co-orbiting GPoA also disproves the assumption that the satellites are randomly distributed.

If we nevertheless group the eight galaxies close to M31's galactic disc in a tentative 'M31 disc plane', a plane fitted to them has a normal vector $\mathbf{n}$\ pointing to $(l,b) = (222^{\circ}.0, -38^{\circ}.1)$, with an uncertainty of $2^{\circ}.9$. The best-fit plane is therefore inclined by $18^\circ$\ to M31's galactic disc. It has only a small offset of $D_{\mathrm{M31}} = 6.7 \pm 3.8$\ kpc from the centre of M31. The RMS height of the galaxies around the best-fit plane is $\Delta = 13.5 \pm 1.0$\ kpc and the axis ratios are $c/a = 0.069 \pm 0.005$\ and $b/a = 0.345 \pm 0.093$.

While these eight galaxies thus lie within a common, thin plane, their line-of-sight velocities do not indicate a preferentially co-orbiting association. This can be seen in Fig. \ref{fig:LGASP_longitude}. In contrast to the GPoA members (red crosses), which have preferentially faster line-of-sight velocities than M31 in one side of M31 and slower line-of-sight velocities on the other, the few galaxies associated to the M31 disc plane for which line-of-sight velocities are known (magenta crosses) do not show a pronounced trend, but a mixture of co- and counter-orbiting objects can not be ruled out. We therefore caution against overrating the possible existence of this second plane of M31 satellites.

A similar plane of galaxies aligned with the Galactic disc of the MW would probably remain undetected due to the difficulty in discovering satellite galaxies obscured by the Galactic disc. The lowest-latitude MW satellite is the Sagittarius dSph at $b = -14^{\circ}$. The nearby (7 kpc) Canis Major over-density is situated at even lower Galactic latitude ($b = -8^{\circ}$), but it might be a substructure in the Galactic disc of the MW and not a MW satellite galaxy \citep{Momany2006}.

\begin{table*}
\begin{minipage}{180mm}
 \small
 \caption{Satellite and Local Group dwarf galaxy planes}
 \label{tab:allplanes}
 \begin{center}
 \begin{tabular}{@{}lcccccc}
 \hline 
Name & VPOSall & VPOS-3 & GPoA & LGP1 & LGP2 & M31 disc plane\\
 \hline
Introduced in & Sect. \ref{sect:MW} & Sect. \ref{sect:MW} & Sect. \ref{sect:M31} & Sect. \ref{subsect:LGP1} & Sect. \ref{subsect:LGP2} & Sect. \ref{sect:other} \\
Type & MW satellites & MW satellites & M31 satellites & non-satellites & non-satellites & M31 satellites \\
 & & & & & & and one non-satellite \\ 
$r_0 \begin{pmatrix} x \\ y \\ z \end{pmatrix}$\ [kpc] & $\begin{pmatrix}  176.4 \pm  0.3 \\ -322.1 \pm  0.3 \\  188.1 \pm  0.6 \end{pmatrix}$  & $\begin{pmatrix}  178.6 \pm  0.2 \\ -321.3 \pm  0.3 \\  178.4 \pm  0.6 \end{pmatrix}$  & $\begin{pmatrix} -197.1 \pm  3.8 \\  322.4 \pm  6.1 \\ -142.7 \pm  3.7 \end{pmatrix}$  & $\begin{pmatrix} -3.3 \pm  3.2 \\ -250.5 \pm  2.9 \\ -54.7 \pm  5.4 \end{pmatrix}$  & $\begin{pmatrix}  525.3 \pm  10.2 \\  267.9 \pm  18.8 \\ -195.5 \pm  8.6 \end{pmatrix}$  & $\begin{pmatrix} -101.6 \pm  5.0 \\  304.4 \pm  10.6 \\ -217.3 \pm  6.2 \end{pmatrix}$ \\ 
$n \begin{pmatrix} l \\ b \end{pmatrix}$\ [$^{\circ}$] & $\begin{pmatrix}  155.6 \\ -3.3 \end{pmatrix}$  & $\begin{pmatrix}  169.5 \\ -2.8 \end{pmatrix}$  & $\begin{pmatrix}  205.8 \\  7.6 \end{pmatrix}$  & $\begin{pmatrix}  220.4 \\ -22.4 \end{pmatrix}$  & $\begin{pmatrix}  242.3 \\ -52.9 \end{pmatrix}$  & $\begin{pmatrix}  222.0 \\ -38.1 \end{pmatrix}$ \\ 
$\Delta n$\ [$^{\circ}$] & 1.12 & 0.43 & 0.79 & 0.41 & 1.72 & 2.87\\ 
$D_{\mathrm{MW}}$\ [kpc] & $ 7.9 \pm 0.3$ & $ 10.4 \pm 0.2$ & $ 30.1 \pm 8.8$ & $ 177.4 \pm 2.1$ & $ 121.5 \pm 17.6$ & $ 77.8 \pm 35.6$\\ 
$D_{\mathrm{M31}}$\ [kpc] & $ 637.3 \pm 13.0$ & $ 509.9 \pm 10.2$ & $ 1.3 \pm 0.6$ & $ 168.1 \pm 4.3$ & $ 132.4 \pm 9.4$ & $ 6.7 \pm 3.8$\\ 
$\Delta$\ [kpc] & $29.3 \pm 0.4$ & $19.9 \pm 0.3$ & $13.6 \pm 0.2$ & $54.8 \pm 1.8$ & $65.5 \pm 3.1$ & $13.5 \pm 1.0$\\ 
$c/a$ & $0.301 \pm 0.004$ & $0.209 \pm 0.002$ & $0.107 \pm 0.005$ & $0.077 \pm 0.003$ & $0.110 \pm 0.004$ & $0.069 \pm 0.005$\\ 
$b/a$ & $0.576 \pm 0.007$ & $0.536 \pm 0.006$ & $0.615 \pm 0.058$ & $0.445 \pm 0.005$ & $0.359 \pm 0.012$ & $0.345 \pm 0.093$\\ 
$N_{\mathrm{members}}$ & 27 & 24 & 19 & 9 & 5 & 8\\ 
 \hline
 \end{tabular}
 \end{center}
 \small \medskip
Parameters of the plane fits discussed in Sects. \ref{sect:MW} to \ref{sect:other}. These are:
$\mathbf{r}_0$: $x$-, $y$- and $z$-position of the centroid of the plane in the coordinate system introduced in Sect. \ref{subsect:dataset}.
$\mathbf{n}$: The direction of the normal vector (minor axis) of the best-fit plane in Galactic longitude $l$\ and latitude $b$.
$\Delta_{\mathrm{n}}$: Uncertainty in the normal direction. This and all other uncertainties were determined by varying the galaxy positions within their uncertainties and then determining the standard deviation in the resulting plane parameters.
$D_{\mathrm{MW}}$\ and $D_{\mathrm{M31}}$: offset of the planes from the MW and M31 position.
$\Delta$: RMS height of the galaxies from the best-fit plane.
$c/a$\ and $b/a$: ratios of the short and intermediate axis to the long axis, determined from the RMS heights in the directions of the three axes.
$N_{\mathrm{members}}$: Number of galaxies associated with the planes used for the fitting. In particular LGP1 and LGP2 might have additional satellite galaxies as members, but these were not included in the plane fits compiled here.
\end{minipage}
\end{table*}

\begin{table*}
\begin{minipage}{180mm}
 \tiny
 \caption{Distances and offsets of galaxies from the MW, M31 and the galaxy planes}
 \label{tab:dwarfdist}
 \begin{center}
 \begin{tabular}{@{}lccccccccc}
 \hline 
Name &  $d_{\mathrm{MW}}$ & $d_{\mathrm{M31}}$ & category & MW VPOSall & MW VPOS-3 & GPoA & LGP1 & LGP2 & M31 disc plane \\
 \hline
The Galaxy &  0.0 & 787.6 & host & $      7.9 \pm 0.3$ & $      10.4 \pm 0.2$ & $      30.1 \pm 8.8$ & $      177.4 \pm 2.1$ & $      121.5 \pm 17.6$ & $      77.8 \pm 35.6$\\ 
Canis Major & 13.6 & 786.3 & MW sat. & $ \bf   0.5 \pm 0.2$ & $ \bf   0.6 \pm 0.3$ & $      16.3 \pm 8.6$ & $      164.1 \pm 2.1$ & $      129.4 \pm 17.9$ & $      66.2 \pm 35.6$\\ 
Sagittarius dSph & 18.2 & 791.7 & MW sat. & $ \bf   22.9 \pm 0.9$ & $ \bf   27.5 \pm 1.0$ & $      48.4 \pm 9.0$ & $      189.3 \pm 2.2$ & $      120.5 \pm 17.2$ & $      85.7 \pm 35.9$\\ 
Segue (I) & 28.0 & 792.4 & MW sat. & $ \bf   5.2 \pm 0.3$ & $ \bf   6.8 \pm 0.5$ & $      7.8 \pm 6.5$ & $      164.2 \pm 2.1$ & $      117.1 \pm 18.3$ & $      72.3 \pm 35.7$\\ 
Ursa Major II & 38.1 & 771.1 & MW sat. & $ \bf   26.3 \pm 1.9$ & $ \bf   23.4 \pm 1.8$ & $      7.1 \pm 6.0$ & $      169.8 \pm 2.0$ & $      106.7 \pm 18.2$ & $      78.1 \pm 34.4$\\ 
Bootes II & 39.4 & 807.0 & MW sat. & $ \bf   17.0 \pm 0.5$ & $ \bf   19.4 \pm 0.3$ & $      30.0 \pm 9.0$ & $      196.3 \pm 2.2$ & $      88.5 \pm 18.1$ & $      105.5 \pm 36.2$\\ 
Segue II & 40.9 & 753.4 & MW sat. & $ \bf   29.2 \pm 0.8$ & $ \bf   25.6 \pm 0.7$ & $      10.9 \pm 7.5$ & $      154.4 \pm 2.0$ & $      140.8 \pm 17.5$ & $      52.5 \pm 33.5$\\ 
Willman 1 & 43.0 & 780.7 & MW sat. & $ \bf   21.6 \pm 2.1$ & $ \bf   19.6 \pm 2.1$ & $      7.0 \pm 5.9$ & $      174.9 \pm 2.1$ & $      95.9 \pm 18.6$ & $      86.9 \pm 34.9$\\ 
Coma Berenices & 44.9 & 802.7 & MW sat. & $ \bf   2.6 \pm 0.3$ & $ \bf   2.8 \pm 0.2$ & $      12.6 \pm 8.2$ & $      184.7 \pm 2.3$ & $      89.4 \pm 18.6$ & $      97.8 \pm 36.0$\\ 
Bootes III & 45.8 & 800.5 & MW sat. & $ \bf   8.9 \pm 0.5$ & $ \bf   12.9 \pm 0.4$ & $      28.3 \pm 8.9$ & $      200.6 \pm 2.3$ & $      79.5 \pm 18.2$ & $      111.5 \pm 35.8$\\ 
LMC & 50.0 & 811.4 & MW sat. & $ \bf   24.1 \pm 0.8$ & $ \bf   16.4 \pm 0.5$ & $      14.8 \pm 8.7$ & $      140.5 \pm 2.3$ & $      167.1 \pm 17.9$ & $      42.1 \pm 32.0$\\ 
SMC & 61.2 & 811.7 & MW sat. & $ \bf   38.0 \pm 1.2$ & $ \bf   32.6 \pm 0.9$ & $      34.2 \pm 9.2$ & $      147.9 \pm 2.3$ & $      175.4 \pm 17.4$ & $      42.7 \pm 32.2$\\ 
Bootes (I) & 63.9 & 820.0 & MW sat. & $ \bf   25.7 \pm 0.7$ & $ \bf   28.8 \pm 0.6$ & $      34.8 \pm 9.1$ & $      212.1 \pm 2.4$ & $      66.1 \pm 18.4$ & $      126.0 \pm 36.6$\\ 
Draco & 75.9 & 754.9 & MW sat. & $ \bf   21.4 \pm 1.4$ & $ \bf   3.6 \pm 0.5$ & $      49.1 \pm 8.4$ & $      231.2 \pm 3.3$ & $      51.2 \pm 17.4$ & $      137.8 \pm 33.3$\\ 
Ursa Minor & 77.9 & 758.3 & MW sat. & $ \bf   32.6 \pm 1.2$ & $ \bf   19.2 \pm 0.6$ & $      26.0 \pm 8.3$ & $      214.6 \pm 2.3$ & $      54.8 \pm 17.9$ & $      126.6 \pm 33.3$\\ 
Sculptor & 86.0 & 765.8 & MW sat. & $ \bf   2.4 \pm 0.6$ & $ \bf   2.7 \pm 0.6$ & $      32.9 \pm 8.7$ & $      133.5 \pm 2.4$ & $      200.4 \pm 16.8$ & $      30.3 \pm 22.7$\\ 
Sextans (I) & 89.1 & 839.0 & MW sat. & $ \bf   2.0 \pm 0.7$ & $ \bf   12.9 \pm 0.5$ & $      37.4 \pm 9.4$ & $      138.3 \pm 2.5$ & $      114.5 \pm 20.0$ & $      62.2 \pm 37.4$\\ 
Ursa Major (I) & 101.7 & 777.2 & MW sat. & $ \bf   53.3 \pm 1.4$ & $      51.4 \pm 1.5$ & $      28.0 \pm 8.5$ & $      176.2 \pm 2.0$ & $      62.2 \pm 19.3$ & $      101.9 \pm 33.9$\\ 
Carina & 106.9 & 842.0 & MW sat. & $ \bf   24.5 \pm 1.5$ & $ \bf   1.3 \pm 0.7$ & $      31.5 \pm 9.6$ & $      83.3 \pm 3.8$ & $      214.7 \pm 19.2$ & $      34.6 \pm 23.3$\\ 
Hercules & 126.0 & 826.6 & MW sat. & $ \bf   71.5 \pm 2.2$ & $      93.6 \pm 4.7$ & $      123.1 \pm 10.3$ & $      306.1 \pm 7.2$ & $      14.3 \pm 10.9$ & $      211.7 \pm 37.0$\\ 
Fornax & 149.4 & 772.6 & MW sat. & $ \bf   16.7 \pm 1.5$ & $ \bf   29.5 \pm 0.9$ & $      14.2 \pm 7.6$ & $      59.7 \pm 5.6$ & $      277.2 \pm 18.4$ & $      66.1 \pm 33.8$\\ 
Leo IV & 154.8 & 901.2 & MW sat. & $ \bf   38.6 \pm 1.0$ & $ \bf   18.0 \pm 0.7$ & $      39.8 \pm 10.0$ & $      164.7 \pm 2.6$ & $      64.7 \pm 21.1$ & $      105.6 \pm 40.0$\\ 
Canes Venatici II & 160.6 & 837.5 & MW sat. & $ \bf   6.1 \pm 1.1$ & $ \bf   1.5 \pm 0.6$ & $      7.0 \pm 5.9$ & $      238.7 \pm 2.5$ & $      19.0 \pm 16.0$ & $      178.2 \pm 35.9$\\ 
Leo V & 178.7 & 915.1 & MW sat. & $ \bf   37.5 \pm 1.2$ & $ \bf   14.2 \pm 0.8$ & $      53.0 \pm 10.3$ & $      164.7 \pm 2.7$ & $      50.7 \pm 20.9$ & $      113.2 \pm 40.3$\\ 
Pisces II & 181.1 & 660.2 & MW sat. & $ \bf   38.0 \pm 1.6$ & $ \bf   4.2 \pm 0.9$ & $      114.3 \pm 7.4$ & $      209.8 \pm 1.7$ & $      161.0 \pm 13.5$ & $      66.8 \pm 29.9$\\ 
Canes Venatici (I) & 217.5 & 863.9 & MW sat. & $ \bf   6.3 \pm 1.7$ & $ \bf   16.7 \pm 0.9$ & $      19.3 \pm 8.9$ & $      286.5 \pm 3.7$ & $      78.2 \pm 21.1$ & $      236.6 \pm 36.3$\\ 
Leo II & 236.0 & 901.5 & MW sat. & $ \bf   26.4 \pm 1.0$ & $ \bf   46.5 \pm 0.9$ & $      98.7 \pm 10.4$ & $      169.8 \pm 2.6$ & $      18.7 \pm 15.5$ & $      138.4 \pm 38.1$\\ 
Leo I & 257.5 & 922.1 & MW sat. & $ \bf   44.4 \pm 1.4$ & $      83.4 \pm 3.0$ & $      166.2 \pm 11.8$ & $      88.0 \pm 3.8$ & $      61.1 \pm 23.7$ & $      61.7 \pm 38.6$\\ 
Andromeda & 787.6 &  0.0 & host & $      637.3 \pm 13.0$ & $      509.9 \pm 10.2$ & $      1.3 \pm 0.6$ & $      168.1 \pm 4.3$ & $      132.4 \pm 9.4$ & $      6.7 \pm 3.8$\\ 
M32 & 809.5 & 22.7 & M31 sat. & $      685.7 \pm 37.6$ & $      548.7 \pm 30.0$ & $ \bf   1.8 \pm 1.3$ & $      165.6 \pm 4.7$ & $      138.1 \pm 9.0$ & $      16.6 \pm 8.4$\\ 
Andromeda IX & 770.0 & 40.5 & M31 sat. & $      643.0 \pm 12.7$ & $      522.2 \pm 10.0$ & $ \bf   30.7 \pm 0.9$ & $      154.7 \pm 4.1$ & $      123.2 \pm 10.2$ & $      8.8 \pm 3.2$\\ 
NGC 205 & 828.2 & 41.6 & M31 sat. & $      668.3 \pm 13.2$ & $      532.9 \pm 10.4$ & $ \bf   2.7 \pm 0.8$ & $      176.3 \pm 4.5$ & $      124.1 \pm 9.0$ & $      4.8 \pm 3.5$\\ 
Andromeda I & 748.8 & 58.4 & M31 sat. & $      597.4 \pm 12.4$ & $      479.9 \pm 9.5$ & $ \bf   0.6 \pm 0.5$ & $      145.1 \pm 4.0$ & $      172.3 \pm 9.4$ & $      34.5 \pm 2.6$\\ 
Andromeda XVII & 731.9 & 70.0 & M31 sat. & $      601.0 \pm 17.1$ & $      477.1 \pm 13.3$ & $ \bf   5.4 \pm 0.6$ & $      195.6 \pm 4.0$ & $      90.8 \pm 10.0$ & $      32.9 \pm 2.2$\\ 
Andromeda XXVII & 832.1 & 74.2 & M31 sat. & $      693.1 \pm 22.8$ & $      551.0 \pm 18.1$ & $ \bf   1.3 \pm 1.1$ & $      204.1 \pm 4.6$ & $      73.1 \pm 9.5$ & $      35.8 \pm 7.1$\\ 
Andromeda III & 751.9 & 75.2 & M31 sat. & $      578.4 \pm 12.5$ & $      456.4 \pm 9.6$ & $ \bf   30.9 \pm 0.5$ & $      162.3 \pm 4.1$ & $      175.2 \pm 8.7$ & $      25.6 \pm 2.4$\\ 
Andromeda XXV & 816.8 & 88.8 & M31 sat. & $      674.9 \pm 22.8$ & $      531.1 \pm 17.7$ & $ \bf   12.8 \pm 1.0$ & $      228.6 \pm 4.9$ & $      45.6 \pm 9.9$ & $      64.4 \pm 6.3$\\ 
Andromeda XXVI & 766.0 & 102.7 & M31 sat. & $      626.4 \pm 20.9$ & $      488.6 \pm 16.0$ & $ \bf   24.4 \pm 0.5$ & $      243.9 \pm 4.8$ & $      30.0 \pm 10.1$ & $      85.5 \pm 3.8$\\ 
Andromeda V & 777.6 & 109.5 & M31 sat. & $      684.1 \pm 16.0$ & $      567.4 \pm 13.1$ & $      81.2 \pm 2.3$ & $      142.5 \pm 4.2$ & $      90.3 \pm 11.6$ & $ \bf   3.3 \pm 2.2$\\ 
Andromeda XI & 738.5 & 110.6 & M31 sat. & $      567.5 \pm 9.5$ & $      456.2 \pm 7.3$ & $ \bf   9.5 \pm 0.8$ & $      122.9 \pm 4.0$ & $      218.9 \pm 9.1$ & $      66.6 \pm 2.4$\\ 
Andromeda XIX & 823.6 & 114.0 & M31 sat. & $      635.1 \pm 55.2$ & $      485.4 \pm 42.5$ & $      86.7 \pm 5.7$ & $      200.9 \pm 5.5$ & $      173.1 \pm 8.1$ & $ \bf   10.0 \pm 2.9$\\ 
Andromeda XXIII & 774.3 & 126.4 & M31 sat. & $      692.3 \pm 25.2$ & $      591.9 \pm 21.5$ & $      120.4 \pm 5.1$ & $      40.2 \pm 6.3$ & $      223.5 \pm 11.7$ & $      123.0 \pm 7.9$\\ 
Andromeda XX & 744.3 & 129.8 & M31 sat. & $      534.4 \pm 20.0$ & $      397.5 \pm 14.7$ & $      110.2 \pm 3.1$ & $      229.2 \pm 4.6$ & $      145.6 \pm 7.3$ & $ \bf   26.7 \pm 2.6$\\ 
Andromeda XIII & 843.5 & 132.1 & M31 sat. & $      653.2 \pm 10.0$ & $      530.4 \pm 8.0$ & $ \bf   7.1 \pm 0.7$ & $      95.6 \pm 4.6$ & $      251.5 \pm 8.5$ & $      106.9 \pm 5.0$\\ 
Andromeda X & 674.6 & 133.8 & M31 sat. & $      589.7 \pm 18.5$ & $      487.3 \pm 15.0$ & $      56.9 \pm 2.6$ & $      137.8 \pm 3.7$ & $      121.0 \pm 11.7$ & $ \bf   9.6 \pm 2.8$\\ 
Andromeda XXI & 831.0 & 134.0 & M31 sat. & $      601.5 \pm 13.4$ & $      442.3 \pm 9.2$ & $      120.1 \pm 1.9$ & $      298.5 \pm 5.1$ & $      40.3 \pm 7.5$ & $      109.0 \pm 5.3$\\ 
Andromeda XXXII & 780.5 & 140.9 & M31 sat. & $      669.2 \pm 26.1$ & $      532.0 \pm 20.2$ & $ \bf   14.4 \pm 1.1$ & $      236.6 \pm 4.9$ & $      9.2 \pm 6.8$ & $      92.6 \pm 5.1$\\ 
NGC 147 & 680.2 & 142.9 & M31 sat. & $      563.4 \pm 14.9$ & $      445.4 \pm 11.5$ & $ \bf   3.9 \pm 0.9$ & $      219.5 \pm 3.7$ & $      46.3 \pm 11.1$ & $      72.4 \pm 2.0$\\ 
Andromeda XXX & 686.6 & 147.7 & M31 sat. & $      589.5 \pm 32.5$ & $      468.4 \pm 25.8$ & $ \bf   6.4 \pm 1.6$ & $      219.6 \pm 4.6$ & $      35.7 \pm 12.0$ & $      75.2 \pm 2.3$\\ 
Andromeda XIV & 798.2 & 161.3 & M31 sat. & $      647.3 \pm 74.3$ & $      525.6 \pm 60.4$ & $ \bf   2.4 \pm 1.1$ & $      75.2 \pm 11.9$ & $      296.1 \pm 21.4$ & $      140.0 \pm 25.2$\\ 
Andromeda XII & 933.0 & 178.6 & M31 sat. & $      762.3 \pm 54.4$ & $      614.5 \pm 43.8$ & $ \bf   1.5 \pm 1.3$ & $      106.8 \pm 7.3$ & $      244.3 \pm 11.3$ & $      110.2 \pm 16.9$\\ 
Andromeda XV & 630.0 & 178.9 & M31 sat. & $      558.2 \pm 35.5$ & $      467.9 \pm 29.9$ & $      61.9 \pm 5.6$ & $      88.9 \pm 6.3$ & $      198.1 \pm 12.4$ & $      70.7 \pm 9.9$\\ 
Andromeda II & 656.3 & 184.1 & M31 sat. & $      540.6 \pm 9.1$ & $      456.2 \pm 7.5$ & $      60.4 \pm 2.1$ & $      60.4 \pm 3.9$ & $      248.0 \pm 11.1$ & $      108.9 \pm 5.5$\\ 
NGC 185 & 620.9 & 187.7 & M31 sat. & $      518.2 \pm 13.2$ & $      412.7 \pm 10.2$ & $ \bf   4.4 \pm 1.6$ & $      204.6 \pm 3.3$ & $      60.8 \pm 11.6$ & $      62.7 \pm 4.5$\\ 
Andromeda XXIX & 733.7 & 188.3 & M31 sat. & $      502.0 \pm 29.3$ & $      361.6 \pm 21.3$ & $      152.4 \pm 7.5$ & $      237.0 \pm 5.6$ & $      177.4 \pm 7.0$ & $ \bf   17.0 \pm 3.5$\\ 
Triangulum & 814.1 & 206.5 & M31 sat. & $      680.7 \pm 11.0$ & $      591.0 \pm 9.7$ & $      128.5 \pm 2.4$ & $      30.2 \pm 5.2$ & $      335.1 \pm 11.1$ & $      215.0 \pm 5.1$\\ 
Andromeda XXIV & 604.8 & 208.2 & M31 sat. & $      545.1 \pm 18.5$ & $      456.4 \pm 15.3$ & $      73.2 \pm 3.5$ & $      128.6 \pm 3.5$ & $      116.6 \pm 12.8$ & $ \bf   7.1 \pm 4.3$\\ 
Andromeda VII & 764.9 & 218.3 & M31 sat. & $      554.3 \pm 17.4$ & $      395.1 \pm 11.4$ & $      132.5 \pm 3.1$ & $      375.4 \pm 7.1$ & $      80.5 \pm 9.9$ & $      215.8 \pm 4.8$\\ 
IC 10 & 798.5 & 252.1 & M31 sat. & $      671.8 \pm 22.1$ & $      526.9 \pm 16.8$ & $ \bf   9.5 \pm 0.9$ & $      303.1 \pm 5.8$ & $      107.6 \pm 13.3$ & $      182.2 \pm 6.3$\\ 
Andromeda XXXI & 760.2 & 263.0 & M31 sat. & $      460.7 \pm 17.3$ & $      289.7 \pm 10.5$ & $      249.8 \pm 7.5$ & $      426.5 \pm 9.1$ & $      43.1 \pm 8.0$ & $      231.1 \pm 5.6$\\ 
LGS 3 & 773.0 & 268.5 & M31 sat. & $      553.9 \pm 10.7$ & $      461.5 \pm 9.1$ & $ \bf   16.1 \pm 1.3$ & $      8.8 \pm 4.7$ & $      386.8 \pm 9.6$ & $      209.5 \pm 5.8$\\ 
Andromeda VI & 785.4 & 269.0 & M31 sat. & $      452.5 \pm 10.3$ & $      311.7 \pm 7.0$ & $      200.0 \pm 3.4$ & $      235.8 \pm 4.4$ & $      229.8 \pm 5.9$ & $ \bf   5.3 \pm 1.9$\\ 
Andromeda XXII & 925.2 & 274.0 & M31 sat. & $      787.7 \pm 58.7$ & $      680.2 \pm 50.7$ & $      131.4 \pm 11.5$ & $      70.4 \pm 19.2$ & $      405.9 \pm 22.8$ & $      284.6 \pm 28.9$\\ 
Andromeda XVI & 480.7 & 323.2 & non-sat. & $      391.5 \pm 18.1$ & $      321.3 \pm 15.0$ & $      8.2 \pm 3.4$ & $ \bf   111.4 \pm 3.9$ & $      211.2 \pm 12.4$ & $      47.7 \pm 13.4$\\ 
Andromeda XXVIII & 660.9 & 367.8 & non-sat. & $      323.4 \pm 42.2$ & $      153.0 \pm 20.1$ & $      356.3 \pm 42.1$ & $      466.7 \pm 37.5$ & $ \bf   8.2 \pm 6.9$ & $      243.4 \pm 22.2$\\ 
IC 1613 & 757.8 & 520.1 & non-sat. & $      388.7 \pm 12.6$ & $      329.6 \pm 10.8$ & $      24.8 \pm 3.2$ & $ \bf   86.4 \pm 5.1$ & $      588.2 \pm 16.0$ & $      343.7 \pm 17.3$\\ 
Phoenix & 414.9 & 867.6 & non-sat. & $      48.0 \pm 3.9$ & $      16.2 \pm 2.2$ & $      14.1 \pm 8.7$ & $ \bf   69.3 \pm 5.2$ & $      526.1 \pm 17.5$ & $      252.0 \pm 39.1$\\ 
NGC 6822 & 451.9 & 897.5 & non-sat. & $      276.7 \pm 7.5$ & $      357.5 \pm 7.8$ & $      487.1 \pm 13.4$ & $      515.7 \pm 7.6$ & $ \bf   31.9 \pm 6.8$ & $      318.3 \pm 37.4$\\ 
Cetus & 755.6 & 680.4 & non-sat. & $      178.0 \pm 4.7$ & $      120.5 \pm 4.0$ & $      179.3 \pm 5.8$ & $ \bf   8.7 \pm 3.7$ & $      615.2 \pm 10.8$ & $      296.7 \pm 20.0$\\ 
Pegasus dIrr & 921.0 & 474.3 & non-sat. & $      377.5 \pm 10.1$ & $      211.6 \pm 6.1$ & $      357.7 \pm 6.4$ & $      291.0 \pm 5.2$ & $      305.6 \pm 6.4$ & $ \bf   10.7 \pm 1.5$\\ 
Leo T & 422.0 & 990.7 & non-sat. & $      138.7 \pm 4.6$ & $      201.8 \pm 5.6$ & $      323.2 \pm 13.6$ & $ \bf   3.8 \pm 3.2$ & $      49.7 \pm 24.6$ & $      33.0 \pm 24.7$\\ 
WLM & 932.7 & 836.2 & non-sat. & $      103.9 \pm 4.8$ & $      27.2 \pm 4.2$ & $      319.4 \pm 8.1$ & $ \bf   24.3 \pm 3.9$ & $      710.0 \pm 13.1$ & $      323.2 \pm 20.5$\\ 
Andromeda XVIII & 1216.7 & 452.5 & non-sat. & $      920.0 \pm 21.7$ & $      691.1 \pm 15.4$ & $      120.2 \pm 5.8$ & $      344.3 \pm 7.8$ & $ \bf   23.8 \pm 6.4$ & $      128.7 \pm 24.9$\\ 
Leo A & 803.0 & 1200.0 & non-sat. & $      338.6 \pm 11.6$ & $      414.8 \pm 13.8$ & $      562.0 \pm 21.2$ & $ \bf   8.3 \pm 4.4$ & $      197.1 \pm 38.1$ & $      119.5 \pm 34.6$\\ 
Aquarius & 1065.5 & 1172.1 & non-sat. & $      448.5 \pm 13.7$ & $      641.8 \pm 13.7$ & $      1022.6 \pm 21.6$ & $      818.0 \pm 13.5$ & $ \bf   95.1 \pm 10.6$ & $      448.8 \pm 30.7$\\ 
Tucana & 882.6 & 1355.7 & non-sat. & $      573.9 \pm 19.6$ & $      528.9 \pm 17.8$ & $      392.2 \pm 18.1$ & $ \bf   39.1 \pm 3.6$ & $      733.9 \pm 21.5$ & $      251.0 \pm 59.2$\\ 
Sagittarius dIrr & 1059.0 & 1356.9 & non-sat. & $      734.3 \pm 34.6$ & $      911.6 \pm 41.1$ & $      1139.3 \pm 51.9$ & $      1000.9 \pm 37.6$ & $ \bf   102.7 \pm 2.2$ & $      675.6 \pm 51.4$\\ 
UGC 4879 & 1367.5 & 1395.2 & non-sat. & $      942.5 \pm 11.6$ & $      959.2 \pm 11.9$ & $      853.6 \pm 10.5$ & $ \bf   1.9 \pm 1.6$ & $      506.7 \pm 40.6$ & $      219.9 \pm 9.4$\\ 
 \hline
 \end{tabular}
 \end{center}
 \small \medskip
LG galaxy distances from the MW ($d_{\mathrm{MW}}$) and from M31 ($d_{\mathrm{M31}}$) in kpc. Category refers to whether a galaxy is considered a host (only the MW and M31), a MW satellite ($d_{\mathrm{MW}} < 300$~kpc), M31 satellite ($d_{\mathrm{M31}} < 300$~kpc) or a non-satellite (both $d_{\mathrm{MW}}$\ and $d_{\mathrm{M31}} > 300$~kpc). The other columns give the distance of each galaxy from the different planes fitted to those galaxies whose offset is printed in boldface in the respective column.
\end{minipage}
\end{table*}

\section{Discussion}
\label{sect:discussion}

\begin{table*}
\begin{minipage}{180mm}
 \small
 \caption{Angles between the dwarf galaxy planes and other features}
 \label{tab:angles}
 \begin{center}
 \begin{tabular}{@{}lccccccc}
 \hline 
  direction  & uncertainty & MW VPOSall & MW VPOS-3 & GPoA & LGP1 & LGP2 & M31 disc plane\\
\hline
MW VPOSall & $1.1^{\circ}$ & --  & $ 14^{\circ} $  & $ 51^{\circ} $  & $ 66^{\circ} $  & $ 85^{\circ} $  & $ 70^{\circ} $ \\ 
MW VPOS-3 & $0.4^{\circ}$ & $ 14^{\circ} $  & --  & $ 38^{\circ} $  & $ 53^{\circ} $  & $ 77^{\circ} $  & $ 59^{\circ} $ \\ 
GPoA & $0.8^{\circ}$ & $ 51^{\circ} $  & $ 38^{\circ} $  & --  & $ 33^{\circ} $  & $ 68^{\circ} $  & $ 48^{\circ} $ \\ 
LGP1 & $0.4^{\circ}$ & $ 66^{\circ} $  & $ 53^{\circ} $  & $ 33^{\circ} $  & --  & $ 35^{\circ} $  & $ 16^{\circ} $ \\ 
LGP2 & $1.7^{\circ}$ & $ 85^{\circ} $  & $ 77^{\circ} $  & $ 68^{\circ} $  & $ 35^{\circ} $  & --  & $ 20^{\circ} $ \\ 
M31 disc plane & $2.9^{\circ}$ & $ 70^{\circ} $  & $ 59^{\circ} $  & $ 48^{\circ} $  & $ 16^{\circ} $  & $ 20^{\circ} $  & -- \\ 
MW disc &   & $ 87^{\circ} $  & $ 87^{\circ} $  & $ 82^{\circ} $  & $ 68^{\circ} $  & $ 37^{\circ} $  & $ 52^{\circ} $ \\ 
M31 disc &   & $ 84^{\circ} $  & $ 73^{\circ} $  & $ 51^{\circ} $  & $ 20^{\circ} $  & $ 23^{\circ} $  & $ 18^{\circ} $ \\ 
MW--M31 line &    & $ 52^{\circ} $  & $ 40^{\circ} $  & $  3^{\circ} $  & $  1^{\circ} $  & $  0^{\circ} $  & $  6^{\circ} $ \\ 
Supergalactic plane & $30^{\circ}$ & $ 72^{\circ} $  & $ 58^{\circ} $  & $ 26^{\circ} $  & $ 17^{\circ} $  & $ 48^{\circ} $  & $ 32^{\circ} $ \\ 
CMB dipole & $3^{\circ}$  & $ 28^{\circ} $  & $ 16^{\circ} $  & $ 21^{\circ} $  & $ 15^{\circ} $  & $  2^{\circ} $  & $  5^{\circ} $ \\ 
LG velocity & $21^{\circ}$  & $ 67^{\circ} $  & $ 71^{\circ} $  & $ 43^{\circ} $  & $ 43^{\circ} $  & $ 27^{\circ} $  & $ 42^{\circ} $ \\ 
Average VPOS orbital pole & $29^{\circ}$  & $ 24^{\circ} $  & $ 14^{\circ} $  & $ 37^{\circ} $  & $ 42^{\circ} $  & $ 64^{\circ} $  & $ 46^{\circ} $ \\ 
Magellanic Stream normal & $15^{\circ}$  & $ 24^{\circ} $  & $ 11^{\circ} $  & $ 27^{\circ} $  & $ 48^{\circ} $  & $ 77^{\circ} $  & $ 57^{\circ} $ \\ 
MW--M31 orbital pole & $55^{\circ}$\footnote{Along the great circle perpendicular to the line connecting the MW and M31.}  & $ 53^{\circ} $  & $ 43^{\circ} $  & $ 22^{\circ} $  & $ 55^{\circ} $  & $ 90^{\circ} $  & $ 70^{\circ} $ \\ 
 \hline
 \end{tabular}
 \end{center}
 \small \medskip
Angles between the different dwarf galaxy planes in the LG and other directions (see Sect. \ref{sect:discussion} for a discussion). Column 1 describes the direction compared in each row, the second column indicates its direction uncertainty and the remaining rows indicate the inclination between it and the dwarf galaxy planes. For vectors (MW-M31 line and velocities) the angle between the vector and the plane is given, for planes the angle between the two normal vectors is given.
\end{minipage}
\end{table*}

The discovery of similar, thin planes of co-rotating satellites around the two major galaxies in the LG, and the additional finding that the non-satellite galaxies in the LG are also confined to two very symmetric planes, poses the question of how all these structures relate to each other. The relative orientations of the different planes is discussed in this section. In addition, the planes are compared to other pronounced structures and directions in and around the LG: the Supergalactic Plane, the motion of the LG with respect to the CMB and the surrounding galaxies, the orbital plane of the MW-M31 system, the Magellanic Stream and the over-density in hypervelocity stars in the MW halo. While not yet fully conclusive, all these comparisons might provide valuable hints leading to a more complete understanding of the origin and dynamics of the dwarf galaxy structures and thus the history of the LG.

The inclinations of the planes relative to each other and with other features are compiled in Table \ref{tab:angles}. In interpreting the orientations, it might help to note that the probability that two randomly oriented planes are inclined by an angle of $\theta$\ or less is $P_{\mathrm{planes}} = 1 - \cos(\theta)$, while the probability of a randomly oriented vector to point to within $\theta$\ or less of a plane is given by $P_{\mathrm{vector}} = \sin(\theta)$. 

\begin{figure}
 \centering
 \includegraphics[width=88mm]{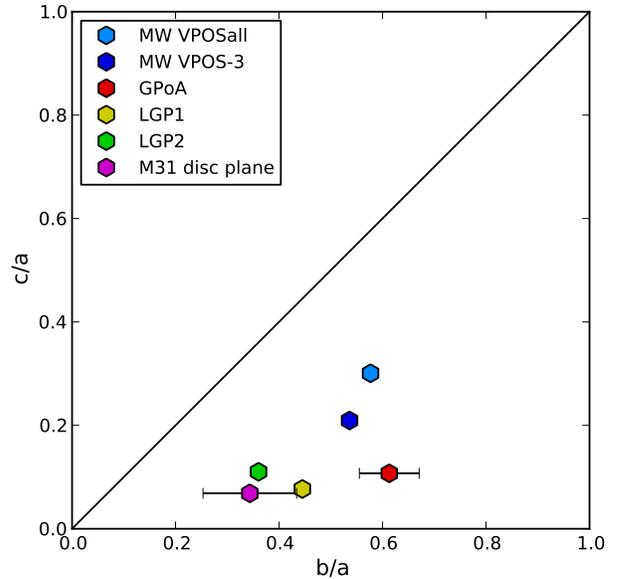}
 \caption{Axis ratios $b/a$\ (intermediate to long) and $c/a$\ (short to long) for the different planes fitted to the LG galaxies. Most planes are very thin, with short axes on the order of one tenth of the long axis. Only the full VPOSall has a considerably larger $c/a$, unless the three outliers are removed from the sample (VPOS-3). The error bars represent the uncertainties as determined by varying the galaxy positions within their distance uncertainties. For most planes, these are smaller than the symbols.
}
 \label{fig:axratios}
\end{figure}

\begin{figure}
 \centering
 \includegraphics[width=88mm]{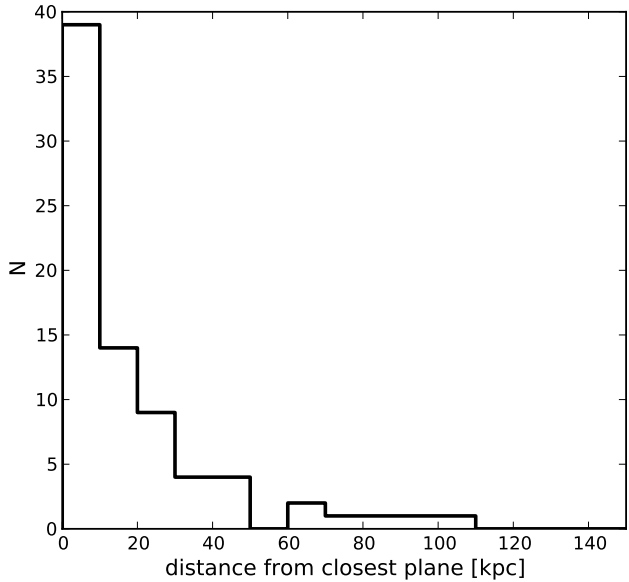}
 \caption{For each of the 76 non-host galaxies in our sample, the distances to each of the 5 LG planes (ignoring the VPOS-3) has been determined and compiled in Table \ref{tab:allplanes}. The minimum of these five distances for each galaxy is the distance to the closest plane. The histogram plots these distances to the closest plane. Almost all LG galaxies (70, 92 per cent) are closer than 50 kpc to one of the planes, only six are more distant.}
 \label{fig:planedist}
\end{figure}

The galaxy planes in the LG have similar axis ratios $c/a \approx 0.1$, with the exception of the VPOSall and VPOS-3, for which this value is 0.3 and 0.2, respectively (Fig. \ref{fig:axratios}). The RMS heights are comparable, too, ranging from 14 to 66 kpc. Almost all (92 per cent) of the galaxies within the 1.5 Mpc radius of the LG are closer than 50 kpc to one of five planes (Fig. \ref{fig:planedist}).

\subsection{Relative orientations of the planes}
\label{subsect:relativeorientation}

\begin{figure*}
 \centering
 \includegraphics[width=180mm]{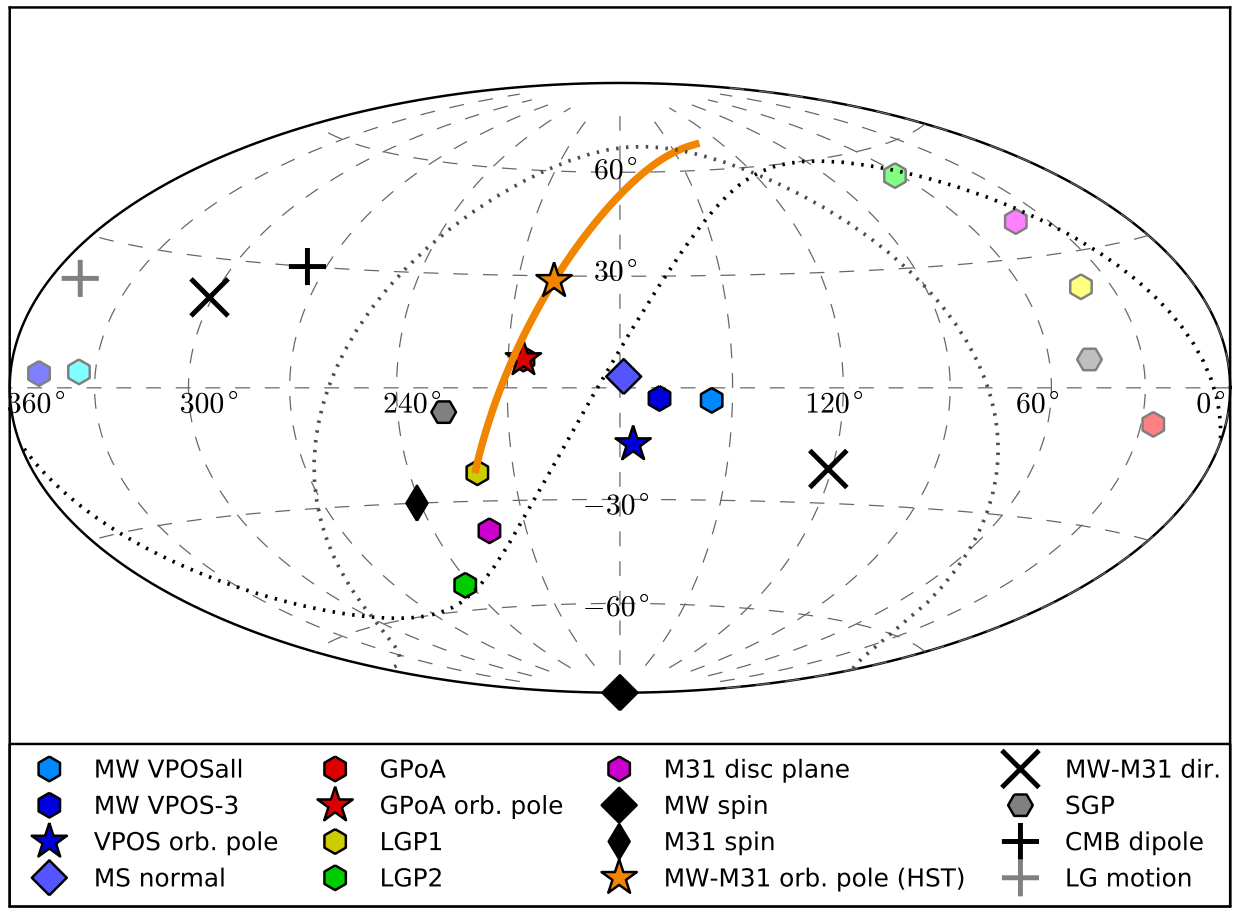}
 \caption{Comparison of plane normal directions (hexagons) with various directions such as the host galaxy spin directions (black diamonds) and the normal direction to the Magellanic Stream as determined in \citet{Pawlowski2012a} (blue diamond) in Galactic coordinates $l$\ and $b$. Each plane has two normal vectors pointing in opposite directions ($180^{\circ}$\ offset). To ease the comparison we de-emphasised the ones outside of the central half of the figure by plotting them in a lighter colour.
Also shown is the normal direction to the Supergalactic Plane (SGP, grey hexagon), which is close to the GPoA and LGP1 normal directions. 
The direction of the line connecting the MW and M31 is indicated by the black $\times$. 
The plus signs indicate the direction of motion of the LG relative to the CMB (black) and the nearby galaxies (grey). The dotted lines are great circles offset by $90^{\circ}$\ from these velocity directions. If a plane normal lies on such a line, the corresponding velocity vector is parallel to the plane.   
The angular momentum directions of the satellite planes (stars) are close to each other, indicating that the VPOS and GPoA preferentially orbit in a similar sense. The most-likely orbital pole of the MW-M31 system (orange star) is prograde with respect to them, too. The great circle segment perpendicular to the MW-M31 direction indicates the 1$\sigma$\ uncertainty of the MW-M31 orbital pole.
 }
 \label{fig:NormalASP}
\end{figure*}

\begin{figure}
 \centering
 \includegraphics[width=88mm]{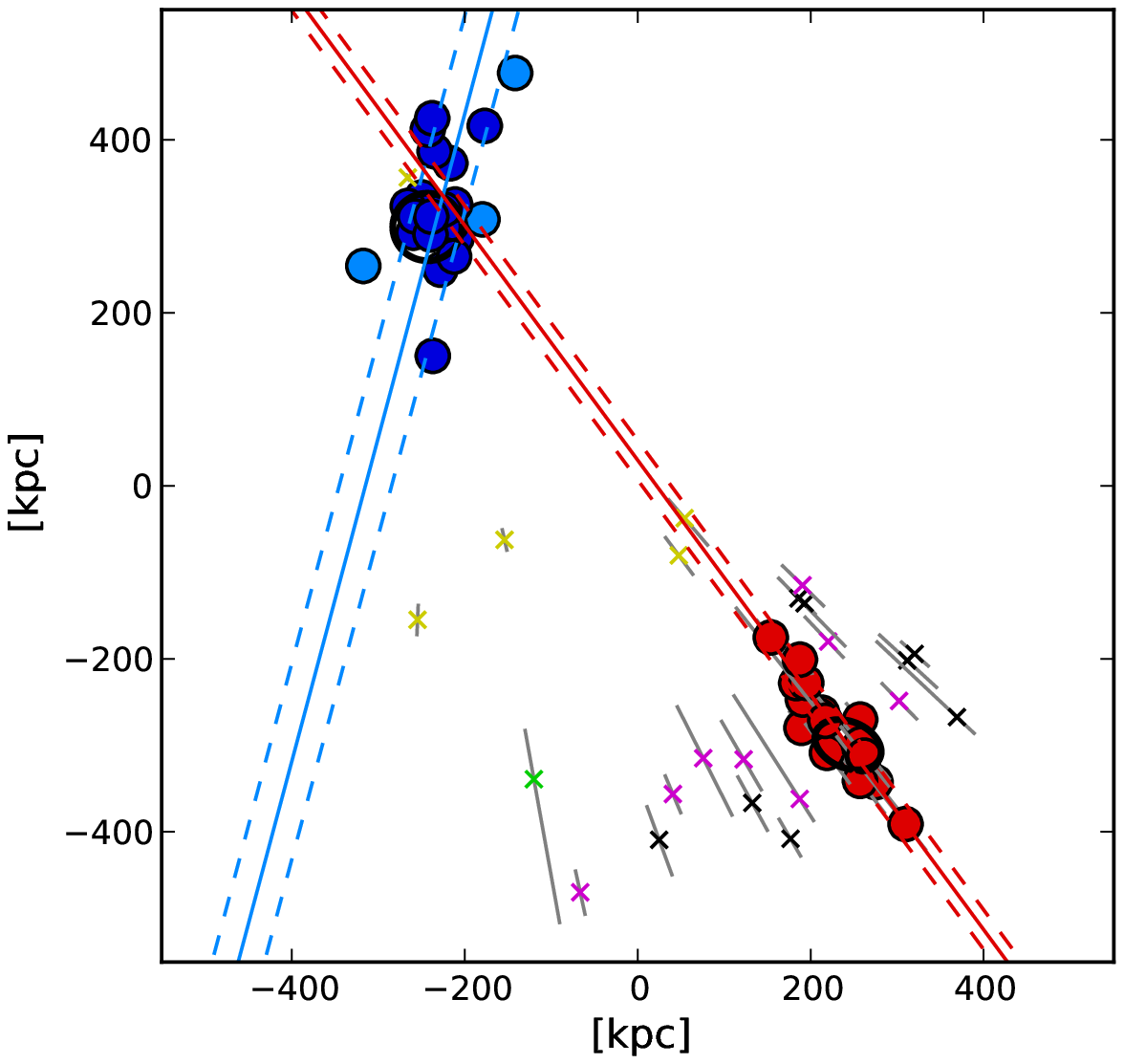}
 \includegraphics[width=88mm]{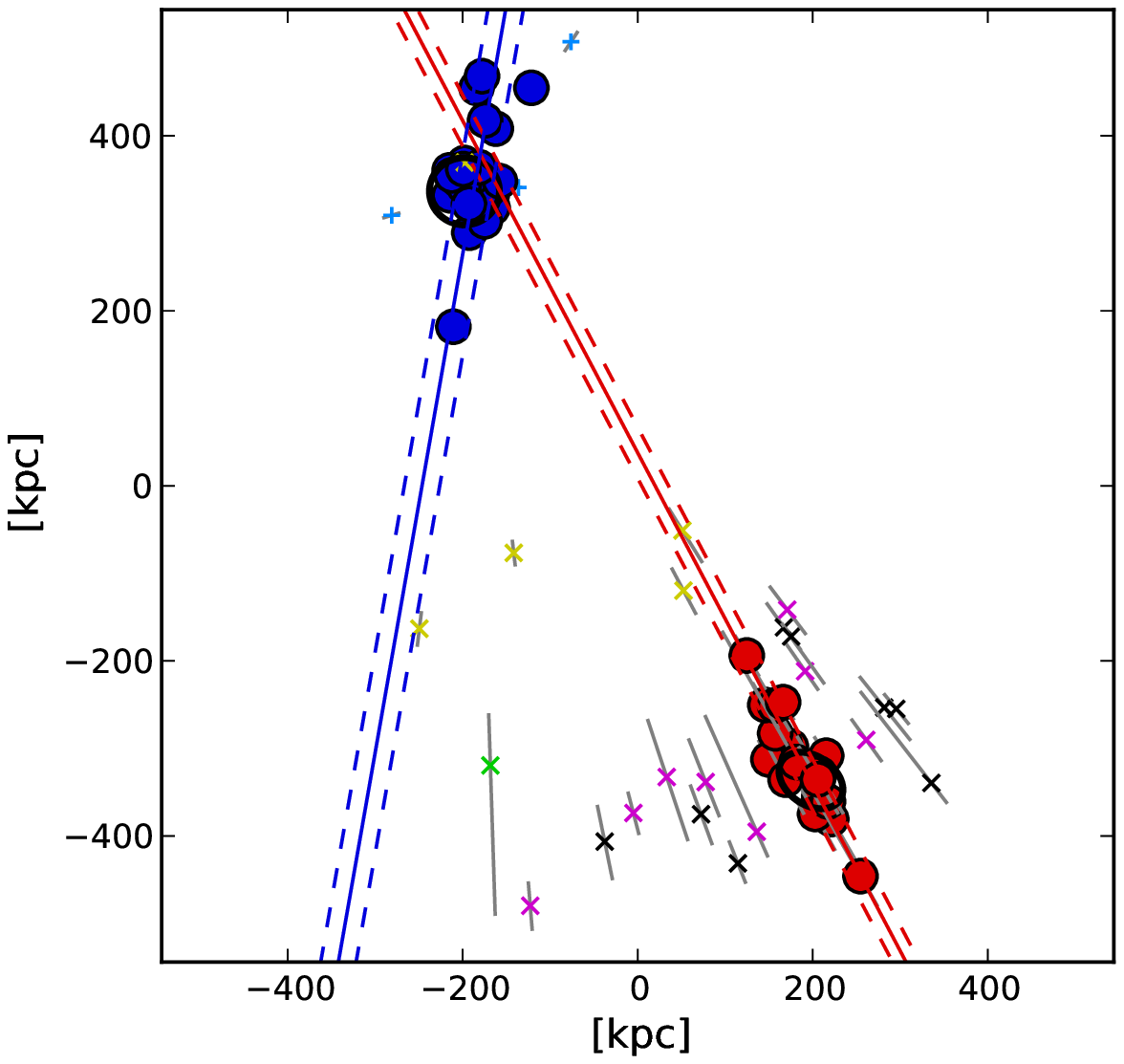}
 \caption{Edge-on view of the satellite galaxy planes around the MW and M31, similar to Fig. \ref{fig:LGP1-LGP2} for the LG planes. As before, galaxies which are members of the VPOS are plotted in blue, GPoA members in red. 
The \textit{upper panel} uses all 27 MW satellites for the VPOSall plane fit, while the \textit{lower panel} represents the VPOS-3 sample, excluding the outliers Leo I, Hercules and Ursa Major (I). The view-direction in both plots is close to the MW spin axis (looking 'downwards' from the MW north): $(l,b) = (260.^{\circ}2, -77.^{\circ}1)$\ (upper panel) and $(l,b) = (269.^{\circ}0, -73.^{\circ}5)$\ (lower panel). The component of the Galactic north points up in both panels and the plots are centred on the midpoint between the MW and M31. The excellent alignment of the MW within the extended GPoA is obvious: the very narrow GPoA (red line) crosses the MW system. The VPOS, in contrast, does not contain M31, but removing the three outliers from the VPOS leads to a considerably smaller angle between the two satellite galaxy planes.
 }
 \label{fig:VPOS-VTDS}
\end{figure}

Figure \ref{fig:NormalASP} illustrates the normal directions to the various planes as compiled in Table \ref{tab:allplanes} in Galactic coordinates. The relative inclinations between the planes and their inclinations with other features are compiled in Table \ref{tab:angles}.

The GPoA normal is inclined by almost $90^{\circ}$\ from the MW-M31 direction, so the GPoA almost contains the line connecting the MW and M31. The GPoA, therefore, is seen edge-on from the MW (inclined by only $3^{\circ}$). As a consequence many MW satellites are close to the GPoA, including the LMC and SMC (Table \ref{tab:dwarfdist}). Furthermore, the GPoA is almost polar with respect to the MW (inclined by $82^{\circ}$\ from the Galactic disc), which is also the case for the VPOS (vast \textit{polar} structure).

The GPoA is inclined by $51^{\circ}$\ to the plane fitted to all MW satellites (VPOSall)\footnote{\citet{Ibata2013} and \citet{Conn2013} discuss that the GPoA is approximately perpendicular to the VPOS.}. It is inclined by only $38^{\circ}$\ to the VPOS-3, the plane fitted after excluding three outliers, which has a normal pointing close to the dominant 4-galaxy-normal peak and the average orbital pole of the MW satellites (Fig. \ref{fig:MW4dwarfnormal}). The two satellite galaxy planes therefore neither align perfectly, nor are they perpendicular to each other. Figure \ref{fig:VPOS-VTDS} (similar to Fig. \ref{fig:LGP1-LGP2} for LGP1 and LGP2) shows the VPOSall (VPOS-3) and the GPoA, from a direction in which both are seen edge on, illustrating the perfect orientation of the GPoA towards the MW.

How do the VPOS and GPoA spin directions compare? The VPOS spin is indicated by the average orbital pole derived from the proper motions of the MW satellites co-orbiting in the VPOS \citep{Pawlowski2013}, which points to $(l,b) = (176^{\circ}.4, -15^{\circ}.0)$\ with a spherical standard distance of $\Delta_{\mathrm{sph}} = 29^{\circ}.3$, which we adopt as its uncertainty. This direction is much closer to the VPOS-3 than the VPOSall normal. For the GPoA, its fortunate edge-on orientation allows to check for a rotational signature using only the line-of-sight velocities of the satellite galaxies. As mentioned before, the line-of-sight velocities reveal that most of the GPoA galaxies co-orbit. \citet{Ibata2013} show (see their figure 3) that in the M31 rest frame the northern satellites in the GPoA recede from the MW, while the southern ones approach. Assuming that the GPoA normal defines its rotation axis (which due to the thinness of the GPoA is a good approximation), the galaxy plane's spin can either point into the direction $(l,b) = (206^{\circ}, 8^{\circ})$\ or the opposite direction $(l,b) = (26^{\circ}, -8^{\circ})$. Looking into the direction of M31 (north up), the northern part of the GPoA recedes relative to M31 and the southern part approaches. Thus, the GPoA spin points to the left (east in Galactic coordinates). Galactic longitude increases towards the east, so the spin direction points to a larger galactic longitude than M31's position ($l_{\mathrm{M31}} = 121.2$). The GPoA spin direction is therefore approximately $(l,b) = (206^{\circ}, 8^{\circ})$.

This direction is indicated with the red star symbol in Fig. \ref{fig:NormalASP}. It is close to the average orbital pole of the VPOS ($37^{\circ}$\ inclined). \textit{Both satellite galaxy planes, the VPOS around the MW and the GPoA around M31, rotate in the same sense, they are prograde with respect to each other}.
Both satellite galaxy plane spins are also approximately perpendicular to the Galactic disc spin of the MW, but they are slightly less inclined and prograde with respect to the spin of M31's galactic disc: the GPoA spin is inclined by $51^{\circ}$, the VPOS average orbital pole by $61^{\circ}$. The VPOSall plane is, however, almost polar with respect to the M31 disc ($84^{\circ}$).

How do the satellite galaxy planes compare to our suggested LGP1 and LGP2? We have already discussed the remarkable symmetry of the LG planes in Sect. \ref{subsect:LGPcompare}, which are both inclined by about $20^{\circ}$\ to the galactic disc of M31 and parallel to the line connecting the MW and M31. They are inclined by $68^{\circ}$\ (LGP1) and $37^{\circ}$\ (LGP2) with respect to the Galactic disc of the MW. Relative to the satellite galaxy planes, LGP1 aligns quite well with the GPoA ($33^{\circ}$), but is more inclined with respect to the VPOSall ($66^{\circ}$) and VPOS-3 ($53^{\circ}$). LGP2 is highly inclined to all satellite galaxy planes (VPOSall: $85^{\circ}$, VPOS-3: $77^{\circ}$, GPoA: $68^{\circ}$). One might therefore suspect that the LGP1 has a larger chance to be related to the satellite galaxy structures than LGP2. This suspicion will find support in Sect. \ref{subsect:magstream}.

\subsection{Orientations relative to the surrounding galaxy distribution and LG velocity}

The galaxies surrounding the LG, with distances on the order of tens of Mpc, preferentially lie in the Supergalactic Plane (SGP), a planar structure approximately perpendicular to the MW disc. The pole of the Supergalactic Coordinate System of \citet{deVaucouleurs1991} points to $(l,b) = (47.^{\circ}.4, 6^{\circ}.3)$. In this coordinate system, the SGP lies approximately along the equator, i.e. along the great circle perpendicular to the pole, which warrants identifying the pole with the normal direction to the SGP. However, depending on the radius within which the SGP orientation is determined the galaxy distribution's minor axis changes by up to $\approx 30^{\circ}$\ from this pole \citep{Lahav2000}, which we therefore adopt as the uncertainty.

The SGP pole is plotted as a grey hexagon in Fig. \ref{fig:NormalASP}. Similar to the VPOSall/VPOS-3, GPoA and LGP1, the SGP is polar with respect to the MW disc. The SGP pole is close to the LGP1 normal, so these two planes are well aligned (inclination only $17^{\circ}$). The GPoA is also oriented similar to the SGP (with an inclination of $26^{\circ}$) and the same is true for the orientation of the M31 disc spin. The VPOSall/VPOS-3 and LGP2 are all inclined by more than $45^{\circ}$\ from the SGP.

The orientation of a plane can also be compared with the direction of motion of its constituents. We here restrict the discussion to the motion of the LG, but will discuss the relative motion of the MW and M31 in Sect. \ref{subsect:M31propmo}.

Interpreting the Cosmic Microwave Background (CMB) dipole anisotropy as a Doppler-shift induced mostly by the peculiar motion of the LG with respect to the CMB rest frame, \citet{Kogut1993} determined that the apparent motion of the LG with respect to the CMB points to $(l,b) = (276^{\circ} \pm 3^{\circ}, 30^{\circ} \pm 3^{\circ})$\ and has an amplitude of $627 \pm 22~\mathrm{km~s}^{-1}$\ \citep{Kogut1993}. Very similar values have been reported by \citet{Bilicki2011}. This interpretation of the CMB dipole as the motion of the LG has been validated for the first time by \citet{Jerjen1993}, who have measured the peculiar motion of the LG with respect to nearby galaxy clusters and found it to be in perfect agreement with the one derived from the CMB.
In Fig. \ref{fig:NormalASP}, the direction of motion of the LG with respect to the CMB is indicated with a black plus sign. The black dotted line is the great circle perpendicular to this direction. If a plane normal lies on this great circle, the CMB velocity vector is parallel to the plane. This is almost the case for the LGP2 normal, indicating that the LG velocity relative to the CMB lies along this galaxy plane (inclined by only $2^{\circ}$). The velocity vector is close to all other galaxy planes, too (see Table \ref{tab:angles}). Relative to the CMB rest frame, the LG moves approximately along the direction of the MW-M31 line and into the direction where the LGP1 and LGP2 intersect. Thus, when the LG is projected such that the two LG planes are seen edge-on, their orientation resembles a Mach cone with regard to the LG's velocity relative to the CMB (see grey arrow in the lower right of Fig. \ref{fig:LGP1-LGP2}).

The CMB dipole indicates the direction of motion of the LG relative to the largest scale in the Universe. It might be more meaningful to compare the LG structure with the velocity of the LG relative to the nearby galaxies. \citet{Tully2008} have determined the motion of the LG within the Local Sheet, the nearby galaxies (distances less than 7 Mpc) which have low relative peculiar velocities. They report that the LG has a low velocity of only $66 \pm 24~\mathrm{km~s}^{-1}$\ with respect to the Local Sheet, which, in Supergalactic Coordinates $L$\ and $B$\ points into the direction of $(L,B) = (150^{\circ} \pm 37^{\circ}, 53^{\circ} \pm 20^{\circ})$. In Galactic coordinates, this corresponds to $(l,b) = (349^{\circ}, 22^{\circ})$\ with a directional uncertainty of $\approx 21^{\circ}$. This direction is indicated by the grey plus sign in Fig. \ref{fig:NormalASP} and again the corresponding great circle is plotted as a dotted line. The velocity of the LG with respect to the nearby galaxies does not align well with any of the planes. It aligns best with LGP2 ($27^{\circ}$\ inclination), and is inclined by more than $40^{\circ}$\ with respect to the other galaxy planes. However, the directional uncertainty is large.

\subsection{The M31 orbital pole}
\label{subsect:M31propmo}

The first direct measurement of the proper motion (PM) of M31 has been presented in a recent series of papers (\citealt*{Sohn2012}, \citealt{vdMarel2012a}, \citealt{vdMarel2012b}). Using \textit{Hubble Space Telescope} (HST) observations, \citet{Sohn2012} have measured the PM in three fields of M31 (in the M31 spheroid, the M31 disc and in the Giant Southern Stream). After correcting the measured PMs for the internal kinematics and averaging over the three fields, \citet{vdMarel2012a} arrive at a heliocentric PM measurement for M31 of $v_{\mathrm{W}} = -162.8 \pm 47.0~\mathrm{km~s}^{-1}$ towards the west and $v_{\mathrm{N}} = -117.2 \pm 45.0~\mathrm{km~s}^{-1}$ towards the north, according to their table 3. This assumes a distance of 770 kpc to M31.

An updated PM estimate for M31 based on the kinematics of its satellite galaxies \citep{vdMarel2008} has also been presented by \citet{vdMarel2012a}. It results in $v_{\mathrm{W}} = -176.1 \pm 144.1~\mathrm{km~s}^{-1}$ and $v_{\mathrm{N}} =  8.4 \pm 85.4~\mathrm{km~s}^{-1}$.
They argue that these values are compatible with the value derived from the HST measurements, and therefore adopt a weighted average of all PMs for their further analysis (including additional PM estimates discussed below). However, the PM estimate based on the line-of-sight velocities of the M31 satellite galaxies is based on the assumption that the satellite galaxy system of M31 on average follows its motion through space, and that the transverse motion of M31 superimposes an apparent solid body rotation onto the \textit{random} line-of-sight velocity field of its satellites. Given the recent discovery of the GPoA \citep{Ibata2013}, a \textit{coherently rotating} plane of M31 satellites, the assumption of underlying random satellite velocities is no longer justified. Here it should be mentioned that the co-rotating GPoA is oriented approximately in north-south direction\footnote{At the position of M31, the direction of Galactic north and Equatorial north differ by less than $3^{\circ}$\ \citep{Brunthaler2007}}, and that it is the north component $v_{\mathrm{N}}$\ of the proper motion which differs most between the PM estimate based on the satellite galaxy line-of-sight velocities and the HST measurement. 

Similarly, the second M31 PM estimate by \citet{vdMarel2008}, which uses the PMs of the M31 satellite galaxies M33 and IC 10 \citep{Brunthaler2005,Brunthaler2007}, simply assigns the satellite galaxy's PM to M31 and then adds the line-of-sight velocity dispersion of the whole M31 satellite galaxy system as uncertainties in all three velocity components. This assumes that the two galaxies are bound to M31. The resulting PM estimates are $v_{\mathrm{W}} = -47.7 \pm 88.2~\mathrm{km~s}^{-1}$\ and $v_{\mathrm{N}} =  70.9 \pm 91.5~\mathrm{km~s}^{-1}$\ for M33 and $v_{\mathrm{W}} = -16.2 \pm 88.0~\mathrm{km~s}^{-1}$ and $v_{\mathrm{N}} =  -47.3 \pm 89.3~\mathrm{km~s}^{-1}$\ for IC 10. As both galaxies are possibly related to our LG planes (LGP1 for IC 10 and LGP2 for M33), they might also be kinematically associated with those planes. Therefore, the assumption that the two satellites are bound to M31 and on average follow its motion through space is not necessarily valid.

In addition to these two methods (satellite line-of-sight velocities and satellite PM's), \citet{vdMarel2012a} also estimate the M31 PM from the line-of-sight velocities of non-satellite LG galaxies. This results in a PM estimate of $v_{\mathrm{W}} = -140.5 \pm 58.0~\mathrm{km~s}^{-1}$ and $v_{\mathrm{N}} =  -102.6 \pm 52.5~\mathrm{km~s}^{-1}$. This method assumes that the galaxies are bound to the LG barycentre, such that they trace the barycentre's motion. Determining the barycentre's motion with respect to the MW then provides the M31 motion with respect to the MW, as only these two galaxies contribute significantly to the barycentre. Interestingly, this third estimate agrees best with the M31 PM from the HST measurement.

We therefore reject those M31 PM estimates based on the galaxy's satellite kinematics as potentially flawed by being based on the invalid assumption that the M31 satellites sample random motions. In the following we only use the M31 PM estimates based on the weighted average of the HST measurements corrected for the internal kinematics. 
Following the coordinate system as introduced in \citet{vdMarel2002}, this corresponds to $\mu_{\alpha} \cos{\delta} = 0.045 \pm 0.013~\mathrm{mas}~\mathrm{yr}^{-1}$\ and $\mu_{\delta} = - 0.032 \pm 0.012~\mathrm{mas}~\mathrm{yr}^{-1}$\ at the distance of 770 kpc assumed by \citet{vdMarel2012a}. M31's heliocentric line-of-sight velocity is $300 \pm 4~\mathrm{km}~\mathrm{s}^{-1}$\ \citep{McConnachie2012}.

The PM includes both M31's space motion as well as the Sun's motion around the MW. The latter consists of the circular velocity of the local standard of rest (LSR) and the Sun's peculiar motion with respect to the LSR. For the LSR circular velocity we adopt $239 \pm 5~\mathrm{km}~\mathrm{s}^{-1}$\ \citep{McMillan2011}\footnote{We have repeated the determination of the MW-M31 orbital pole direction using a circular velocity of the LSR of $220~\mathrm{km~s}^{-1}$. The resulting orbital pole direction differs by only $4^{\circ}$\ relative to the pole determined for the higher LSR circular velocity and is thus well within the uncertainties.}. For the three components of the Sun's motion with respect to the LSR we adopt the values by \citet*{Schoenrich2010}: $(U,V,W) = (11.10~\mathrm{km}~\mathrm{s}^{-1}, 12.24~\mathrm{km}~\mathrm{s}^{-1}, 7.25~\mathrm{km}~\mathrm{s}^{-1})$\ for the three coordinates, i.e. radially inwards to the Galactic Centre, in the direction of Galacic rotation and towards the MW north. These are the same values used by \citet{vdMarel2012a}.

With this information we determine the three components of the M31 velocity with respect to the MW in our coordinate system. We randomly select the values of the two PM directions, M31's line-of-sight velocity and distance from Gaussian distributions centred on the most-likely values and having a width of the reported uncertainties. We keep the Sun's velocity components fixed as the uncertainties are negligible compared to those in the PM. We draw 10,000 sets of values and for each of them determine M31's velocity vector in our Cartesian coordinate system. 

Finally, we determine the orbital plane of the MW-M31 system for each of the resulting 10,000 velocity vectors. This assumes that the dynamics of the LG are governed by the two major galaxies only. The orientation of the orbital plane is described by the orbital pole of the MW-M31 orbit, which is the direction of the orbital angular momentum. It is determined by taking the cross-product of the MW-M31 position vector and the MW-M31 velocity vector. The orbital pole is thus perpendicular to both the position and the velocity vector. As the position of M31 is well known, the orbital pole of the MW-M31 system is constrained to the great circle perpendicular to this direction. The scatter in the 10,000 generated velocity vector directions, representing the PM uncertainty, thus results in an uncertain direction of the orbital pole along this great circle.

In our spherical coordinate system the resulting orbital pole points to $(l, b) = (199^\circ, 29^\circ)$\ (orange star in Fig. \ref{fig:NormalASP}). Each of the 10,000 velocity vectors results in a different orbital pole direction along the same great circle. One $\sigma$\ (68.3 per cent) of them are found within an angle of $\approx 55^\circ$\ from the average orbital pole. This region is marked with an orange line in Fig. \ref{fig:NormalASP}. The uncertainty is large because a perfectly radial orbit of M31 is allowed within the velocity uncertainty, such that in principle all orbital pole directions along the great circle are possible.

As is apparent from Fig. \ref{fig:NormalASP}, the most-likely orbital pole derived from the HST PM measurement indicates an almost polar orbit with respect to the Galactic disc of the MW, similar to most satellite and dwarf galaxy planes. The orbital pole points into the same direction as the normal vector (and spin direction) of the GPoA around M31 ($22^\circ$). Its $1\sigma$\ uncertainty extends towards the normal vector of LGP1 ($55^\circ$\ inclined to most-likely orbital pole). The spin of the MW satellites orbiting in the VPOS ($49^\circ$) as well as the galactic disc spin of M31 itself ($71^\circ$) are within $90^{\circ}$\ of its direction. \textit{The spin of the satellite galaxy planes around the MW (VPOS) and M31 (GPoA), the spin of M31 itself and the most-likely MW-M31 orbital angular momentum are all prograde with respect to each other. The VPOS and GPoA spin as well as the most-likely MW-M31 orbital pole, the MS normal ($32^{\circ}$\ inclined) and the LGP1 normal direction are confined to a region of $\approx 30^\circ$\ radius.} This might hint at a similar orbital sense of the LGP1 member galaxies.
On a larger scale, the normal of the SGP is close to the great circle segment indicating the uncertainty of the MW-M31 orbital pole, but inclined by $45^{\circ}$ from its most-likely direction. Within its uncertainty the MW-M31 orbital plane approximately aligns with the SGP.

\subsection{The Magellanic Stream}
\label{subsect:magstream}

\begin{figure*}
 \centering
 \includegraphics[width=180mm]{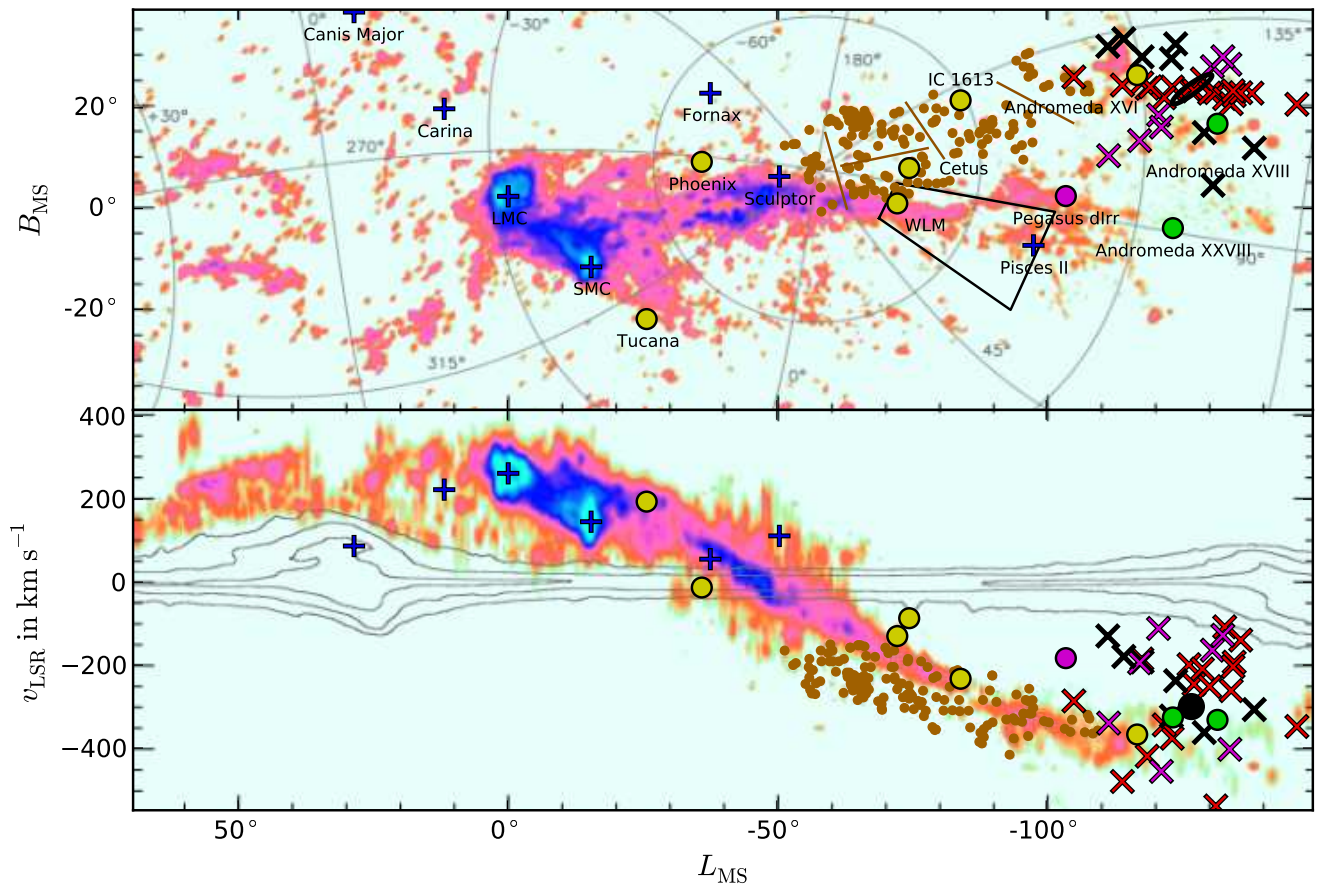}
 \caption{
 Comparison of the Magellanic Stream (MS) and the LG dwarf galaxies.  
The black ellipse indicates the position and orientation of the galactic disc of M31, all other symbols and colours for the galaxies are the same as in Figs. \ref{fig:LGASP} and \ref{fig:LGASP_longitude}.
They have been overlaid onto a map of the Magellanic Stream as published in \citet{Nidever2010} (their fig. 8, colour scale inverted for better visibility of the galaxy points). The \textit{upper panel} shows the position of the MS and the LG galaxies in the Magellanic Stream Coordinate system of \citet{Nidever2008}. In the \textit{lower panel}, the line-of-sight velocities $v_{\mathrm{GSR}}$\ along the MS, as measured from the Sun, are plotted against the MS longitude $L_{\mathrm{MS}}$. Overlaid onto the figure by \citet{Nidever2010} are the velocities of those LG galaxies which are found within the limits of the upper panel, i.e. which have a MS latitude $B_{\mathrm{MS}}$\ between $-40^{\circ}$\ and $+40^{\circ}$. For some LG galaxies, for example Pisces II, no velocity measurements are available, they are therefore not included in the lower panel.
The brown dots indicate the positions and velocities of HVCs as discovered by \citet{Westmeier2008}. The  brown lines separate the HVCs into the five groups or filaments discussed by \citet{Westmeier2008}, most of which are elongated approximately parallel to the MS, the GPoA and the VPOS. See Sect. \ref{subsect:magstream} for a discussion. The black wedge indicates the mirrored direction to the over-density of hypervelocity stars in the MW halo (Sect. \ref{subsect:HVS}).
 }
 \label{fig:MagStream}
\end{figure*}

The Magellanic Stream (\citealt{Wannier1972}, \citealt*{Mathewson1974}) is a gaseous stream in the southern hemisphere, starting at the position of the LMC and SMC and extending over $150^{\circ}$\ towards the approximate position of M31 on the sky \citep{Nidever2010}. The distance of the MS from the Sun has not yet been successfully measured as no stellar counterpart of the stream has been discovered yet \citep{Guhathakurta1998}. 

Currently, there are several competing scenarios for the origin of the MS. It might have formed by the stripping of the Magellanic Clouds' gas on their orbit around the MW, either by tidal forces \citep*[e.g.][]{Connors2006} or by ram-pressure stripping due to the MW's hot halo gas \citep[e.g.][]{Mastropietro2005}. However, their large velocities derived from proper motion measurements \citep{Kallivayalil2013} indicate that the Magellanic Clouds can not have completed many orbits around the MW, which poses a challenge to these stream formation models \citep[but see][]{Mastropietro2009}. Depending on the MW potential and the circular velocity of the LSR, the proper motion measurements indicate that the Magellanic Clouds might even be on their first infall towards the MW. This lead \citet{Besla2010, Besla2012} to suggest that the MS might have been formed by the tidal interaction of LMC and SMC before they were accreted onto the MW. However, as pointed out by \citet{Pawlowski2012a}, the strong alignment of the MS, the LMC and SMC and their orbits with the VPOS around the MW would be an unlikely coincidence if the Magellanic Clouds would be unrelated to this structure and falling in towards the MW for the first time. One possible explanation for this coincidence is provided by the suggestion that the LMC and SMC \citep{Yang2010}, as well as other MW satellites in the VPOS \citep{Fouquet2012}, are TDGs which stem from a major merger in M31 \citep{Hammer2010, Hammer2013}, or that the MW satellites may be TDGs formed out of a long-past encounter between a larger LMC progenitor and the MW \citep{Pawlowski2011}.

In Fig. \ref{fig:MagStream}, we plot the positions and velocities of the LG galaxies on top of the Magellanic Stream map published as fig. 8 in \citet{Nidever2010}. For this, the galaxy positions have been transformed to the Magellanic Stream Coordinate System introduced by \citet{Nidever2008}, in which the Magellanic Stream lies along the equator and the position of the LMC on the equator defines the zero-point of the Magellanic Stream Longitude, $L_{\mathrm{MS}}$. The upper panel plots all LG galaxies which lie along the Magellanic Stream and no more than $40^{\circ}$\ away in Magellanic Stream Latitude $B_{\mathrm{MS}}$. The MW south pole lies in the centre of the plot, at $L_{\mathrm{MS}} \approx -55^{\circ}$. The lower panel plots the line-of-sight velocity of the galaxies with respect to the local standard of rest, overlaid on the same line-of-sight velocity measured along the Magellanic Stream. Only those galaxies which fall into the region shown in the upper panel, and which have measured velocities, are included. Thus, all galaxies in the lower panel lie close to the Magellanic Stream in projection. The symbols and colours again indicate the plane membership of the respective galaxies as in the previous plots.

The satellite galaxy structure around the MW approximately aligns with the MS, as has already been noticed by \citet{LyndenBell1976} and is discussed in detail in \citet{Pawlowski2012a}. The VPOSall and the MS are inclined by $24^{\circ}$\, and the VPOS-3 aligns even better ($11^{\circ}$).

The position and orientation of M31 is indicated by the black ellipse in Fig. \ref{fig:MagStream}. The MS approximately connects the LMC/SMC with M31, in the projected position (upper panel of Fig. \ref{fig:MagStream}), where M31 is offset to the 'north' by about $20^{\circ}$, but also in the line-of-sight velocities (lower panel), where M31 almost coincides with the position of a 'bump' to slightly less-negative velocities close to the tip of the MW. 
However, not only M31 coincides with the MS. The GPoA around M31 (red crosses) is oriented almost parallel to the MS (approximately horizontal in the upper panel of Fig. \ref{fig:MagStream}). The inclination between the GPoA and the MS is only $27^{\circ}$, and both are oriented polar with respect to the MW.

In particular the LG galaxies associated with LGP1 (yellow points) are close to the MS in projection (upper panel) and also follow the MS velocity trend (lower panel). The LGP1 member Phoenix, which is at the same time very close to the VPOSall/VPOS-3 planes and the GPoA, also lies along the MS equator.
The potential LGP1 member LSG 3 (leftmost red cross) is close to the MS velocity at its position, too. The other two potential LGP1 members (the two leftmost black crosses) M33 and Andromeda XXII \citep[which might in turn be a satellite of M33,][]{Chapman2013} deviate by about 200 km~s$^{-1}$\ from the MS's velocity at their projected position.

Similarly, the two LGP2 members (green symbols) which are close to the MS within the region plotted in the upper panel follow the MS's velocity trend. This is also true for the three M31 satellites which are potential LGP2 members: IC 10 (the rightmost red cross), Andromeda VII (rightmost black cross) and Andromeda XXI (third black cross from the right).

In addition, those M31 satellites (magenta symbols) which are close to the disc plane of M31 (black ellipse indicates its orientation) seem to connect the MS (starting with the Pegasus dIrr at the MS equator) with M31 in projected position (upper panel). It is worth mentioning that several HI clouds lie in the same direction and form a connection between the MS equator and M31, too.

We also include the compact high-velocity clouds (HVCs) detected by \citet{Westmeier2008} in Fig. \ref{fig:MagStream}. These HVCs are thought to be of common origin and associated to the MS because they lie close to the MS and their velocities closely follow those of the MS (see the small brown dots in both panels of Fig. \ref{fig:MagStream}). The HVCs are found at $B_{\mathrm{MS}} > 0^{\circ}$, so they lie in a similar region like some of the galaxies associated with LGP1, in particular Cetus, IC 1613 and LGS 3. \citet{Westmeier2008} discuss the possibility that the HVCs could be compact condensations within a more extended stream of mainly ionized gas associated with the MS. They also report that the HVCs can be grouped into five 'filaments'. We indicate these groups by plotting lines separating the groups in Fig. \ref{fig:MagStream}. Most of the groups are elongated approximately parallel to the MS, as already mentioned by \citet{Westmeier2008}. We find that they are at the same time approximately parallel to the GPoA. One example for this is the rightmost HVC filament, which extends the GPoA to the left in the upper panel of Fig. \ref{fig:MagStream}. 

The numerous agreements in position, orientation and velocity hint at an intimate connection between the MS, the Westmeier-HVCs, the VPOS around the MW, the GPoA around M31, the LGP1 and possibly even LGP2. A physical connection of the MS with these structures would imply a much larger extend of the MS than previously assumed. The decrease in the gas column density along the MS might then not only be due to a decrease in the gas mass along the stream, but also due to an increase in the stream's distance.
The slightly more-negative velocity of the MS and the HVCs compared to the LG galaxies in the same direction might be caused by the acceleration of more nearby gas towards the MW by the MW potential. This effect would be enhanced because more nearby gas clouds are more easily detected due to the $1/r^{2}$-behaviour of the flux density.

\subsection{Hypervelocity stars}
\label{subsect:HVS}

A hypervelocity star (HVS) is defined as a star which has such a large velocity that is can not be bound to the MW. The known HVSs are mostly of spectral type B and they are not distributed isotropically around the MW. There is a significant over-density in the direction of the constellation of Leo \citep{Abadi2009,Brown2009}. The over-density lies between Galactic longitudes $l$\ of $240^{\circ}$\ to $270^{\circ}$, and Galactic latitudes $b$\ of $75^{\circ}$\ down to at least $45^{\circ}$\ \citep*[see for example Figures 4 and 5 of][]{Brown2012}. The over-density might continue to lower Galactic latitudes, but the area covered by the SDSS, from which the target stars for the HVS survey of \citet{Brown2012} are selected, ends there. By comparing with the distribution of survey stars, \citet{Brown2012} demonstrate that the anisotropy is primarily in Galactic longitude, not in Galactic latitude. Therefore, the HVS over-density seems to be a polar structure, raising the question of whether it is aligned with the VPOS.

The normal to the VPOSall points to $(l,b) \approx (155^{\circ}, 0^{\circ})$, so the satellite galaxy structure runs approximately along the great-circle defined by the Galactic longitudes which are $90^{\circ}$\ offset from the normal direction: $l \approx 65^{\circ}$\ and $l \approx 245^{\circ}$. Similarly, the normal to the VPOS-3 points to $(l,b) \approx (170^{\circ}, 0^{\circ})$, so the corresponding great-circle is defined by Galactic longitudes $l \approx 80^{\circ}$\ and $l \approx 260^{\circ}$.
\textit{Therefore, the HVS over-density between $l = 240^{\circ}$\ and $l = 270^{\circ}$\ does indeed lie within the polar structure around the MW.} It aligns somewhat better with the VPOS-3 than with the fit to all MW satellites, which is also the case for several other features such as the MS and the MW satellite orbital poles.

HVSs are commonly assumed to be ejected by the disruption of a binary star system by the super-massive black hole in the centre of the MW \citep{Hills1988}. As this mechanism does not predict a strongly anisotropic distribution of HVSs, additional formation scenarios have been developed \citep[see for example][and references therein]{Brown2012}. Of particular interest in the context of the LG dwarf galaxy structures is the suggestion that the tidal disruption of a dwarf galaxy near the centre of the MW can contribute stars with high velocities to the MW halo \citep*{Abadi2009,Piffl2011}. Due to their common origin and orbital direction, these HVSs would cluster in a common direction.

If the HVS formation is related to objects such as dwarf galaxies falling in towards the MW, this parent object might in turn have been related to the planar galaxy structures, in particular the VPOS, LGP1 or LGP2. To get a crude estimate of the possible parent object's infall direction, we mirror the current positions of the HVS over-density on the sky. This assumes that the orbit of the parent object and the ejected HVSs is perfectly radial, which is not exactly the case. However, high eccentricities are beneficial for the creation of faster HVSs (\citealt*{Teyssier2009}, \citealt{Piffl2011}). The absence of an observed remnant of the parent object, a known problem for the tidal HVS scenario \citep{Piffl2011, Brown2012}, might be another indication for an almost radial orbit, as a close encounter can essentially destroy the infalling object during the first perigalactic passage. Nevertheless, a slightly non-radial orbit will result in an angle between the approaching and the departing path which is different from $180^{\circ}$. Thus, mirroring the HVS positions provides only a very general direction from which the HVS progenitor might have fallen in.

The mirrored direction to the HVS over-density lies between Galactic longitudes $l$\ of $60^{\circ}$\ to $90^{\circ}$, and Galactic latitudes $b$\ of $-75^{\circ}$\ to at least $-45^{\circ}$, with the possibility that it extends beyond this latitude. This region is highlighted by the black wedge in the upper panel of Fig. \ref{fig:MagStream}. It lies along a part of the Magellanic Stream and close to the region of the infalling HVC of \citet{Westmeier2008}. It is within about $50^{\circ}$\ from the position of M31. Its proximity to the LGP1, as indicated by the nearby LGP1-galaxies WLM and Cetus, is consistent with the possibility that a parent object might have fallen in along this structure. 

However, the tidal scenario for the HVS origin has serious difficulties, such as the spread in the HVS ejection times, and might therefore be unable to explain the formation of the observed HVSs. The alignment of the over-density with the MW VPOS might then simply be coincidental. Currently none of the competing scenarios for the origin of the HVS anisotropy are without difficulties \citep{Brown2012}. When investigating the tidal and possibly other scenarios for the HVS over-density, it might therefore be worthwhile to consider the constraints provided by the dwarf galaxy structures in the LG.

\section{Possible origins of the found structures in the LG}
\label{sect:origins}

\begin{figure}
 \centering
 \includegraphics[width=70mm]{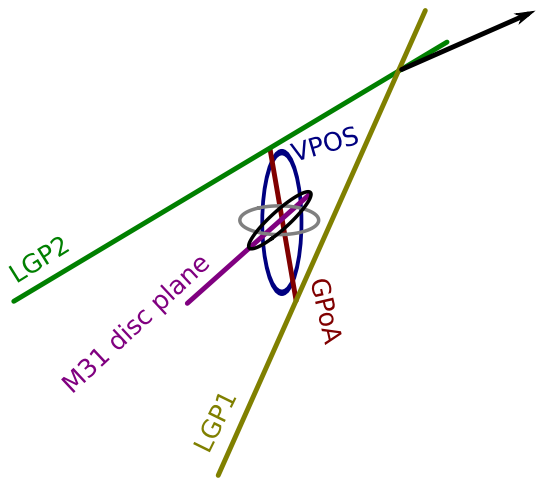}
 \caption{
Cartoon of the LG structure (compare to Fig. \ref{fig:LGP1-LGP2}). The positions and orientations of the galactic discs of the MW (grey) and of M31 (black) are indicated by the ellipses in the centre. Looking along the MW-M31 line, most planes in the LG are seen approximately edge-on, the only exception is the VPOS plane (blue), which is inclined relative to this view. The arrow indicates the direction of motion of the LG relative to the CMB.
 }
 \label{fig:cartoon}
\end{figure}

What could be the origin of the planar galaxy structures? The presently proposed scenarios can be broadly classified into two types which we discuss in the following. They are either based on the accretion of primordial dwarf galaxies or on the formation of phase-space correlated second-generation dwarf galaxies. However, we make no claim to be  complete in this discussion of possible origins of the planar structures because modifications and entirely different explanations might arise in the future.

To recapitulate: most currently known LG galaxies are distributed either in vast discs of satellite galaxies about the two major hosts, the MW and M31, or in two symmetric planes that are approximately equidistant from the hosts and inclined relative to each other by $35^{\circ}$. Fig. \ref{fig:cartoon} depicts this situation schematically.

\subsection{Primordial dwarf galaxies}

The majority of (satellite) dwarf galaxies in the Universe are often believed to be primordial dwarf galaxies which reside in dark matter (sub-) haloes. Their expected distribution is usually studied based on dark matter simulations within the $\Lambda$CDM framework. 
Most studies to date have focussed on investigating the overall flatness of a satellite galaxy distribution in the attempt to explain the VPOS, or more generally the flattening of the MW satellite galaxy system \citep{Zentner2005,Libeskind2005,Deason2011}. They are therefore not immediately applicable to the GPoA, which only consists of a subset of M31 satellite galaxies. On the one hand, the overall M31 satellite system is only moderately flattened and therefore more consistent with cosmological expectations. However, the GPoA is thinner than the VPOSall (but not so much thinner than the VPOS-3) and most of its members evidentially co-orbit, which both indicates a common origin for about half of the M31 satellites. Many of the other M31 satellite galaxies are found close to the galactic disc plane of M31. Making things worse, not even the relatively high frequency of galaxy pairs in the MW and M31 satellite system is expected by the current galaxy formation models based on $\Lambda$CDM \citep{Fattahi2013}.

Studies reporting putative agreement between the flattening of the MW satellite galaxy system and cosmological simulations need to be interpreted carefully. One example is the recent claim by \citet{Wang2013} that 5-10 per cent of simulated satellite systems can be as flat as the MW satellite system. Their study investigates satellite systems derived from two different kinds of simulations: the six high-resolution Aquarius simulations \citep{Springel2008}, and the larger-scale Millennium-II simulation \citep{BoylanKolchin2009}. \citet{Wang2013} populate these dark-matter-only simulations using a semi-analytic galaxy formation model and an abundance matching technique. The flattening of the 11 model satellites with the largest stellar mass is then compared to that of the 11 most luminous MW satellites using two different measures: the axis ratios $c/a$, and the ratio of the RMS thickness of the best-fit satellite plane, $r_{\mathrm{thick}}$, to a cutoff radius, $r_{\mathrm{cut}}$, fixed at = 250 kpc.
None of the six high-resolution simulations is able to reproduce the observed flattening in $c/a$, even when accounting for a 16.5 per cent sky obscuration region due to the MW disc\footnote{\citet{Wang2013} state that they use a 33 per cent occulted sky fraction by removing all satellites within an angle $\theta_{\mathrm{crit}} = 9^{\circ}.5$\ of the artificial MW disc. However, this angle corresponds to an obscuration of only 16.5 per cent ($\sin \theta_{\mathrm{crit}} = 0.165$), which is also in better agreement with their statement that on average 2.6 satellites were replaced per halo because they were within the obscuration region. A 33 per cent obscured region implies $\theta_{\mathrm{crit}} = 19^{\circ}.3$, but this would be inconsistent with the inclusion of the Sagittarius dwarf galaxy in the MW satellite sample due to its low Galactic latitude of $b = -14^{\circ}.1$.}. 

Their second comparison using $r_{\mathrm{thick}} / r_{\mathrm{cut}}$\ to measure the flattening can only yield meaningful results if the satellite systems have the same radial profile. A compact system will have a smaller $r_{\mathrm{thick}}$\ than a more extended one even if both have the same isotropic angular distribution. The flattening, as measured by dividing $r_{\mathrm{thick}}$\ by a cut-off radius which is not determined for each individual satellite system but fixed at the MW value ($r_{\mathrm{cut}} = 250$\,kpc), is therefore biased towards assigning a more extreme flattening to a radially more concentrated distribution. Therefore, an apparent agreement of the simulated and observed flattening as measured via $r_{\mathrm{thick}} / r_{\mathrm{cut}}$\ is no indication that MW-like satellite systems are present in the simulation \citep[see also][]{Kang2005,Metz2007}. Figure 5 of \citet{Wang2013} indicates that those simulations where $r_{\mathrm{thick}} / r_{\mathrm{cut}}$\ comes close to the value derived for the MW satellites indeed have a more concentrated radial profile. While \citet{Wang2013} state that the MW satellite population is '[...] flatter than most of the simulations', their figure 10 reveals that it is in fact flatter than \textit{all} six simulated satellite populations. Thus, despite a bias towards a stronger flattening none of the six high-resolution simulations reproduces the same $r_{\mathrm{thick}} / r_{\mathrm{cut}}$\ as the 11 bright MW satellites.

Nevertheless, \citet{Wang2013} state that 5-10 per cent of the simulated satellite systems are as flat as the MW system. This number is entirely based on the satellites derived from 1686 MW-like haloes of the Millennium-II simulation. In particular the comparison based on $r_{\mathrm{thick}} / r_{\mathrm{cut}}$\ also suffers from the different radial profiles of the simulated satellite systems. That modelled satellite systems result in $r_{\mathrm{thick}} / r_{\mathrm{cut}}$-values similar to those for the observed satellites is not informative because the differences in the radial distributions are unaccounted for. An additional, major problem with the Millennium-II results is the comparably low resolution of the simulation. \citet{Wang2013} had to include satellites 'within' unresolved sub-haloes in their analysis in order to arrive at an approximate agreement in the radial profile of the top 11 satellites between the low- and the high-resolution simulations. However, \citet{Wang2013} themselves mention the disadvantage of following unresolved satellites, stating that their spatial distribution is uncertain and model dependent because their orbits cannot be tracked within the $N$-body simulation. They also explain: 'A position is assigned to these galaxies by tracking the most bound particle of the host subhalo from the time it was last resolved. This position is unlikely to be a very accurate estimate of the true orbit of the satellite [...]'. The determination of the satellite flattening is entirely based on the satellite positions. It is therefore very questionable whether the flattening determined from the Millennium-II haloes has any informative value concerning the comparison with the flattening of the MW satellite system.

\citet{Starkenburg2013} perform a similar analysis also based on the Aquarius simulations, but investigate the flattening of the galaxy \textit{directions}, without considering their radial distance from the host. They report that the distributions of bright satellites in all six simulations is less flattened than the distribution of the 12 brightest MW satellites. Their models contain a factor of about 2 to 4 more bright satellites ($M_{\mathrm{V}} < -8.5$) than the MW, indicating a more fundamental mismatch between the models and the observed situation. When randomly sampling the observed number of satellites (12 in their case) from their model satellites, there is a low probability that such a sub-sample can reproduce the observed flattening. However, such a random sampling lacks a physical motivation as it removes otherwise expected satellites from the distribution.

To account for the coherent orbital directions of MW satellites within the VPOS \citep{Metz2008,Pawlowski2013}, it has been suggested that some of them might have been accreted onto the MW together as a group \citep{LiHelmi2008,DOnghiaLake2008,Deason2011}. Such a common origin would leave an imprint in the form of a common orbital angular momentum and would therefore also be in principle applicable to the GPoA. However, observed dwarf associations are much more extended than structures as thin as the VPOS or the even thinner GPoA, which therefore can not be formed by accretion of dwarf associations \citep{Metz2009b}. High-resolution simulations also indicate that the 11 most-massive satellites are not accreted in groups but individually \citep{Wang2013}.
In regard to the LG planes LGP1 and LGP2, group infall can also not be considered an explanation. The scenario is based on the idea that the galaxies were close together before being accreted onto their host, so would not disperse along a plane of 1-2 Mpc diameter.

On LG scales the influence of the filamentary distribution of dark matter haloes might become important.  
However, the dark matter filaments found in numerical simulations are too extended to resemble structures such as the LGP1 and LGP2, which in addition have axis ratios indicating a planar rather than a filament-like shape. The size of filaments at present time is comparable to the virial radius of the host galaxy (about 300  kpc for the MW and M31) and therefore too large to explain any of the thin planar structures that have heights of only a few tens of kpc \citep{Vera-Ciro2011}. Analysing the simulated LG equivalent from the Constrained Local Universe Simulation, \citet{Libeskind2011} demonstrate that signatures for a preferred direction of infall of sub-haloes is detected on scales down to the virial radius of a main halo. This still implies a much larger size scale than the RMS height of the planar galaxy structures in the LG and in addition a preferred infall direction does not imply that the majority or even entirety of sub-haloes is accreted from it.
The similar size between the filaments and the host galaxy haloes therefore results in a near-isotropic accretion of dark matter sub-haloes onto host haloes \citep{Lovell2011,Pawlowski2012b}. Unless all the baryonic matter is confined to a filament which are one to two orders of magnitude thinner than their dark matter counterparts, for which there is no evidence, the accretion along cosmic filaments is therefore unable to account for the dwarf galaxy structures in the LG.

Even if the intrinsic distribution of primordial dwarf galaxies is not sufficiently flat to resemble the LG planes, there could still be effects which cause the distribution to become thinner. \citet{Pasetto2009} investigate the tidal forces exerted on the LG by nearby galaxies groups (within 4.5 Mpc). They demonstrate that the planar distribution of LG galaxies discovered by \citet{Pasetto2007}, which is close to our plane fit for all non-satellite galaxies, is compatible with the current external force field. The population of non-satellite LG galaxies might have been tidally compressed in the direction perpendicular to the plane during the past 9 Gyr. However, the analysis of \citet{Pasetto2009} is based on the orbits of the external galaxies as derived from a minimum action method, which implies major uncertainties. In addition, their sample of external galaxies is limited to only six groups within 4.5 Mpc. 
Concerning the apparent two-plane structure in the LG, it is unclear whether the influence of tidal compression could be responsible, but the small inclinations between the plane fitted to all non-satellite galaxies and both LGP1 and LGP2 (Fig. \ref{fig:LG4dwarfnormal}) indicates that both planes also align approximately with the plane of tidal compression by \citet{Pasetto2009}. However, even if the effect described by \citet{Pasetto2009} is responsible for the planar arrangement of the non-satellite LG galaxies, this does not provide any information about the nature of the galaxies, as the effect would influence both primordial and second-generation galaxies formed via dynamical processes at later times.

\subsection{Second-generation dwarf galaxies}

The tidal forces acting during galaxy collisions involving disc galaxies can expel matter from the galactic discs, resulting in the formation of long tidal tails. These tails contain stars from the progenitor galaxy and large amounts of gas. New stellar systems, super star clusters and tidal dwarf galaxies (TDGs), can form from this tidal debris, reaching masses of up to $10^{10}~\mathrm{M}_{\sun}$ \citep*{Bournaud2006,Wetzstein2007,Bournaud2008,Fouquet2012}. TDGs formed in a common tidal tail will either orbit in a coherent thin plane around their progenitor or its interaction partner \citep*{Pawlowski2011}, or they will be expelled together with the tidal debris to larger distances. TDGs therefore suggest themselves as a natural origin for coherent planes of dwarf galaxies. A detailed discussion of the TDG scenario with emphasis on the VPOS can be found in \citet{Pawlowski2012a}, but we like to mention that TDGs can be long-lived \citep{Kroupa1997,Recchi2007,Duc2011,Casas2012} and that ancient TDGs show remarkable similarities with dwarf elliptical galaxies \citep{Dabringhausen2013}.

A tidal tail has to be of comparable thickness to that of the observed galaxy planes, otherwise it could not be responsible for forming that structure (as argued before for the group infall and filamentary accretion scenarios). In general, a tidal tail expelled from a galactic disc has a similar height as the disc. It can be as thin as several kpc only.
The RMS height of LGP1 is 55 kpc, or only 36 kpc for the extended galaxy sample excluding the outlier Andromeda XVI. For LGP2, the RMS height is 66 kpc, or only 6 kpc excluding the most-distant outlier. These heights should be considered to be upper limits of the structures' extend. The reason is that the plane fit does not take a possible curvature in the galaxy distribution into account. A tidal tail could be bend towards the major galaxies by their gravitational potential if it does not run through the centre of mass of the system. This is the case for both LGP1 and LGP2, which are offset from the major masses in the LG (MW and M31). Therefore, we consider the LG plane heights to be comparable to the value derived for the VPOSall (29 kpc, 20 kpc for the VPOS-3) and possibly even the GPoA (14 kpc). The LG planes are consistent with being tidal tails approximately connecting the MW and M31.

\citet{LyndenBell1976} has first suggested that the planarity of the MW satellite distribution might be explained by second-generation galaxies. He speculated that some of the MW satellites close to the plane defined by the MS are objects that were torn out of a hypothetical Greater Magellanic galaxy, the major surviving part of which today is the LMC. This interpretation was revisited by \citet{Kroupa1997} and \citet{Casas2012} by demonstrating that the high dynamical mass-to-light ratios of the MW dwarf spheroidal satellites may be explained by significant tidal influences, although non-Newtonian explanations appear more likely \citep{McGaugh2010}. The possibility that the satellites may be second-generation galaxies has also been discussed in \citet{Pawlowski2011}, who demonstrated that tidal debris indeed follows a planar distribution and that both co- and counter-orbiting debris can be formed. However, this scenario would require a past encounter of the MW and the LMC-progenitor, which might be difficult to reconcile with the large proper motion of the LMC (\citealt{Kallivayalil2013}, but see the discussion in \citealt{Pawlowski2012a}).

Alternatively, the LMC itself might be of tidal origin. The most-sophisticated scenario to date involving the formation of TDGs in the LG has been presented by \citet{Hammer2010}. They suggest that M31 experienced a major merger which started about 9 Gyr ago. Their numerical models demonstrate that such a merger can reproduce many of the features observed in M31, including the bulge, the thin and thick disc, the 10-kpc ring and the giant stream. After constraining the merger to reproduce these features, additional agreements with the observed LG became apparent. During the merger, which involved young and thus very gas-rich galaxies (gas fraction of 60 per cent or even more), a large number of TDGs are formed. Many of these can be expected to orbit the merger-remnant and the modelled tidal debris indeed reproduces the orientation and rotation of the GPoA around M31 \citep{Hammer2013}. In addition, parts of the tidal tail developing during the first pericentre of the merger escape from M31's potential. Their direction of motion points towards the MW. 
This led \citet{Yang2010} to suggest that the Magellanic Clouds might be TDGs originating from the M31 merger, a scenario which also explains the large angular momentum of the Magellanic Clouds, their velocities being the sum of the relative MW-M31 velocity and the additional velocity by the expelled tidal tail. Following up on this, \citet{Fouquet2012} investigated whether the whole VPOS around the MW might have been formed by TDGs expelled towards the MW. They conclude that a link between the VPOS and a major merger at the location of M31 is plausible. The tidal compression by the external distribution of galaxies investigated by \citet{Pasetto2009} might have supported such an alignment by 'bending' the tidal tail towards the MW.

A common origin of the two satellite planes VPOS (around the MW) and GPoA (around M31) would also imply the existence of at least one tidal tail connecting the two major galaxies. This might be LGP1, which exhibits a number of consistencies with this scenario. It is parallel to the line connecting the MW with M31 (it is a necessary requirement for the tidal tail to be close to both galaxies, to M31 because it is the tail's origin, and to the MW if TDGs accreted from the tail are to form the VPOS). It might start at M31's position (see Fig. \ref{fig:LGP1faceon}). As discussed in Sect. \ref{subsect:magstream}, the LGP1 members also lie close to the MS in projection and have similar velocities, which might indicate that both the LGP1 and the MS are different parts of the same, larger tidal tail which is being accreted onto the MW. In this scenario, the MS could be a mixture of a part of the tidal tail connecting M31 with the MW and gas expelled from the Magellanic Clouds via ram-pressure stripping and tidal interactions. According to \citet{Raychaudhury1989}, M31 was closer to the orbital plane of the LMC in the past, which would further align the VPOS, LGP1 and GPoA \citep[see also][]{Yang2010}. Finally, a relative movement of the MW with respect to the tidal tail changes the direction from which the TDGs are accreted with time. This would result in a spread of the orbital directions of accreted objects, which would widen the accreted debris structure and could possibly explain the wider extend of the VPOS compared to the GPoA and the spread (or 2-peak shape) of the 4-galaxy-normal directions as seen in Fig. \ref{fig:MW4dwarfnormal} for the MW satellites.

An alternative to this M31 merger model is the possibility that TDGs formed in a past fly-by encounter between the early MW and M31 about 10 Gyr ago \citep{Pawlowski2012a}. Under the assumption of Milgromian dynamics, the MW-M31 system must have had a past, close encounter between the two galaxies about 7-11 Gyr ago \citep{Zhao2013}. This is consistent with the expected formation age of the VPOS discussed in \citet{Pawlowski2012a}. Tidal debris formed in such a fly-by encounter can connect the two departing galaxies for a long time after the encounter \citep{Pawlowski2011}. This scenario is in qualitative agreement with the existence of the LG planes being parallel to the line connecting the MW and M31. Within the uncertainties of the proper motion measurement for M31, the MW-M31 orbital plane and LGP1 have a very similar orientation. This is in agreement with the expectation that the large-scale tidal debris of such an encounter are confined to the orbital plane of the interacting galaxies. 
It is helpful for the development of extended tidal tails if the orbital angular momentum of the encounter and the spin angular momentum of the galactic disc are prograde and well-aligned, which would be the case if the MW-M31 orbital pole aligns with the LGP1 normal. Similarly, that both LG planes are inclined by only $20^{\circ}$\ to the galactic disc of M31 might be another indication for a tidal origin in M31. As tidal debris preferentially co-orbits in the direction defined by the angular momentum of the encounter the prograde orbital sense of the VPOS and the GPoA is consistent with a tidal debris origin, while a number of apparently counter-orbiting satellites can also be expected in a TDG scenario \citep{Pawlowski2011}.

A similar scenario has been proposed by \citet{Sawa2005}. In their model, the primordial MW and M31 had a pericentric passage about 10 Gyr ago, with a minimum distance of less then 150 kpc.  
Instead of TDG formation, they hypothesise that extended gas around the proto-galaxies was compressed by the encounter, resulting in the condensation of gas clouds of which some evolved into dwarf galaxies. Assuming that these galaxies are distributed in the orbital plane of the interaction, \citet{Sawa2005} predict a M31 proper motion (in galactic coordinates) of 
\begin{displaymath}
(\mu_{\mathrm{l}}, \mu_{\mathrm{b}})_{\mathrm{predicted}} = (38 \pm 16~\mu \mathrm{as~yr}^{-1}, -49 \pm 5~\mu \mathrm{as~yr}^{-1}).
\end{displaymath}
In their study they adopt a circular velocity of the LSR of $220~\mathrm{km~s}^{-1}$. Transforming the HST proper motion of M31 by \citet{vdMarel2012a}, which was discussed in Sect. \ref{subsect:M31propmo}, to Galactic coordinates according to \citet{Brunthaler2007} results in 
\begin{displaymath}
(\mu_{\mathrm{l}}, \mu_{\mathrm{b}})_{\mathrm{measured}} = (46 \pm 13~\mu \mathrm{as~yr}^{-1}, -30 \pm 13~\mu \mathrm{as~yr}^{-1}).
\end{displaymath}
These values are very similar to those predicted by \citet{Sawa2005}. The predicted and measured values of $\mu_{\mathrm{l}}$\ overlap well and those of $\mu_{\mathrm{b}}$\ almost agree within the respective uncertainties. The scenario proposed by \citet{Sawa2005} should therefore also be considered and investigated further.

If many of the LG dwarf galaxies turn out to be TDGs or similar second-generation objects, in accordance with the initial suggestion \citep{LyndenBell1976}, this would imply that near-field cosmology has chosen a wrong assumption when investigating the MW and M31 satellite galaxy system as \textit{purely} tracing dark matter sub-haloes. If the majority of satellite galaxies are of tidal origin, this would disastrously worsen the 'missing satellites problem'. It would render much of the research results obtained based on this particular interpretation of the MW satellite system highly questionable and might even result in a paradigm shift in our understanding of gravity (\citealt{Kroupa2012a}, \citealt*{Kroupa2012b}). The implications of the TDG scenario are therefore extremely far-reaching, illustrating that the question of the origin of the dwarf galaxy planes must not be taken lightly.

\subsection{Outlook}

The discovery of the satellite and LG dwarf galaxy planes currently poses a riddle to the field of galaxy formation. None of the currently proposed explanations is without problems or has already addressed all issues (see the previous section). Even the debate about the origin of the VPOS, the longest-known structure, is far from reaching a consensus \citep{Kroupa2010,Deason2011,Lovell2011,Libeskind2011,Pawlowski2012a,Pawlowski2012b,Fouquet2012,Kroupa2012a,Wang2013}. However, this debate illustrates the current dilemma well: within the prevailing cold dark matter based cosmology, a thin structure of co-orbiting satellites such as the VPOS is unexpected. The promising alternative is the formation of TDGs in galaxy interactions, which naturally explains co-orbiting, planar structures. But it faces the problem that within the dark matter paradigm, TDGs should appear free of dark matter \citep{Barnes1992,Wetzstein2007}. The high mass-to-light ratios derived from the velocity dispersions of the MW satellites seem to contradict this, unless the velocity dispersions are seriously over-estimated \citep[e.g.][]{McConnachie2010}, the underlying assumption that the galaxies are bound systems is invalid \citep{Kroupa1997, Klessen1998,Casas2012} or the underlying dynamics is non-Newtonian \citep{Angus2008,FamaeyMcGaugh2012,McGaugh2013}. The additional information provided by the newly discovered planar structures in the LG will help the search for a consistent solution. What could be the next steps in this regard?

First of all, it will be necessary to investigate whether the found planar distributions of the non-satellite LG galaxies are indeed coherent dynamical structures, or mere chance alignments that arise due to the low number of known objects (and that happen to be very symmetric and aligned with the MS and possibly the MW-M31 orbit by chance, too). In particular the upcoming searches for MW satellite and LG dwarf galaxies in the southern hemisphere will test this by providing a more complete census of the LG dwarf galaxy population \citep{Jerjen2010,Jerjen2012}. If the position on the sky of a newly-discovered dwarf galaxy is given, the plane parameters listed in Table \ref{tab:allplanes} (centroid position and plane normal) can be used predict the distances to the galaxies, assuming it to be a plane member. Additionally, very important dynamical information could be provided not only by line-of-sight velocity measurements, but also by proper motion measurements of LG galaxies.

A different route of investigation could address the question whether such planes are common throughout the Universe. This might prove difficult to investigate, as a full analysis of dwarf galaxy systems requires knowledge of all three spatial coordinates. Furthermore, the observational studies would need to be deep in order to discover faint dwarf galaxies. At the same time, such observations would have to cover a wide field around the host galaxies (to discover VPOS/GPoA analogues) or even a whole galaxy group (to search for LGP1/LGP2 analogues). 
In addition, as distance determinations are not precise enough to allow the investigation of the full three dimensional dwarf galaxy distribution, only projected distributions can be analysed. Thus, not only a few but a large sample of LG-like galaxy groups has to be studied to properly estimate of the abundance of planar dwarf galaxy distributions which will only be discovered unambiguously if seen close to edge-on. Therefore, even if similar planar distributions are common in the Universe it is no surprise that they have not yet been discovered.

In addition to observational investigations, both the dynamics and possible formation scenarios of dwarf galaxy planes need to be investigated numerically. This includes (but is not restricted to) modelling the formation of planar structures of TDGs in galaxy mergers and fly-by interactions, investigating possible orbits for the non-satellite LG galaxies that preserve the distribution's planarity over time and testing whether a realistic treatment of baryons in high-resolution cosmological simulations could result in planar galaxy distribution.

\section{Conclusions}
\label{sect:conclusions}

The MW is surrounded by a vast polar structure (VPOS) of satellite galaxies, a thin plane with a RMS height of only 29 kpc (for the VPOSall, only 20 kpc for the VPOS-3 which excludes only three outliers) that is oriented perpendicular to the Galactic disc. The satellite galaxy orbital poles indicate that most MW satellites (8 of 11) co-orbit in the VPOS \citep{Metz2008,Pawlowski2013}, and also the young halo globular clusters and streams in the MW haloes are aligned with it \citep{Pawlowski2012a}.
A similar structure has recently been discovered at very high significance in the M31 satellite galaxy system \citep{Ibata2013,Conn2013}. This Great Plane of Andromeda (GPoA) consists of up to 19 of the 34 known M31 satellites, has a RMS height of only 14 kpc and is seen edge-on from the MW. This favourable orientation reveals that most satellites within the GPoA co-orbit.
Thus, planar structures of satellite galaxies have been found around both major LG galaxies, which constitute the two only satellite galaxy systems for which precise three-dimensional positions are available.

The non-satellite galaxies in the LG as a whole are only mildly flattened, but they can be split into two sub-samples which have intriguing properties. All but one of the 15 non-satellite LG galaxies lie within one of two LG planes (LGP1 and LGP2), which are inclined relative to each other by $35^{\circ}$. Both planes are thin (RMS heights of 55 and 66 kpc, LGP1 might be as thin as 36 kpc if Andromeda XVI is considered a member of the GPoA) and have very symmetric orientations. They are inclined by only $\approx 20^{\circ}$\ with respect to the galactic disc of M31, are both parallel to the line connecting the M31 with the MW and have similar offsets from both major galaxies. In addition, the LG galaxies apparently follow a common, arc-like trend in radial distance from the midpoint between the MW and M31.

Comparing the orientations of the VPOS, GPoA, LGP1 and LGP2 with other prominent features observed around the MW indicates possible connections. On the largest scales, the LGP1 and the GPoA are closely aligned with the Supergalactic Plane. The LG velocity with respect to the CMB lies within most of the planar structures and approximately points towards the tip of the wedge formed by LGP1 and LGP2.

On LG scales, the VPOS and GPoA are are inclined by $51^{\circ}$\ (for the VPOSall, $38^{\circ}$\ for the VPOS-3) and their satellites preferentially co-orbit in the same direction, which is also prograde with respect to the orbital sense of the MW-M31 system as deduced from the M31 proper motion. The most-likely orbital plane of the MW-M31 system is closely aligned to the GPoA, but due to the large proper motion uncertainties the orbital plane is also consistent with being aligned with the LGP1.

The Magellanic Stream (MS) might be the link between the VPOS and the GPoA, which would imply a larger extend of the gaseous structure than commonly assumed. It is aligned with both satellite galaxy planes (inclined by less than $30^{\circ}$\ to each) and approximately connects the Magellanic Clouds (which lie and orbit within the VPOS) with M31, both in projected position and in line-of-sight velocity. The non-satellite galaxies which we suggest as members of the LGP1 intriguingly follow the same trend. This is also true for high-velocity clouds probably associated with the MS, which themselves can be separated into filament-like groups that are oriented approximately parallel to the VPOS, the MS and the GPoA. The hypervelocity star over-density observed in the MW halo also aligns with the VPOS. Some theories suggest that the over-density was formed when a dwarf galaxy on a highly eccentric orbit was disrupted near the centre of the MW. A crude estimate places the possible origin of such a dwarf galaxy in the general direction of the MS.

We are therefore lead to consider the $\approx 40^{\circ}$-wide region extending between the Magellanic Clouds and M31 to be the 'direction of decision' for scenarios which intend to explain the formation and mutual orientation in position and velocity space of the satellite and non-satellite structures. The correlated, planar structures of galaxies in the LG are unexpected in the common galaxy formation theories which assume that essentially all galaxies are primordial, dark matter dominated objects. The structures may be a natural occurrence if the LG was shaped by a major galaxy interaction, which expelled tidal debris to large distances. However, if the majority of the LG galaxies are tidal dwarf galaxies formed from the tidal debris, then our understanding of galaxy formation, near-field cosmology and possibly even gravitational dynamics is in need of major revisions \citep{Kroupa2012a,Kroupa2012b}.

\section*{Acknowledgements}
MP acknowledges support through DFG reseach grant KR 1635/18-2 in the frame of the DFG Priority Programme 1177, \textit{Witnesses of Cosmic History: Formation and evolution of galaxies, black holes, and their environment}. HJ acknowledges the financial support from the Australian Research Council through the Discovery Project grant DP120100475. The authors also acknowledge financial support of a DAAD/Go8 travel grant.
We thank Tobias Westmeier for providing the data of the HVCs plotted in Fig. \ref{fig:MagStream}.

\bibliographystyle{mn2e}
\bibliography{./LGplanesref}

\begin{thebibliography}{}

\bibitem[\protect\citeauthoryear{{Abadi}, {Navarro} \& {Steinmetz}}{{Abadi}
  et~al.}{2009}]{Abadi2009}
{Abadi} M.~G.,  {Navarro} J.~F.,    {Steinmetz} M.,  2009, \apjl, 691, L63

\bibitem[\protect\citeauthoryear{{Angus}}{{Angus}}{2008}]{Angus2008}
{Angus} G.~W.,  2008, \mnras, 387, 1481

\bibitem[\protect\citeauthoryear{{Angus}, {Diaferio} \& {Kroupa}}{{Angus}
  et~al.}{2011}]{Angus2011}
{Angus} G.~W.,  {Diaferio} A.,    {Kroupa} P.,  2011, \mnras, 416, 1401

\bibitem[\protect\citeauthoryear{{Barnes} \& {Hernquist}}{{Barnes} \&
  {Hernquist}}{1992}]{Barnes1992}
{Barnes} J.~E.,  {Hernquist} L.,  1992, \araa, 30, 705

\bibitem[\protect\citeauthoryear{{Besla}, {Kallivayalil}, {Hernquist}, {van der
  Marel}, {Cox} \& {Kere{\v s}}}{{Besla} et~al.}{2010}]{Besla2010}
{Besla} G.,  {Kallivayalil} N.,  {Hernquist} L.,  {van der Marel} R.~P.,  {Cox}
  T.~J.,    {Kere{\v s}} D.,  2010, \apjl, 721, L97

\bibitem[\protect\citeauthoryear{{Besla}, {Kallivayalil}, {Hernquist}, {van der
  Marel}, {Cox} \& {Kere{\v s}}}{{Besla} et~al.}{2012}]{Besla2012}
{Besla} G.,  {Kallivayalil} N.,  {Hernquist} L.,  {van der Marel} R.~P.,  {Cox}
  T.~J.,    {Kere{\v s}} D.,  2012, \mnras, 421, 2109

\bibitem[\protect\citeauthoryear{{Bilicki}, {Chodorowski}, {Jarrett} \&
  {Mamon}}{{Bilicki} et~al.}{2011}]{Bilicki2011}
{Bilicki} M.,  {Chodorowski} M.,  {Jarrett} T.,    {Mamon} G.~A.,  2011, \apj,
  741, 31

\bibitem[\protect\citeauthoryear{{Bournaud}}{{Bournaud}}{2010}]{Bournaud2010}
{Bournaud} F.,  2010, Advances in Astronomy, 2010

\bibitem[\protect\citeauthoryear{{Bournaud} \& {Duc}}{{Bournaud} \&
  {Duc}}{2006}]{Bournaud2006}
{Bournaud} F.,  {Duc} P.-A.,  2006, \aap, 456, 481

\bibitem[\protect\citeauthoryear{{Bournaud}, {Duc} \& {Emsellem}}{{Bournaud}
  et~al.}{2008}]{Bournaud2008}
{Bournaud} F.,  {Duc} P.-A.,    {Emsellem} E.,  2008, \mnras, 389, L8

\bibitem[\protect\citeauthoryear{{Bovill} \& {Ricotti}}{{Bovill} \&
  {Ricotti}}{2011}]{Bovill2011}
{Bovill} M.~S.,  {Ricotti} M.,  2011, \apj, 741, 18

\bibitem[\protect\citeauthoryear{{Boylan-Kolchin}, {Bullock} \&
  {Kaplinghat}}{{Boylan-Kolchin} et~al.}{2011}]{BoylanKolchin2011}
{Boylan-Kolchin} M.,  {Bullock} J.~S.,    {Kaplinghat} M.,  2011, \mnras, 415,
  L40

\bibitem[\protect\citeauthoryear{{Boylan-Kolchin}, {Springel}, {White},
  {Jenkins} \& {Lemson}}{{Boylan-Kolchin} et~al.}{2009}]{BoylanKolchin2009}
{Boylan-Kolchin} M.,  {Springel} V.,  {White} S.~D.~M.,  {Jenkins} A.,
  {Lemson} G.,  2009, \mnras, 398, 1150

\bibitem[\protect\citeauthoryear{{Brown}, {Geller} \& {Kenyon}}{{Brown}
  et~al.}{2012}]{Brown2012}
{Brown} W.~R.,  {Geller} M.~J.,    {Kenyon} S.~J.,  2012, \apj, 751, 55

\bibitem[\protect\citeauthoryear{{Brown}, {Geller}, {Kenyon} \&
  {Bromley}}{{Brown} et~al.}{2009}]{Brown2009}
{Brown} W.~R.,  {Geller} M.~J.,  {Kenyon} S.~J.,    {Bromley} B.~C.,  2009,
  \apjl, 690, L69

\bibitem[\protect\citeauthoryear{{Brunthaler}, {Reid}, {Falcke}, {Greenhill} \&
  {Henkel}}{{Brunthaler} et~al.}{2005}]{Brunthaler2005}
{Brunthaler} A.,  {Reid} M.~J.,  {Falcke} H.,  {Greenhill} L.~J.,    {Henkel}
  C.,  2005, Science, 307, 1440

\bibitem[\protect\citeauthoryear{{Brunthaler}, {Reid}, {Falcke}, {Henkel} \&
  {Menten}}{{Brunthaler} et~al.}{2007}]{Brunthaler2007}
{Brunthaler} A.,  {Reid} M.~J.,  {Falcke} H.,  {Henkel} C.,    {Menten} K.~M.,
  2007, \aap, 462, 101

\bibitem[\protect\citeauthoryear{{Casas}, {Arias}, {Pe{\~n}a Ram{\'{\i}}rez} \&
  {Kroupa}}{{Casas} et~al.}{2012}]{Casas2012}
{Casas} R.~A.,  {Arias} V.,  {Pe{\~n}a Ram{\'{\i}}rez} K.,    {Kroupa} P.,
  2012, \mnras, 424, 1941

\bibitem[\protect\citeauthoryear{{Chapman}, {Widrow}, {Collins}, {Dubinski},
  {Ibata}, {Rich}, {Ferguson}, {Irwin}, {Lewis}, {Martin}, {McConnachie},
  {Pe{\~n}arrubia} \& {Tanvir}}{{Chapman} et~al.}{2013}]{Chapman2013}
{Chapman} S.~C.,  {Widrow} L.,  {Collins} M.~L.~M.,  {Dubinski} J.,  {Ibata}
  R.~A.,  {Rich} M.,  {Ferguson} A.~M.~N.,  {Irwin} M.~J.,  {Lewis} G.~F.,
  {Martin} N.,  {McConnachie} A.,  {Pe{\~n}arrubia} J.,    {Tanvir} N.,  2013,
  \mnras, 430, 37

\bibitem[\protect\citeauthoryear{{Conn}, {Ibata}, {Lewis}, {Parker}, {Zucker},
  {Martin}, {McConnachie}, {Irwin}, {Tanvir}, {Fardal}, {Ferguson}, {Chapman}
  \& {Valls-Gabaud}}{{Conn} et~al.}{2012}]{Conn2012}
{Conn} A.~R.,  {Ibata} R.~A.,  {Lewis} G.~F.,  {Parker} Q.~A.,  {Zucker} D.~B.,
   {Martin} N.~F.,  {McConnachie} A.~W.,  {Irwin} M.~J.,  {Tanvir} N.,
  {Fardal} M.~A.,  {Ferguson} A.~M.~N.,  {Chapman} S.~C.,    {Valls-Gabaud} D.,
   2012, \apj, 758, 11

\bibitem[\protect\citeauthoryear{{Conn}, {Lewis}, {Ibata}, {Parker}, {Zucker},
  {McConnachie}, {Martin}, {Irwin}, {Tanvir}, {Fardal} \& {Ferguson}}{{Conn}
  et~al.}{2011}]{Conn2011}
{Conn} A.~R.,  {Lewis} G.~F.,  {Ibata} R.~A.,  {Parker} Q.~A.,  {Zucker} D.~B.,
   {McConnachie} A.~W.,  {Martin} N.~F.,  {Irwin} M.~J.,  {Tanvir} N.,
  {Fardal} M.~A.,    {Ferguson} A.~M.~N.,  2011, \apj, 740, 69

\bibitem[\protect\citeauthoryear{{Conn}, {Lewis}, {Ibata}, {Parker}, {Zucker},
  {McConnachie}, {Martin}, {Valls-Gabaud}, {Tanvir}, {Irwin}, {Ferguson} \&
  {Chapman}}{{Conn} et~al.}{2013}]{Conn2013}
{Conn} A.~R.,  {Lewis} G.~F.,  {Ibata} R.~A.,  {Parker} Q.~A.,  {Zucker} D.~B.,
   {McConnachie} A.~W.,  {Martin} N.~F.,  {Valls-Gabaud} D.,  {Tanvir} N.,
  {Irwin} M.~J.,  {Ferguson} A.~M.~N.,    {Chapman} S.~C.,  2013, ArXiv
  e-prints

\bibitem[\protect\citeauthoryear{{Connors}, {Kawata} \& {Gibson}}{{Connors}
  et~al.}{2006}]{Connors2006}
{Connors} T.~W.,  {Kawata} D.,    {Gibson} B.~K.,  2006, \mnras, 371, 108

\bibitem[\protect\citeauthoryear{{Dabringhausen} \& {Kroupa}}{{Dabringhausen}
  \& {Kroupa}}{2013}]{Dabringhausen2013}
{Dabringhausen} J.,  {Kroupa} P.,  2013, \mnras, 429, 1858

\bibitem[\protect\citeauthoryear{{de Vaucouleurs}}{{de
  Vaucouleurs}}{1958}]{deVaucouleurs1958}
{de Vaucouleurs} G.,  1958, \apj, 128, 465

\bibitem[\protect\citeauthoryear{{de Vaucouleurs}, {de Vaucouleurs}, {Corwin}
  Jr., {Buta}, {Paturel} \& {Fouqu{\'e}}}{{de Vaucouleurs}
  et~al.}{1991}]{deVaucouleurs1991}
{de Vaucouleurs} G.,  {de Vaucouleurs} A.,  {Corwin} Jr. H.~G.,  {Buta} R.~J.,
  {Paturel} G.,    {Fouqu{\'e}} P.,  1991, {Third Reference Catalogue of Bright
  Galaxies. Volume I: Explanations and references. Volume II: Data for galaxies
  between 0$^{h}$ and 12$^{h}$. Volume III: Data for galaxies between 12$^{h}$
  and 24$^{h}$.}

\bibitem[\protect\citeauthoryear{{Deason}, {McCarthy}, {Font}, {Evans},
  {Frenk}, {Belokurov}, {Libeskind}, {Crain} \& {Theuns}}{{Deason}
  et~al.}{2011}]{Deason2011}
{Deason} A.~J.,  {McCarthy} I.~G.,  {Font} A.~S.,  {Evans} N.~W.,  {Frenk}
  C.~S.,  {Belokurov} V.,  {Libeskind} N.~I.,  {Crain} R.~A.,    {Theuns} T.,
  2011, \mnras, 415, 2607

\bibitem[\protect\citeauthoryear{{D'Onghia} \& {Lake}}{{D'Onghia} \&
  {Lake}}{2008}]{DOnghiaLake2008}
{D'Onghia} E.,  {Lake} G.,  2008, \apjl, 686, L61

\bibitem[\protect\citeauthoryear{{Dubinski} \& {Carlberg}}{{Dubinski} \&
  {Carlberg}}{1991}]{Dubinski1991}
{Dubinski} J.,  {Carlberg} R.~G.,  1991, \apj, 378, 496

\bibitem[\protect\citeauthoryear{{Duc} et~al.}{{Duc} et~al.}{2011}]{Duc2011}
{Duc} P.-A. et al.,  2011, \mnras, 417, 863

\bibitem[\protect\citeauthoryear{{Duc} \& {Mirabel}}{{Duc} \&
  {Mirabel}}{1998}]{Duc1998}
{Duc} P.-A.,  {Mirabel} I.~F.,  1998, \aap, 333, 813

\bibitem[\protect\citeauthoryear{{Elmegreen}, {Kaufman} \&
  {Thomasson}}{{Elmegreen} et~al.}{1993}]{Elmegreen1993}
{Elmegreen} B.~G.,  {Kaufman} M.,    {Thomasson} M.,  1993, \apj, 412, 90

\bibitem[\protect\citeauthoryear{{Famaey} \& {McGaugh}}{{Famaey} \&
  {McGaugh}}{2012}]{FamaeyMcGaugh2012}
{Famaey} B.,  {McGaugh} S.~S.,  2012, Living Reviews in Relativity, 15, 10

\bibitem[\protect\citeauthoryear{{Fattahi}, {Navarro}, {Starkenburg}, {Barber}
  \& {McConnachie}}{{Fattahi} et~al.}{2013}]{Fattahi2013}
{Fattahi} A.,  {Navarro} J.~F.,  {Starkenburg} E.,  {Barber} C.~R.,
  {McConnachie} A.~W.,  2013, \mnras, 431, L73

\bibitem[\protect\citeauthoryear{{Fouquet}, {Hammer}, {Yang}, {Puech} \&
  {Flores}}{{Fouquet} et~al.}{2012}]{Fouquet2012}
{Fouquet} S.,  {Hammer} F.,  {Yang} Y.,  {Puech} M.,    {Flores} H.,  2012,
  \mnras, 427, 1769

\bibitem[\protect\citeauthoryear{{Gott} III \& {Thuan}}{{Gott} \&
  {Thuan}}{1978}]{Gott1978}
{Gott} III J.~R.,  {Thuan} T.~X.,  1978, \apj, 223, 426

\bibitem[\protect\citeauthoryear{{Guhathakurta} \& {Reitzel}}{{Guhathakurta} \&
  {Reitzel}}{1998}]{Guhathakurta1998}
{Guhathakurta} P.,  {Reitzel} D.~B.,  1998, in {Zaritsky} D.,  ed., Galactic
  Halos Vol.~136 of Astronomical Society of the Pacific Conference Series,
  {Local Group Suburbia: Red Giants in M31's Outer Spheroid and a Search for
  Stars in the Magellanic Stream}.
p.~22

\bibitem[\protect\citeauthoryear{{Hammer}, {Yang}, {Fouquet}, {Pawlowski},
  {Kroupa}, {Puech}, {Flores} \& {Wang}}{{Hammer} et~al.}{2013}]{Hammer2013}
{Hammer} F.,  {Yang} Y.,  {Fouquet} S.,  {Pawlowski} M.~S.,  {Kroupa} P.,
  {Puech} M.,  {Flores} H.,    {Wang} J.,  2013, \mnras

\bibitem[\protect\citeauthoryear{{Hammer}, {Yang}, {Wang}, {Puech}, {Flores} \&
  {Fouquet}}{{Hammer} et~al.}{2010}]{Hammer2010}
{Hammer} F.,  {Yang} Y.~B.,  {Wang} J.~L.,  {Puech} M.,  {Flores} H.,
  {Fouquet} S.,  2010, \apj, 725, 542

\bibitem[\protect\citeauthoryear{{Hartwick}}{{Hartwick}}{2000}]{Hartwick2000}
{Hartwick} F.~D.~A.,  2000, \aj, 119, 2248

\bibitem[\protect\citeauthoryear{{Hills}}{{Hills}}{1988}]{Hills1988}
{Hills} J.~G.,  1988, \nat, 331, 687

\bibitem[\protect\citeauthoryear{{Ibata}, {Lewis}, {Conn}, {Irwin},
  {McConnachie}, {Chapman}, {Collins}, {Fardal}, {Ferguson}, {Ibata}, {Mackey},
  {Martin}, {Navarro}, {Rich}, {Valls-Gabaud} \& {Widrow}}{{Ibata}
  et~al.}{2013}]{Ibata2013}
{Ibata} R.~A.,  {Lewis} G.~F.,  {Conn} A.~R.,  {Irwin} M.~J.,  {McConnachie}
  A.~W.,  {Chapman} S.~C.,  {Collins} M.~L.,  {Fardal} M.,  {Ferguson}
  A.~M.~N.,  {Ibata} N.~G.,  {Mackey} A.~D.,  {Martin} N.~F.,  {Navarro} J.,
  {Rich} R.~M.,  {Valls-Gabaud} D.,    {Widrow} L.~M.,  2013, \nat, 493, 62

\bibitem[\protect\citeauthoryear{{Jerjen}}{{Jerjen}}{2010}]{Jerjen2010}
{Jerjen} H.,  2010, Advances in Astronomy, 2010

\bibitem[\protect\citeauthoryear{{Jerjen}}{{Jerjen}}{2012}]{Jerjen2012}
{Jerjen} H.,  2012, \pasa, 29, 383

\bibitem[\protect\citeauthoryear{{Jerjen}, {Freeman} \& {Binggeli}}{{Jerjen}
  et~al.}{1998}]{Jerjen1998}
{Jerjen} H.,  {Freeman} K.~C.,    {Binggeli} B.,  1998, \aj, 116, 2873

\bibitem[\protect\citeauthoryear{{Jerjen} \& {Tammann}}{{Jerjen} \&
  {Tammann}}{1993}]{Jerjen1993}
{Jerjen} H.,  {Tammann} G.~A.,  1993, \aap, 276, 1

\bibitem[\protect\citeauthoryear{{Kallivayalil}, {van der Marel}, {Besla},
  {Anderson} \& {Alcock}}{{Kallivayalil} et~al.}{2013}]{Kallivayalil2013}
{Kallivayalil} N.,  {van der Marel} R.~P.,  {Besla} G.,  {Anderson} J.,
  {Alcock} C.,  2013, \apj, 764, 161

\bibitem[\protect\citeauthoryear{{Kang}, {Mao}, {Gao} \& {Jing}}{{Kang}
  et~al.}{2005}]{Kang2005}
{Kang} X.,  {Mao} S.,  {Gao} L.,    {Jing} Y.~P.,  2005, \aap, 437, 383

\bibitem[\protect\citeauthoryear{{Karachentsev}, {Kashibadze}, {Makarov} \&
  {Tully}}{{Karachentsev} et~al.}{2009}]{Karachentsev2009}
{Karachentsev} I.~D.,  {Kashibadze} O.~G.,  {Makarov} D.~I.,    {Tully} R.~B.,
  2009, \mnras, 393, 1265

\bibitem[\protect\citeauthoryear{{Keller}, {Mackey} \& {Da Costa}}{{Keller}
  et~al.}{2012}]{Keller2012}
{Keller} S.~C.,  {Mackey} D.,    {Da Costa} G.~S.,  2012, \apj, 744, 57

\bibitem[\protect\citeauthoryear{{Klessen} \& {Kroupa}}{{Klessen} \&
  {Kroupa}}{1998}]{Klessen1998}
{Klessen} R.~S.,  {Kroupa} P.,  1998, \apj, 498, 143

\bibitem[\protect\citeauthoryear{{Klypin}, {Kravtsov}, {Valenzuela} \&
  {Prada}}{{Klypin} et~al.}{1999}]{Klypin1999}
{Klypin} A.,  {Kravtsov} A.~V.,  {Valenzuela} O.,    {Prada} F.,  1999, \apj,
  522, 82

\bibitem[\protect\citeauthoryear{{Koch} \& {Grebel}}{{Koch} \&
  {Grebel}}{2006}]{Koch2006}
{Koch} A.,  {Grebel} E.~K.,  2006, \aj, 131, 1405

\bibitem[\protect\citeauthoryear{{Kogut} et~al.}{{Kogut} et~al.}{1993}]{Kogut1993}
{Kogut} A. et al.,  1993, \apj, 419, 1

\bibitem[\protect\citeauthoryear{{Kroupa}}{{Kroupa}}{1997}]{Kroupa1997}
{Kroupa} P.,  1997, \na, 2, 139

\bibitem[\protect\citeauthoryear{{Kroupa}}{{Kroupa}}{2012}]{Kroupa2012a}
{Kroupa} P.,  2012, \pasa, 29, 395

\bibitem[\protect\citeauthoryear{{Kroupa}, {Famaey}, {de Boer},
  {Dabringhausen}, {Pawlowski}, {Boily}, {Jerjen}, {Forbes}, {Hensler} \&
  {Metz}}{{Kroupa} et~al.}{2010}]{Kroupa2010}
{Kroupa} P.,  {Famaey} B.,  {de Boer} K.~S.,  {Dabringhausen} J.,  {Pawlowski}
  M.~S.,  {Boily} C.~M.,  {Jerjen} H.,  {Forbes} D.,  {Hensler} G.,    {Metz}
  M.,  2010, \aap, 523, A32+

\bibitem[\protect\citeauthoryear{{Kroupa}, {Pawlowski} \& {Milgrom}}{{Kroupa}
  et~al.}{2012}]{Kroupa2012b}
{Kroupa} P.,  {Pawlowski} M.,    {Milgrom} M.,  2012, International Journal of
  Modern Physics D, 21, 30003

\bibitem[\protect\citeauthoryear{{Kroupa}, {Theis} \& {Boily}}{{Kroupa}
  et~al.}{2005}]{Kroupa2005}
{Kroupa} P.,  {Theis} C.,    {Boily} C.~M.,  2005, \aap, 431, 517

\bibitem[\protect\citeauthoryear{{Kunkel} \& {Demers}}{{Kunkel} \&
  {Demers}}{1976}]{Kunkel1976}
{Kunkel} W.~E.,  {Demers} S.,  1976, in {Dickens} R.~J.,  {Perry} J.~E.,
  {Smith} F.~G.,   {King} I.~R.,  eds, The Galaxy and the Local Group Vol.~182
  of Royal Greenwich Observatory Bulletins, {The Magellanic Plane}.
p.~241

\bibitem[\protect\citeauthoryear{{Lahav}, {Santiago}, {Webster}, {Strauss},
  {Davis}, {Dressler} \& {Huchra}}{{Lahav} et~al.}{2000}]{Lahav2000}
{Lahav} O.,  {Santiago} B.~X.,  {Webster} A.~M.,  {Strauss} M.~A.,  {Davis} M.,
   {Dressler} A.,    {Huchra} J.~P.,  2000, \mnras, 312, 166

\bibitem[\protect\citeauthoryear{{Li} \& {Helmi}}{{Li} \&
  {Helmi}}{2008}]{LiHelmi2008}
{Li} Y.,  {Helmi} A.,  2008, \mnras, 385, 1365

\bibitem[\protect\citeauthoryear{{Libeskind}, {Frenk}, {Cole}, {Helly},
  {Jenkins}, {Navarro} \& {Power}}{{Libeskind} et~al.}{2005}]{Libeskind2005}
{Libeskind} N.~I.,  {Frenk} C.~S.,  {Cole} S.,  {Helly} J.~C.,  {Jenkins} A.,
  {Navarro} J.~F.,    {Power} C.,  2005, \mnras, 363, 146

\bibitem[\protect\citeauthoryear{{Libeskind}, {Knebe}, {Hoffman},
  {Gottl{\"o}ber}, {Yepes} \& {Steinmetz}}{{Libeskind}
  et~al.}{2011}]{Libeskind2011}
{Libeskind} N.~I.,  {Knebe} A.,  {Hoffman} Y.,  {Gottl{\"o}ber} S.,  {Yepes}
  G.,    {Steinmetz} M.,  2011, \mnras, 411, 1525

\bibitem[\protect\citeauthoryear{{Libeskind}, {Yepes}, {Knebe},
  {Gottl{\"o}ber}, {Hoffman} \& {Knollmann}}{{Libeskind}
  et~al.}{2010}]{Libeskind2010}
{Libeskind} N.~I.,  {Yepes} G.,  {Knebe} A.,  {Gottl{\"o}ber} S.,  {Hoffman}
  Y.,    {Knollmann} S.~R.,  2010, \mnras, 401, 1889

\bibitem[\protect\citeauthoryear{{Lovell}, {Eke}, {Frenk} \&
  {Jenkins}}{{Lovell} et~al.}{2011}]{Lovell2011}
{Lovell} M.~R.,  {Eke} V.~R.,  {Frenk} C.~S.,    {Jenkins} A.,  2011, \mnras,
  413, 3013

\bibitem[\protect\citeauthoryear{{Lynden-Bell}}{{Lynden-Bell}}{1976}]{LyndenBe%
ll1976}
{Lynden-Bell} D.,  1976, \mnras, 174, 695

\bibitem[\protect\citeauthoryear{{Lynden-Bell}}{{Lynden-Bell}}{1982}]{LyndenBe%
ll1982}
{Lynden-Bell} D.,  1982, The Observatory, 102, 202

\bibitem[\protect\citeauthoryear{{Mackey} \& {van den Bergh}}{{Mackey} \& {van
  den Bergh}}{2005}]{Mackey05}
{Mackey} A.~D.,  {van den Bergh} S.,  2005, \mnras, 360, 631

\bibitem[\protect\citeauthoryear{{Majewski}}{{Majewski}}{1994}]{Majewski1994}
{Majewski} S.~R.,  1994, \apjl, 431, L17

\bibitem[\protect\citeauthoryear{{Martin}, {Ibata}, {Bellazzini}, {Irwin},
  {Lewis} \& {Dehnen}}{{Martin} et~al.}{2004}]{Martin2004}
{Martin} N.~F.,  {Ibata} R.~A.,  {Bellazzini} M.,  {Irwin} M.~J.,  {Lewis}
  G.~F.,    {Dehnen} W.,  2004, \mnras, 348, 12

\bibitem[\protect\citeauthoryear{{Martin} et~al.}{{Martin} et~al.}{2011}]{Martin2013}
{Martin} N.~F. et al.,  2013, \apj, 772, 15

\bibitem[\protect\citeauthoryear{{Mastropietro}}{{Mastropietro}}{2009}]{Mastro%
pietro2009}
{Mastropietro} C.,  2009, in {Van Loon} J.~T.,  {Oliveira} J.~M.,  eds, IAU
  Symposium Vol.~256 of IAU Symposium, {Modeling a high velocity LMC: The
  formation of the Magellanic Stream}.
pp 117--121

\bibitem[\protect\citeauthoryear{{Mastropietro}, {Moore}, {Mayer}, {Wadsley} \&
  {Stadel}}{{Mastropietro} et~al.}{2005}]{Mastropietro2005}
{Mastropietro} C.,  {Moore} B.,  {Mayer} L.,  {Wadsley} J.,    {Stadel} J.,
  2005, \mnras, 363, 509

\bibitem[\protect\citeauthoryear{{Mateo}}{{Mateo}}{1998}]{Mateo1998}
{Mateo} M.~L.,  1998, \araa, 36, 435

\bibitem[\protect\citeauthoryear{{Mathewson}, {Cleary} \& {Murray}}{{Mathewson}
  et~al.}{1974}]{Mathewson1974}
{Mathewson} D.~S.,  {Cleary} M.~N.,    {Murray} J.~D.,  1974, \apj, 190, 291

\bibitem[\protect\citeauthoryear{{McConnachie}}{{McConnachie}}{2012}]{McConnac%
hie2012}
{McConnachie} A.~W.,  2012, \aj, 144, 4

\bibitem[\protect\citeauthoryear{{McConnachie} \& {C{\^o}t{\'e}}}{{McConnachie}
  \& {C{\^o}t{\'e}}}{2010}]{McConnachie2010}
{McConnachie} A.~W.,  {C{\^o}t{\'e}} P.,  2010, \apjl, 722, L209

\bibitem[\protect\citeauthoryear{{McConnachie} \& {Irwin}}{{McConnachie} \&
  {Irwin}}{2006}]{McConnachie2006}
{McConnachie} A.~W.,  {Irwin} M.~J.,  2006, \mnras, 365, 902

\bibitem[\protect\citeauthoryear{{McConnachie}, {Irwin}, {Ferguson}, {Ibata},
  {Lewis} \& {Tanvir}}{{McConnachie} et~al.}{2005}]{McConnachie2005}
{McConnachie} A.~W.,  {Irwin} M.~J.,  {Ferguson} A.~M.~N.,  {Ibata} R.~A.,
  {Lewis} G.~F.,    {Tanvir} N.,  2005, \mnras, 356, 979

\bibitem[\protect\citeauthoryear{{McConnachie} et~al.}{{McConnachie} et~al.}{2009}]{McConnachie2009}
{McConnachie} A.~W. et al.,  2009, \nat, 461, 66

\bibitem[\protect\citeauthoryear{{McGaugh} \& {Milgrom}}{{McGaugh} \&
  {Milgrom}}{2013}]{McGaugh2013}
{McGaugh} S.,  {Milgrom} M.,  2013, \apj, 766, 22

\bibitem[\protect\citeauthoryear{{McGaugh} \& {Wolf}}{{McGaugh} \&
  {Wolf}}{2010}]{McGaugh2010}
{McGaugh} S.~S.,  {Wolf} J.,  2010, \apj, 722, 248

\bibitem[\protect\citeauthoryear{{McMillan}}{{McMillan}}{2011}]{McMillan2011}
{McMillan} P.~J.,  2011, \mnras, 414, 2446

\bibitem[\protect\citeauthoryear{{Metz}, {Kroupa} \& {Jerjen}}{{Metz}
  et~al.}{2007}]{Metz2007}
{Metz} M.,  {Kroupa} P.,    {Jerjen} H.,  2007, \mnras, 374, 1125

\bibitem[\protect\citeauthoryear{{Metz}, {Kroupa} \& {Jerjen}}{{Metz}
  et~al.}{2009}]{Metz2009}
{Metz} M.,  {Kroupa} P.,    {Jerjen} H.,  2009, \mnras, 394, 2223

\bibitem[\protect\citeauthoryear{{Metz}, {Kroupa} \& {Libeskind}}{{Metz}
  et~al.}{2008}]{Metz2008}
{Metz} M.,  {Kroupa} P.,    {Libeskind} N.~I.,  2008, \apj, 680, 287

\bibitem[\protect\citeauthoryear{{Metz}, {Kroupa}, {Theis}, {Hensler} \&
  {Jerjen}}{{Metz} et~al.}{2009}]{Metz2009b}
{Metz} M.,  {Kroupa} P.,  {Theis} C.,  {Hensler} G.,    {Jerjen} H.,  2009,
  \apj, 697, 269

\bibitem[\protect\citeauthoryear{{Momany}, {Zaggia}, {Gilmore}, {Piotto},
  {Carraro}, {Bedin} \& {de Angeli}}{{Momany} et~al.}{2006}]{Momany2006}
{Momany} Y.,  {Zaggia} S.,  {Gilmore} G.,  {Piotto} G.,  {Carraro} G.,  {Bedin}
  L.~R.,    {de Angeli} F.,  2006, \aap, 451, 515

\bibitem[\protect\citeauthoryear{{Moore}, {Ghigna}, {Governato}, {Lake},
  {Quinn}, {Stadel} \& {Tozzi}}{{Moore} et~al.}{1999}]{Moore1999}
{Moore} B.,  {Ghigna} S.,  {Governato} F.,  {Lake} G.,  {Quinn} T.,  {Stadel}
  J.,    {Tozzi} P.,  1999, \apjl, 524, L19

\bibitem[\protect\citeauthoryear{{Nidever}, {Majewski} \& {Burton}}{{Nidever}
  et~al.}{2008}]{Nidever2008}
{Nidever} D.~L.,  {Majewski} S.~R.,    {Burton} W.~B.,  2008, \apj, 679, 432

\bibitem[\protect\citeauthoryear{{Nidever}, {Majewski}, {Butler Burton} \&
  {Nigra}}{{Nidever} et~al.}{2010}]{Nidever2010}
{Nidever} D.~L.,  {Majewski} S.~R.,  {Butler Burton} W.,    {Nigra} L.,  2010,
  \apj, 723, 1618

\bibitem[\protect\citeauthoryear{{Pasetto} \& {Chiosi}}{{Pasetto} \&
  {Chiosi}}{2007}]{Pasetto2007}
{Pasetto} S.,  {Chiosi} C.,  2007, \aap, 463, 427

\bibitem[\protect\citeauthoryear{{Pasetto} \& {Chiosi}}{{Pasetto} \&
  {Chiosi}}{2009}]{Pasetto2009}
{Pasetto} S.,  {Chiosi} C.,  2009, \aap, 499, 385

\bibitem[\protect\citeauthoryear{{Pawlowski}, {Kroupa}, {Angus}, {de Boer},
  {Famaey} \& {Hensler}}{{Pawlowski} et~al.}{2012b}]{Pawlowski2012b}
{Pawlowski} M.~S.,  {Kroupa} P.,  {Angus} G.,  {de Boer} K.~S.,  {Famaey} B.,
   {Hensler} G.,  2012b, \mnras, 424, 80

\bibitem[\protect\citeauthoryear{{Pawlowski}, {Kroupa} \& {de
  Boer}}{{Pawlowski} et~al.}{2011}]{Pawlowski2011}
{Pawlowski} M.~S.,  {Kroupa} P.,    {de Boer} K.~S.,  2011, \aap, 532, A118

\bibitem[\protect\citeauthoryear{{Pawlowski} \& {Kroupa}}{{Pawlowski} \& {Kroupa}}{2013}]{Pawlowski2013}
{Pawlowski} M.~S.,  {Kroupa} P.,  2013, submitted to MNRAS

\bibitem[\protect\citeauthoryear{{Pawlowski}, {Pflamm-Altenburg} \&
  {Kroupa}}{{Pawlowski} et~al.}{2012a}]{Pawlowski2012a}
{Pawlowski} M.~S.,  {Pflamm-Altenburg} J.,    {Kroupa} P.,  2012a, \mnras, 423,
  1109

\bibitem[\protect\citeauthoryear{{Piffl}, {Williams} \& {Steinmetz}}{{Piffl}
  et~al.}{2011}]{Piffl2011}
{Piffl} T.,  {Williams} M.,    {Steinmetz} M.,  2011, \aap, 535, A70

\bibitem[\protect\citeauthoryear{{Raychaudhury} \&
  {Lynden-Bell}}{{Raychaudhury} \& {Lynden-Bell}}{1989}]{Raychaudhury1989}
{Raychaudhury} S.,  {Lynden-Bell} D.,  1989, \mnras, 240, 195

\bibitem[\protect\citeauthoryear{{Recchi}, {Theis}, {Kroupa} \&
  {Hensler}}{{Recchi} et~al.}{2007}]{Recchi2007}
{Recchi} S.,  {Theis} C.,  {Kroupa} P.,    {Hensler} G.,  2007, \aap, 470, L5

\bibitem[\protect\citeauthoryear{{Sawa} \& {Fujimoto}}{{Sawa} \&
  {Fujimoto}}{2005}]{Sawa2005}
{Sawa} T.,  {Fujimoto} M.,  2005, \pasj, 57, 429

\bibitem[\protect\citeauthoryear{{Sch{\"o}nrich}, {Binney} \&
  {Dehnen}}{{Sch{\"o}nrich} et~al.}{2010}]{Schoenrich2010}
{Sch{\"o}nrich} R.,  {Binney} J.,    {Dehnen} W.,  2010, \mnras, 403, 1829

\bibitem[\protect\citeauthoryear{{Sohn}, {Anderson} \& {van der Marel}}{{Sohn}
  et~al.}{2012}]{Sohn2012}
{Sohn} S.~T.,  {Anderson} J.,    {van der Marel} R.~P.,  2012, \apj, 753, 7

\bibitem[\protect\citeauthoryear{{Springel}, {Wang}, {Vogelsberger}, {Ludlow},
  {Jenkins}, {Helmi}, {Navarro}, {Frenk} \& {White}}{{Springel}
  et~al.}{2008}]{Springel2008}
{Springel} V.,  {Wang} J.,  {Vogelsberger} M.,  {Ludlow} A.,  {Jenkins} A.,
  {Helmi} A.,  {Navarro} J.~F.,  {Frenk} C.~S.,    {White} S.~D.~M.,  2008,
  \mnras, 391, 1685

\bibitem[\protect\citeauthoryear{{Starkenburg}, {Helmi}, {De Lucia}, {Li},
  {Navarro}, {Font}, {Frenk}, {Springel}, {Vera-Ciro} \& {White}}{{Starkenburg}
  et~al.}{2013}]{Starkenburg2013}
{Starkenburg} E.,  {Helmi} A.,  {De Lucia} G.,  {Li} Y.-S.,  {Navarro} J.~F.,
  {Font} A.~S.,  {Frenk} C.~S.,  {Springel} V.,  {Vera-Ciro} C.~A.,    {White}
  S.~D.~M.,  2013, \mnras, 429, 725

\bibitem[\protect\citeauthoryear{{Strigari}, {Bullock}, {Kaplinghat}, {Simon},
  {Geha}, {Willman} \& {Walker}}{{Strigari} et~al.}{2008}]{Strigari2008}
{Strigari} L.~E.,  {Bullock} J.~S.,  {Kaplinghat} M.,  {Simon} J.~D.,  {Geha}
  M.,  {Willman} B.,    {Walker} M.~G.,  2008, \nat, 454, 1096

\bibitem[\protect\citeauthoryear{{Teyssier}, {Johnston} \& {Shara}}{{Teyssier}
  et~al.}{2009}]{Teyssier2009}
{Teyssier} M.,  {Johnston} K.~V.,    {Shara} M.~M.,  2009, \apjl, 707, L22

\bibitem[\protect\citeauthoryear{{Tollerud}, {Geha}, {Vargas} \&
  {Bullock}}{{Tollerud} et~al.}{2013}]{Tollerud2013}
{Tollerud} E.~J.,  {Geha} M.~C.,  {Vargas} L.~C.,    {Bullock} J.~S.,  2013,
  \apj, 768, 50

\bibitem[\protect\citeauthoryear{{Tully}}{{Tully}}{2013}]{Tully2013}
{Tully} R.~B.,  2013, \nat, 493, 31

\bibitem[\protect\citeauthoryear{{Tully}, {Shaya}, {Karachentsev}, {Courtois},
  {Kocevski}, {Rizzi} \& {Peel}}{{Tully} et~al.}{2008}]{Tully2008}
{Tully} R.~B.,  {Shaya} E.~J.,  {Karachentsev} I.~D.,  {Courtois} H.~M.,
  {Kocevski} D.~D.,  {Rizzi} L.,    {Peel} A.,  2008, \apj, 676, 184

\bibitem[\protect\citeauthoryear{{van der Marel}, {Alves}, {Hardy} \&
  {Suntzeff}}{{van der Marel} et~al.}{2002}]{vdMarel2002}
{van der Marel} R.~P.,  {Alves} D.~R.,  {Hardy} E.,    {Suntzeff} N.~B.,  2002,
  \aj, 124, 2639

\bibitem[\protect\citeauthoryear{{van der Marel}, {Besla}, {Cox}, {Sohn} \&
  {Anderson}}{{van der Marel} et~al.}{2012}]{vdMarel2012b}
{van der Marel} R.~P.,  {Besla} G.,  {Cox} T.~J.,  {Sohn} S.~T.,    {Anderson}
  J.,  2012, \apj, 753, 9

\bibitem[\protect\citeauthoryear{{van der Marel}, {Fardal}, {Besla}, {Beaton},
  {Sohn}, {Anderson}, {Brown} \& {Guhathakurta}}{{van der Marel}
  et~al.}{2012}]{vdMarel2012a}
{van der Marel} R.~P.,  {Fardal} M.,  {Besla} G.,  {Beaton} R.~L.,  {Sohn}
  S.~T.,  {Anderson} J.,  {Brown} T.,    {Guhathakurta} P.,  2012, \apj, 753, 8

\bibitem[\protect\citeauthoryear{{van der Marel} \& {Guhathakurta}}{{van der
  Marel} \& {Guhathakurta}}{2008}]{vdMarel2008}
{van der Marel} R.~P.,  {Guhathakurta} P.,  2008, \apj, 678, 187

\bibitem[\protect\citeauthoryear{{Vera-Ciro}, {Sales}, {Helmi}, {Frenk},
  {Navarro}, {Springel}, {Vogelsberger} \& {White}}{{Vera-Ciro}
  et~al.}{2011}]{Vera-Ciro2011}
{Vera-Ciro} C.~A.,  {Sales} L.~V.,  {Helmi} A.,  {Frenk} C.~S.,  {Navarro}
  J.~F.,  {Springel} V.,  {Vogelsberger} M.,    {White} S.~D.~M.,  2011,
  \mnras, 416, 1377

\bibitem[\protect\citeauthoryear{{Wang}, {Frenk} \& {Cooper}}{{Wang}
  et~al.}{2013}]{Wang2013}
{Wang} J.,  {Frenk} C.~S.,    {Cooper} A.~P.,  2013, \mnras, 429, 1502

\bibitem[\protect\citeauthoryear{{Wannier} \& {Wrixon}}{{Wannier} \&
  {Wrixon}}{1972}]{Wannier1972}
{Wannier} P.,  {Wrixon} G.~T.,  1972, \apjl, 173, L119

\bibitem[\protect\citeauthoryear{{Westmeier} \& {Koribalski}}{{Westmeier} \&
  {Koribalski}}{2008}]{Westmeier2008}
{Westmeier} T.,  {Koribalski} B.~S.,  2008, \mnras, 388, L29

\bibitem[\protect\citeauthoryear{{Wetzstein}, {Naab} \& {Burkert}}{{Wetzstein}
  et~al.}{2007}]{Wetzstein2007}
{Wetzstein} M.,  {Naab} T.,    {Burkert} A.,  2007, \mnras, 375, 805

\bibitem[\protect\citeauthoryear{{Yang} \& {Hammer}}{{Yang} \&
  {Hammer}}{2010}]{Yang2010}
{Yang} Y.,  {Hammer} F.,  2010, \apjl, 725, L24

\bibitem[\protect\citeauthoryear{{Zentner}, {Kravtsov}, {Gnedin} \&
  {Klypin}}{{Zentner} et~al.}{2005}]{Zentner2005}
{Zentner} A.~R.,  {Kravtsov} A.~V.,  {Gnedin} O.~Y.,    {Klypin} A.~A.,  2005,
  \apj, 629, 219

\bibitem[\protect\citeauthoryear{{Zhao}, {Famaey}, {L{\"u}ghausen} \&
  {Kroupa}}{{Zhao} et~al.}{2013}]{Zhao2013}
{Zhao} H.,  {Famaey} B.,  {L{\"u}ghausen} F.,    {Kroupa} P.,  2013, A\&A
  accepted

\end{thebibliography}

\label{lastpage}

\end{document}